\documentclass[12pt]{article}
\usepackage{a4wide,epsfig,amsmath,amssymb,cite,scalefnt,graphicx}
\numberwithin{equation}{section}
\allowdisplaybreaks
\usepackage{tikz}
\usetikzlibrary{intersections}
\usetikzlibrary{calc}
\usetikzlibrary{shapes}

\newcommand{\vsq}{\node[fill=black,rectangle,inner sep=2pt,minimum size=0pt]}
\usepackage{tabularx}
\usepackage{graphics}
\usepackage{array}

\def\Jcal{{\cal I}''}

\def\be{\begin{equation}}
\def\ee{\end{equation}}
\def\nn{\nonumber\\}

\def\msbar{{\overline{\rm MS}}}

\def\Kcal{{\cal{K}}}
\def\dtil{{\tilde {\delta}}^{IR}}
\def\Rbar{\overline{R}}

\def\eps{\epsilon}

\def\frakk[#1#2{{{#1}\over{#2}}}

\input amssym.def
\input amssym
\def\be{\begin{equation}}
\def\ee{\end{equation}}
\def\nn{\nonumber\\}

\def\cirk{\,{\raise1pt \hbox{${\scriptscriptstyle \circ}$}}\,}

\def \olr{{\raise6.5pt\hbox{$\leftrightarrow  \! \! \! \! \!$}}}

\def\hbar{\bar h}

\def\ncal{{\cal{I}}}
\def\Ical{{\kappa}}
\font\ninerm=cmr9 \font\ninesy=cmsy9
\font\eightrm=cmr8 \font\sixrm=cmr6
\font\eighti=cmmi8 \font\sixi=cmmi6
\font\eightsy=cmsy8 \font\sixsy=cmsy6
\font\eightbf=cmbx8 \font\sixbf=cmbx6
\font\eightit=cmti8
\def\eightpoint{\def\rm{\fam0\eightrm}
  \textfont0=\eightrm \scriptfont0=\sixrm \scriptscriptfont0=\fiverm
  \textfont1=\eighti  \scriptfont1=\sixi  \scriptscriptfont1=\fivei
  \textfont2=\eightsy \scriptfont2=\sixsy \scriptscriptfont2=\fivesy
  \textfont3=\tenex   \scriptfont3=\tenex \scriptscriptfont3=\tenex
  \textfont\itfam=\eightit  \def\it{\fam\itfam\eightit}%
  \textfont\bffam=\eightbf  \scriptfont\bffam=\sixbf
  \scriptscriptfont\bffam=\fivebf  \def\bf{\fam\bffam\eightbf}%
  \normalbaselineskip=9pt
  \setbox\strutbox=\hbox{\vrule height7pt depth2pt width0pt}%
  \let\big=\eightbig  \normalbaselines\rm}
\catcode`@=11 %
\def\eightbig#1{{\hbox{$\textfont0=\ninerm\textfont2=\ninesy
  \left#1\vbox to6.5pt{}\right.\n@@space$}}}
\def\vfootnote#1{\insert\footins\bgroup\eightpoint
  \interlinepenalty=\interfootnotelinepenalty
  \splittopskip=\ht\strutbox %
  \splitmaxdepth=\dp\strutbox %
  \leftskip=0pt \rightskip=0pt \spaceskip=0pt \xspaceskip=0pt
  \textindent{#1}\footstrut\futurelet\next\fo@t}
\catcode`@=12 %
\def\today{\number\day\ \ifcase\month\or January\or February\or March\or
April\or May\or June\or July\or
August\or September\or October\or November\or December\fi, \number\year}

\input epsf

\begin{document}

\begin{titlepage}
\begin{flushright}
LTH1393
\end{flushright}
\date{}
\vspace*{3mm}

\begin{center}
{\Huge An alternative formulation of infra-red counterterms}\\[12mm]
{\bf I.~Jack}\\
%\end{center}

\vspace{5mm}
Dept. of Mathematical Sciences,
University of Liverpool, Liverpool L69 3BX, UK\\

\end{center}

\vspace{3mm}

\begin{abstract}
 We present an alternative procedure for defining infra-red counterterms within dimensional regularisation for use in the $R^*$ procedure. The counterterms are given by simple closed expressions, and lead to the standard $\msbar$ UV counterterms.
\end{abstract}

\vfill

\end{titlepage}
\section{Introduction}
 In order to simplify diagrammatic computations of renormalisation group functions, one often wants to consider massless propagators and also minimise the number of external momenta, but these practical expedients tend to introduce infra-red (IR) divergences. One can postpone the issue by judicious use of infra-red rearrangement\cite{vladim}, where one carefully chooses momenta to ``nulllify'' (i.e. set to zero) with the reassurance that the final result for the ultraviolet (UV) counterterm is unaffected by the details of this choice; but as one increases the loop order, at some point IR divergences become unavoidable. For a scalar $\phi^4$ theory, this appears to be at five-loop order. An algorithm has been developed\cite{chet1,chet2,chet3,chet3a} for defining IR counterterms within dimensional regularisation. This algorithm can be reduced to a fairly simple prescription for computing the IR counterterm from a related UV counterterm, using a procedure somewhat analogous to the standard $\Rbar$ procedure for obtaining a UV counterterm for a given divergent diagram from the momentum integrals for the diagram and its divergent subdiagrams. The method has been used in classic computations such as the five-loop and six-loop anomalous dimensions in $\phi^4$ theory\cite{chet4,chet5,chet6, panzer,chet7}; and the computation of the five-loop QCD $\beta$-function\cite{chet8,herz1}. More recently the issue of IR divergences has been discussed in the Hopf algebra context\cite{herz2,herz3}. Here we shall present an alternative definition of IR counterterms which is arguably even simpler than the afore-mentioned approach. The current drawbacks are that we happened upon it by trial and error, so we do not yet have a general proof of its validity; and it only shows its full simplicity for the case of IR-divergent diagrams with a double propagator (in fact we do not yet have a proposal for the case of a triple propagator). We emphasise that our alternative IR (which we shall abbreviate to AIR) counterterms are different from their standard counterparts, but give the same results when used in UV counterterm computations.

The contents are organised as follows: in Section 2 we give a brief pedagogical introduction to the standard treatment of IR counterterms. In Section 3 we introduce our alternative proposal for the majority of the IR counterterms (those with double propagators), together with some simple examples at two and three loops. In Section 4 we go on to present some harder examples at four loops. In Section 5 we give some indications of how the AIR counterterms work in general, and make some remarks in the Conclusions regarding a complete proof. In Appendix A we give a full list of our AIR counterterms up to four loops, together with their standard counterparts. In Appendix B we show how the AIR counterterms give standard results at five loops. Finally in Appendix C we consider IR countertems for structures with only single propagators, which also leads to a variant of our results for double propagators.

\section{The $R^*$ procedure}
The standard $R^*$ procedure for extracting infra-red counterterms within the context of dimensional regularisation and using them to regulate infra-red divergences was developed in Refs.~\cite{chet1,chet2} and refined in Ref.~\cite{chet3}. It has been widely used, for instance in the computation of the five-loop $\beta$-function for $\phi^4$ theory; its application here was well explained in Ref.~\cite{klein}; see also Ref.~\cite{larin} for a pedagogical description with lots of examples. Before describing the $R^*$ operation, it will be useful to introduce the standard procedures and notation of regularisation more generally. As usual we use dimensional regularisation so that we are working in $d=4-\epsilon$ dimensions and the divergences appear as poles in $\epsilon$. For a graph $G$ with only UV (and not IR) divergences, $R$ denotes the operation of adding counterterm graphs to $G$ in order to obtain a finite result, so that
\be
RG=\sum_{\Gamma}\Delta_{\Gamma}(G/\Gamma),
\ee
where $\Gamma$ is a divergent subgraph of $G$, $\Delta_{\Gamma}$ is the corresponding counterterm and $G/\Gamma$ is the result of shrinking $\Gamma$ to a point within $G$. We then denote by $\Kcal$ the operation of extracting the poles in $\epsilon$, so that we have by definition
\be
\Kcal R G=0.
\label{Rdef}
\ee

It is convenient to define the so-called ``incomplete $R$ operation'', denoted $\Rbar$, which omits the graph $G$ itself from the set $\Gamma$. We then have an expression for the UV counterterm for $G$ itself,
\be
\Delta_G=-\Kcal \Rbar G+O(1).
\label{Deldef}
\ee
Of course the standard minimal subtraction procedure is to omit the $O(1)$ terms here.
For a graph with IR as well as UV divergences, we define $R^*$ as a similar procedure but incorporating IR as well as UV counterterms. We then have in analogy with Eqs.~\eqref{Rdef}, \eqref{Deldef}
\begin{subequations}\label{rstar}
\begin{align}
\Kcal R^* G=&\,0,\label{rstar:a}\\
\Delta_G=&\,-\Kcal \Rbar^* G+O(1).\label{rstar:b}
\end{align}
\end{subequations}
Once again, the $O(1)$ terms are omitted in standard minimal subtraction. The procedure introduced in Refs.~\cite{chet1,chet2} and also followed in Ref.~\cite{klein} was to deduce the IR counterterms by examining simple superficially UV-convergent reference diagrams containing the IR divergences in question.  For instance in the lowest-order case, IR divergences come from the double propagator 
\be
\Ical_1=\tikz[scale=0.6,baseline=(vert_cent.base)]{
  \node (vert_cent) {\hspace{-13pt}$\phantom{-}$};
  \draw (0,-.4)--(0,.4);
        \draw (0,.5) circle [radius=0.1];
\draw (0,-.5) circle [radius=0.1];
\filldraw (0,0) circle [radius=0.1];
}
\ee
and the corresponding counterterm is represented
\be
\ncal_1=\left(\,
\tikz[scale=0.6,baseline=(vert_cent.base)]{
  \node (vert_cent) {\hspace{-13pt}$\phantom{-}$};
  \draw (0,-.4)--(0,.4);
        \draw (0,.5) circle [radius=0.1];
\draw (0,-.5) circle [radius=0.1];
\filldraw (0,0) circle [radius=0.1];
}\,\right)_{IR}
\ee
with the double propagator represented by a filled circle. Here we introduce our general notation that the IR-divergent structure will be denoted by a $\kappa$ with a subscript, and the corresponding IR counterterm will be denoted by $()_{IR}$ or by $\ncal$ with a subscript. The assumption made in
Refs.~\cite{chet1,chet2, klein} was that the IR counterterms, like the UV counterterms, contain only poles in $\epsilon$. We shall follow that assumption  in this Section for the case of the standard IR counterterms, but relax it in subsequent Sections when we introduce our AIR counterterms. A simple UV-convergent diagram containing $\kappa_1$ is 
\be
\tikz[scale=0.6,baseline=(vert_cent.base)]{
  \node (vert_cent) {\hspace{-13pt}$\phantom{-}$};
  \draw (-0.1,0)--(0.1,0)
        (0.7,0) ++(0:0.6cm) arc (0:360:0.6cm and 0.4cm)
        (1.3,0)--(1.5,0);
        \filldraw  (0.7,0.4) circle [radius=0.1];
}
\ee
and Eq.~\eqref{rstar:a} then takes the form
\be
\Kcal R^*\left(\tikz[scale=0.6,baseline=(vert_cent.base)]{
  \node (vert_cent) {\hspace{-13pt}$\phantom{-}$};
  \draw (-0.1,0)--(0.1,0)
        (0.7,0) ++(0:0.6cm) arc (0:360:0.6cm and 0.4cm)
        (1.3,0)--(1.5,0);
        \filldraw  (0.7,0.4) circle [radius=0.1];
}\right)= \Kcal\left[\tikz[scale=0.6,baseline=(vert_cent.base)]{
  \node (vert_cent) {\hspace{-13pt}$\phantom{-}$};
  \draw (-0.1,0)--(0.1,0)
        (0.7,0) ++(0:0.6cm) arc (0:360:0.6cm and 0.4cm)
        (1.3,0)--(1.5,0);
        \filldraw  (0.7,0.4) circle [radius=0.1];
}\right]+\left(\,
\tikz[scale=0.6,baseline=(vert_cent.base)]{
  \node (vert_cent) {\hspace{-13pt}$\phantom{-}$};
  \draw (0,-.4)--(0,.4);
        \draw (0,.5) circle [radius=0.1];
\draw (0,-.5) circle [radius=0.1];
\filldraw (0,0) circle [radius=0.1];
}\,\right)_{IR}\tikz[scale=0.6,baseline=(vert_cent.base),scale=1]{
	%\draw[help lines] (-3,-3) grid (3,3);
	\node (vert_cent) {\hspace{-13pt}$\phantom{-}$};
	\draw (-1.5,0)--(-1,0) (1,0)--(1.5,0);
\draw (0:1cm) arc (0:-180:1cm) ;
}=0.
\label{finite}
\ee
Of course Eq.~\eqref{finite} may be rearranged to give an expression for the IR counterterm, and the same procedure applied at higher orders, giving
\begin{subequations}\label{IRdiags}
\begin{align}
\ncal_1=(\Ical_1)_{IR}=\left(\,
\tikz[scale=0.6,baseline=(vert_cent.base)]{
  \node (vert_cent) {\hspace{-13pt}$\phantom{-}$};
  \draw (0,-.4)--(0,.4);
        \draw (0,.5) circle [radius=0.1];
\draw (0,-.5) circle [radius=0.1];
\filldraw (0,0) circle [radius=0.1];
}\,\right)_{IR}
=&\,-p^2\Kcal\left[\tikz[scale=0.6,baseline=(vert_cent.base)]{
  \node (vert_cent) {\hspace{-13pt}$\phantom{-}$};
  \draw (-0.1,0)--(0.1,0)
        (0.7,0) ++(0:0.6cm) arc (0:360:0.6cm and 0.4cm)
        (1.3,0)--(1.5,0);
        \filldraw  (0.7,0.4) circle [radius=0.1];
}\right],\\
\ncal_{2a}=(\Ical_{2a})_{IR}=\left(\tikz[scale=0.6,baseline=(vert_cent.base),scale=1]{
\draw(0,-1.4) -- (0,0);
\draw [bend right=-40] (0,0) to (-0.1,1.4);
\draw [bend left=-40] (0,0) to (0.1,1.4);
\filldraw  (0.0,-0.75) circle [radius=0.1];
\draw  (0.0,1.5) circle [radius=0.1];
\draw  (0.0,-1.5) circle [radius=0.1];
}\right)_{IR}=&\,
-p^2\Kcal\Bigl[\tikz[scale=0.6,baseline=(vert_cent.base),scale=1]{
	%\draw[help lines] (-3,-3) grid (3,3);
	\node (vert_cent) {\hspace{-13pt}$\phantom{-}$};
	\draw (-1.5,0)--(-1,0) (1,0)--(1.5,0);
	\draw (0,0) circle[radius=1cm];
	\draw [bend right=40] (-1,0) to (0,1);
\filldraw (45:1cm) circle [radius=0.1];
}+\Delta\left(\tikz[scale=0.6,baseline=(vert_cent.base)]{
  \node (vert_cent) {\hspace{-13pt}$\phantom{-}$};
  \filldraw (1.3,0) circle [radius=0.1]; \filldraw (0.1,0) circle [radius=0.1];    
  \draw  (0.7,0) ++(0:0.6cm) arc (0:360:0.6cm and 0.4cm);
}\right)\tikz[scale=0.6,baseline=(vert_cent.base)]{
  \node (vert_cent) {\hspace{-13pt}$\phantom{-}$};
  \draw (-0.1,0)--(0.1,0)
        (0.7,0) ++(0:0.6cm) arc (0:360:0.6cm and 0.4cm)
        (1.3,0)--(1.5,0);
        \filldraw  (0.7,0.4) circle [radius=0.1];
}\\
&+\left(\tikz[scale=0.6,baseline=(vert_cent.base)]{
  \node (vert_cent) {\hspace{-13pt}$\phantom{-}$};
  \draw (0,-.4)--(0,.4);
        \draw (0,.5) circle [radius=0.1];
\draw (0,-.5) circle [radius=0.1];
\filldraw (0,0) circle [radius=0.1];
}\,\right)_{IR}\left\{\tikz[scale=0.6,baseline=(vert_cent.base),scale=1]{
	%\draw[help lines] (-3,-3) grid (3,3);
	\node (vert_cent) {\hspace{-13pt}$\phantom{-}$};
	\draw (-1.5,0)--(-1,0) (1,0)--(1.5,0);
	\draw [bend right=40] (-1,0) to (0,1);
\draw [bend left=40] (-1,0) to (0,1);
\draw (0:1cm) arc (0:-180:1cm) ;
}+\Delta\left(\tikz[scale=0.6,baseline=(vert_cent.base)]{
  \node (vert_cent) {\hspace{-13pt}$\phantom{-}$};
 \filldraw (1.3,0) circle [radius=0.1]; 
\filldraw (0.1,0) circle [radius=0.1];
  \draw      (0.7,0) ++(0:0.6cm) arc (0:360:0.6cm and 0.4cm);
}\right)\tikz[scale=0.6,baseline=(vert_cent.base),scale=1]{
	%\draw[help lines] (-3,-3) grid (3,3);
	\node (vert_cent) {\hspace{-13pt}$\phantom{-}$};
	\draw (-1.5,0)--(-1,0) (1,0)--(1.5,0);
\draw (0:1cm) arc (0:-180:1cm) ;
}\right\}\Bigr],
\\
\ncal_{2b}=(\Ical_{2b})_{IR}=\left(\,\tikz[scale=0.6,baseline=(vert_cent.base),scale=1]{
\filldraw  (0,0.5) circle [radius=0.1];
\draw (-1,0) circle [radius=0.1];
\draw (1,0) circle [radius=0.1];
\draw (0,0) circle [radius=0.1];
\draw (0,1) --(0,0.1);
\draw [bend right=40] (0,1) to (-1,0.1);
\draw [bend left=40] (0,1) to (1,0.1); 
}\,\right)_{IR}=&\,-\tfrac12p^4\Kcal\left[\tikz[scale=0.6,baseline=(vert_cent.base)]{
  \node (vert_cent) {\hspace{-13pt}$\phantom{-}$};
  \draw (-0.1,0)--(0.1,0)
        (0.7,0) ++(0:0.6cm) arc (0:360:0.6cm and 0.4cm)
        (1.3,0)--(1.5,0);
\filldraw  (0.7,0) circle [radius=0.1];
\draw (0.7,-0.4) --(0.7,0.4);
}+\left(\,
\tikz[scale=0.6,baseline=(vert_cent.base)]{
  \node (vert_cent) {\hspace{-13pt}$\phantom{-}$};
  \draw (0,-.4)--(0,.4);
        \draw (0,.5) circle [radius=0.1];
\draw (0,-.5) circle [radius=0.1];
\filldraw (0,0) circle [radius=0.1];
}\,\right)_{IR}\tikz[scale=0.6,baseline=(vert_cent.base)]{
  \node (vert_cent) {\hspace{-13pt}$\phantom{-}$};
  \draw (-0.1,0)--(0.1,0)
        (0.7,0) ++(0:0.6cm) arc (0:360:0.6cm and 0.4cm)
        (1.3,0)--(1.5,0);
        \filldraw  (0.7,0.4) circle [radius=0.1];
\filldraw  (0.7,-0.4) circle [radius=0.1];
}\right],
\end{align}
\end{subequations}
where $p$ is the external momentum.
The third term in the expression for $\ncal_{2a}$ is immediately zero due to the loop with no incoming momentum (we shall display such trivially zero diagrams in the first few examples but suppress them later). Here in general the reference diagram for an IR counterterm with two attachment points (represented by open circles) is constructed by joining the attachment points with a propagator and routing an external momentum through the attachment points. Similarly the reference diagram for an IR counterterm with three attachment points is constructed by joining the three attachment points with propagators and routing an external momentum through any pair of the attachment points. The one-loop UV counterterm is given by 
\be
\Delta\left(\tikz[scale=0.6,baseline=(vert_cent.base)]{
  \node (vert_cent) {\hspace{-13pt}$\phantom{-}$};
   \filldraw (1.3,0) circle [radius=0.1]; \filldraw (0.1,0) circle [radius=0.1];
    \draw    (0.7,0) ++(0:0.6cm) arc (0:360:0.6cm and 0.4cm);
}\right)=-\frac{2}{\epsilon}.
\label{UVone}
\ee
We find (using results given for bubble-type Feynman integrals given in Eqs.~\eqref{Ldef}, \eqref{Lres})
\begin{subequations}\label{IR12}
\begin{align}
\ncal_1=&\,\tfrac{2}{\epsilon},\label{IR12:a}\\
\ncal_{2a}=\ncal_{2b}\equiv \ncal_2=&\,\tfrac{2}{\epsilon^2}+\tfrac{1}{\epsilon}.\label{IR12:b}
\end{align}
\end{subequations}
The results for $\ncal_{2a}$, $\ncal_{2b}$ were first obtained in Refs.~\cite{chet2}, \cite{chet1} respectively.
The complete set of IR counterterms required for the computation of the five-loop $\beta$-function in scalar $\phi^4$ theory was derived using similar methods in Ref.~\cite{klein}, and the results are summarised here in Eq.~\eqref{IRold}\footnote{The expression for $\ncal_{2a}$ given in Ref.~\cite{klein} differs from that given in Eq.~\eqref{IR12:b} and Ref.~\cite{chet2}. This is due to the omission of the terms involving $\ncal_1$ in the expression for $\ncal_{2a}$ in Ref.~\cite{klein}. This is compensated by a corresponding omission of $\ncal_1$ in the expressions for those five-loop UV counterterms which involve $\ncal_{2a}$, such as their Diagram 103. This seems to be at variance with their treatment of $\ncal_1$ in other diagrams, and of course does not lead to the equality of $\ncal_{2a}$ and $\ncal_{2b}$ which, as we are about to explain, is predicted by the general results of Ref.~\cite{chet3}.}. 
The equality of the two-loop IR counterterms apparent in Eq.~\eqref{IR12:b} was explained in Refs.~\cite{chet3, herz} and further generalised in Ref.~\cite{herz}. It was shown that the IR counterterm may be evaluated by considering the associated {\it contracted vacuum graph} (CVG), obtained by identifying the open circles in the IR counterterms as depicted in Eq.~\eqref{IRdiags} (the terminology ``contracted vacuum graph'' was first introduced in Ref.~\cite{herz} though the concept is implicit in Ref~\cite{chet3}). The CVG essentially arises in considering a Taylor expansion of the IR-divergent subdiagram in its external momenta, and taking its zeroth term. The two two-loop IR-divergent subdiagrams $\Ical_{2a}$ and $\Ical_{2b}$ have the same CVG 
\be
\tikz[scale=0.6,baseline=(vert_cent.base),scale=1]{
	%\draw[help lines] (-3,-3) grid (3,3);
	\node (vert_cent) {\hspace{-13pt}$\phantom{-}$};
	\draw (0,0) circle[radius=1cm];
	\draw [bend right=-40] (-60:1) to (60:1);
\filldraw  (-1,0) circle [radius=0.1];
}
\ee
and hence the same IR counterterm, $\ncal_2$. In general we denote IR-divergent subdiagrams with the same CVG by using a common number in the subscript, but different letters $a$, $b$, etc. The letter will henceforth be dropped from the IR counterterm (as we have done in Eq.~\eqref{IR12}) since these are identical for a common CVG. The CVG is typically a logarithmically superficially UV-divergent diagram corresponding to one of the 1PI graphs contributing to the $\beta$-function. Furthermore its momentum integral is zero since it is scaleless, having no external momenta. Eq.~\eqref{rstar:a} then gives a simple recursive relation between IR and UV counterterms:
%As such its UV divergence may be evaluated using the standard procedure of infra-red rearrangement, by feeding a momentum in at one of the %vertices and extracting it at another - and thus ``nullifying'' the momenta at the remaining vertices. One way to explain the procedure for extracting %the IR counterterm from the CVG is to consider evaluating the UV divergence using the particular (somewhat singular-looking) IR rearrangement where %the external momentum is inserted and extracted at the same point, at one end of the double propagator. In this simple two-loop case we have
\begin{subequations}\label{IRtwo}
\begin{align}
0=\Kcal R^*\left[\tikz[scale=0.6,baseline=(vert_cent.base)]{
  \node (vert_cent) {\hspace{-13pt}$\phantom{-}$};
   \draw     (0.7,0) ++(0:0.6cm) arc (0:360:0.6cm and 0.4cm);
\filldraw  (1.3,0) circle [radius=0.1];
\filldraw  (0.1,0) circle [radius=0.1];
}\right]=&\,\Kcal\Bigl[\Delta\left(\tikz[scale=0.6,baseline=(vert_cent.base)]{
  \node (vert_cent) {\hspace{-13pt}$\phantom{-}$};
  \filldraw (1.3,0) circle [radius=0.1]; \filldraw (0.1,0) circle [radius=0.1];
    \draw    (0.7,0) ++(0:0.6cm) arc (0:360:0.6cm and 0.4cm);
}\right)+\left(\,
\tikz[scale=0.6,baseline=(vert_cent.base)]{
  \node (vert_cent) {\hspace{-13pt}$\phantom{-}$};
  \draw (0,-.4)--(0,.4);
        \draw (0,.5) circle [radius=0.1];
\draw (0,-.5) circle [radius=0.1];
\filldraw (0,0) circle [radius=0.1];
}\,\right)_{IR}\Bigr],\label{IRtwo:a}\\
0=\Kcal R^*\left[\tikz[scale=0.6,baseline=(vert_cent.base),scale=1]{
	%\draw[help lines] (-3,-3) grid (3,3);
	\node (vert_cent) {\hspace{-13pt}$\phantom{-}$};
	\draw (0,0) circle[radius=1cm];
	\draw [bend right=-40] (-180:1) to (-60:1);
\filldraw  (60:1) circle [radius=0.1];
}\right]=&\,\Kcal\Bigl[\Delta\left(\tikz[scale=0.6,baseline=(vert_cent.base),scale=1]{
	%\draw[help lines] (-3,-3) grid (3,3);
	\node (vert_cent) {\hspace{-13pt}$\phantom{-}$};
	\draw (0,0) circle[radius=1cm];
	\draw [bend right=-40] (-180:1) to (-60:1);
\filldraw  (60:1) circle [radius=0.1];
}\right)+\left(\tikz[scale=0.6,baseline=(vert_cent.base),scale=1]{
\draw(0,-1.4) -- (0,0);
\draw (0,1.5) circle [radius=0.1];
\draw (0,-1.5) circle [radius=0.1];
\draw [bend right=-40] (0,0) to (-0.1,1.4);
\draw [bend left=-40] (0,0) to (0.1,1.4);
\filldraw  (0.0,-0.75) circle [radius=0.1];
}\right)_{IR}+\Delta\left(\tikz[scale=0.6,baseline=(vert_cent.base)]{
  \node (vert_cent) {\hspace{-13pt}$\phantom{-}$};
  \filldraw (1.3,0) circle [radius=0.1]; \filldraw (0.1,0) circle [radius=0.1];
    \draw    (0.7,0) ++(0:0.6cm) arc (0:360:0.6cm and 0.4cm);
}\right)\tikz[scale=0.6,baseline=(vert_cent.base),scale=1]{
	%\draw[help lines] (-3,-3) grid (3,3);
	\node (vert_cent) {\hspace{-13pt}$\phantom{-}$};
	\draw (0,0) circle[radius=1cm];
\filldraw  (-1,0) circle [radius=0.1];
\filldraw  (0:1) circle [radius=0.1];
}\nn
&+\left(\,
\tikz[scale=0.6,baseline=(vert_cent.base)]{
  \node (vert_cent) {\hspace{-13pt}$\phantom{-}$};
  \draw (0,-.4)--(0,.4);
        \draw (0,.5) circle [radius=0.1];
\draw (0,-.5) circle [radius=0.1];
\filldraw (0,0) circle [radius=0.1];
}\,\right)_{IR}\left\{\tikz[scale=0.6,baseline=(vert_cent.base),scale=1]{
	%\draw[help lines] (-3,-3) grid (3,3);
	\node (vert_cent) {\hspace{-13pt}$\phantom{-}$};
	\draw [bend right=-40] (-180:1) to (-60:1);
\draw (-180:1cm) arc (-180:-60:1cm) ;
}+\Delta\left(\tikz[scale=0.6,baseline=(vert_cent.base)]{
  \node (vert_cent) {\hspace{-13pt}$\phantom{-}$};
   \filldraw (1.3,0) circle [radius=0.1]; \filldraw (0.1,0) circle [radius=0.1];
     \draw   (0.7,0) ++(0:0.6cm) arc (0:360:0.6cm and 0.4cm);
}\right)\tikz[scale=0.6,baseline=(vert_cent.base),scale=1]{
	%\draw[help lines] (-3,-3) grid (3,3);
}\right\}\Bigr].\label{IRtwo:b}
\end{align}
\end{subequations}
 The third and fourth diagrams on the right-hand side of Eq.~\eqref{IRtwo:b} are scale-invariant and again zero. Quoting standard results from Eq.~\eqref{divs}, we have
\be
\Delta\left[\tikz[scale=0.6,baseline=(vert_cent.base)]{
  \node (vert_cent) {\hspace{-13pt}$\phantom{-}$};
  \draw (0.7,0) ++(0:0.6cm) arc (0:360:0.6cm and 0.4cm);
\filldraw  (1.3,0) circle [radius=0.1];
\filldraw  (0.1,0) circle [radius=0.1];
}\right]=\Delta_{1,1}=\frac{2}{\epsilon},\quad \Delta\left[\tikz[scale=0.6,baseline=(vert_cent.base),scale=1]{
	%\draw[help lines] (-3,-3) grid (3,3);
	\node (vert_cent) {\hspace{-13pt}$\phantom{-}$};
	\draw (0,0) circle[radius=1cm];
	\draw [bend right=-40] (-180:1) to (-60:1);
\filldraw  (60:1) circle [radius=0.1];
}\right]=\Delta_{2,2}=\tfrac{2}{\eps^2}-\tfrac{1}{\eps}.
\ee
Here, $\Delta$ is our notation for a UV counterterm and the first label in the subscripts is  the loop order, the second refers to the numbering in Appendix A of Ref.~\cite{klein}. The results for the UV counterterms are listed in Eq.~\eqref{divs}.
Inserting into Eq.~\eqref{IRtwo},
we easily reproduce the results for $\ncal_1$ and then $\ncal_{2}$ from Eq.~\eqref{IR12}. We have explicitly shown the diagrams which are manifestly zero; once one is accustomed to suppressing these, the method just described gives an efficient means of computing IR counterterms from the UV counterterm for the associated CVG. 
%It is not clear from our method why all the other IR counterterms with the same CVG (such as $\ncal_{2b}$ in the case of $\ncal_{2a}$) should be i%dentical, but this is explained in Ref.~\cite{herz}. 
So far we have
\be
\ncal_1=-\Delta_{1,1}, \quad \ncal_2=-\Delta_{2,2}+(\ncal_1)^2.
\label{iruv}
\ee
\section{Alternative IR counterterms and simple examples} 
In this section we introduce our alternative proposal for the IR counterterms. We claim that these AIR counterterms, when used in the standard loop-by-loop computation of the UV counterterms within minimal subtraction, lead to the same results as produced by the standard IR counterterms described in the previous Section. The main advantage of our AIR counterterms is that they may be presented in a simple closed form; whereas the computation of the standard IR counterterms requires the use of subtractions as we see in Eq.~\eqref{IRdiags}, or the derivation of a relation with the UV counterterms as in Eq.~\eqref{iruv}. The majority of the IR divergences occur in structures with a double propagator, and we shall focus attention on these here and elsewhere in the main body of the text. We shall return to the case of IR divergent structures with no double propagators in Appendix C; however, currently we have no alternative form for IR counterterms associated with triple or even more propagators. So for the time being, we confine our attention to the case where the $L$-loop CVG corresponding to an IR divergence has one double propagator (the case of CVGs with two double propagators is considered in Appendix A and the results are listed in Tables ~\ref{IRres3} and \ref{IRres4}). Suppose we sever this double propagator at its centre, so we obtain a logarithmically divergent $(L-1)$-loop graph $G^{\rm{sev}}$ with two external lines. We then feed in an external momentum $p$ along these external lines, and write down the Feynman integral for this graph in the form 
\be
F_{G^{\rm{sev}}} (1/p^2)^{\frac12(L-1)\epsilon}.
\label{Fdef}
\ee
 Our claim is that an alternative form for the IR counterterms corresponding to this CVG is given by
\be
F_{G^{\rm{sev}}}C_L,
\label{Xdef}
\ee
where
\be
C_L=\frac{2F_L}{\epsilon\Gamma\left(\tfrac{d}{2}-1\right)}\left(1-\frac{(L-1)}{L\left(\tfrac{d}{2}-1\right)}\right),
\label{cdef}
\ee
with
\be
F_L=\left[\Gamma\left(\tfrac{d}{2}-1\right)\right]^L.
\ee
We observe that in particular we have the simple result
\be
C_1=\tfrac{2}{\epsilon}.
\label{C1}
\ee
 A point to emphasise here is that the standard IR counterterms are (at least in all explicit computations of which we are aware) defined by a form of minimal subtraction; in other words one discards finite parts in the evaluation of these IR counterterms. Here, on the other hand, it is important to include all terms in the expansion of Eq.~\eqref{Xdef} which contribute to poles in the final result, for whatever UV counterterm is being calculated\footnote{There is a possible parallel here with recent computations\cite{jag} using a $\overline{\hbox{MaxS}}$ scheme where all terms in the $\epsilon$ expansion are absorbed into the renormalisation constants, giving results identical with the $\widetilde{\hbox{MOM}}$ scheme where only finite terms are subtracted.}. We have included the factor $F_L$ in $C_L$ since it suppresses the appearance of terms like $\gamma$ and $\zeta_2$  in our new counterterms (and also products of powers of $\gamma$ and a single $\zeta_2$)\footnote{We cannot rule out the appearance of powers of $\zeta_2$ which cannot be distinguished from $\zeta_4$, $\zeta_6$ etc.}; though it is equally possible to omit it without altering the final result for the UV counterterm. A similar factor may be included in the standard IR counterterms, though oddly here the effect is to complicate the countertems. In either case, the fact that this optional factor may be added without altering the final result for the UV counterterm is a consequence of the particular combinations of graphs appearing in the $R^*$ process. Unless explicitly stated otherwise, we assume henceforth that the factor of $F_L$ is indeed included. In Appendix A we give further details; in Tables~\ref{IRres12}, \ref{IRres3} and \ref{IRres4} we give a list of IR divergent subdiagrams relevant for our calculations, together with their CVGs, the corresponding severed graph $G^{\rm{sev}}$ and an expression for the IR counterterm. Then in Eq.~\eqref{IRold} we give the standard results for IR counterterms, and in Eq.~\eqref{irdiffs} we give explicit results for the differences between our IR counterterms and the standard ones. It may be observed from Eq.~\eqref{irdiffs} that the leading and next-to-leading poles in each of our new IR counterterms agree with the standard versions.

Our new proposal has been essentially developed by trial and error; in fact it emerged from an attempt to treat the infra-red counterterms as propagator corrections, as was done at lowest order (but in configuration space) in Ref.~\cite{JO1}. However, we have checked it by verifying that it gives the same result as the standard method in numerous examples. In particular, we have confirmed that it reproduces the known results in all the five-loop diagrams contributing to the scalar $\phi^4$ $\beta$-function which were evaluated using the $R^*$ method in Ref.~\cite{klein}. These comparisons are relegated to Appendix B, but we carry out several simple lower-loop examples in somewhat more detail in the main text. In fact the two and three-loop calculations (which require at most one and two-loop IR counterterms) are somewhat trivial. We immediately see from the first line of Table~\ref{IRres12} and Eq.~\eqref{C1} that (using a prime to distinguish our alternative versions of counterterms) our one-loop IR counterterm $\ncal_1'$ is identical to the standard $\ncal_1$ in Eq.~\eqref{IR12:a}. Furthermore, from the second line of Table~\ref{IRres12} and using Eqs.~\eqref{cdef} and \eqref{Lres}, we see that our two-loop version $\ncal'_{2}$ is given by 
\be
\ncal'_{2}=L(1,1)C_2=\tfrac{2}{\epsilon^2}+\tfrac{1}{\epsilon}+\tfrac12+\left(\tfrac14-\tfrac12\zeta_3\right)\epsilon+\ldots
\label{ir2p}
\ee
Comparing with Eq.~\eqref{IR12:b} we see that $\ncal'_{2}$ agrees with $\ncal_{2}$
as far as the poles are concerned (a trivial example of our earlier assertion that the leading and next-to-leading poles agree between the two methods) with a finite difference as recorded in Eq.~\eqref{irdiffs:b}.  Nevertheless there is still something to check. As we have said, it is in general necessary to include finite and $O(\epsilon)$ parts in our IR counterterms. This will only become essential in four and higher loop diagrams, where we potentially require three- and higher-loop IR counterterms. In these higher-loop examples, the simple poles in our IR counterterms differ from the standard versions, and this is compensated by contributions from finite parts of our new counterterms. Therefore for consistency we should check that the inclusion of the finite term in our two-loop counterterm $\ncal_{2}'$ also leads to the standard results for three-loop UV counterterms. A more subtle point is that, as we mentioned earlier, the simple form of our IR counterterms (in particular the absence of $\gamma$ and $\zeta_2$ which we already see in Eq.~\eqref{ir2p}) is due to our inclusion of the factor $F_L$. We would like to be able to check our assertion that the inclusion of $F_L$ has no effect on the final UV counterterms. For these reasons we present a number of two and three-loop examples in some detail. We do this in the remainder of this section, where we also take the opportunity to give a full explanation of the $R^*$ process. However the reader who wishes to skip these somewhat trivial computations and who is sufficiently familiar with the $R^*$ process is invited to turn immediately to the next Section. 

We start with a rather simple two-loop example. Since $\ncal'_1=\ncal_1$, there is no difference between our computation and the standard one. Nevertheless we include it both for completeness and to enable a check that the final result is unchanged if we omit the factor $F_1$. We consider the UV counterterm $\Delta_{2,2}$.  The diagram is depicted on the LHS of Eq.~\eqref{exdiaga}. Note that $\Delta_{2,1}$, the counterterm for the other two-loop diagram in $\phi^4$ theory, does not contain a simple pole and therefore does not contribute to the $\beta$-function, so we do not consider this diagram further. $\Delta_{2,2}$  may be evaluated using IR rearrangement without introducing IR divergences, by feeding in a momentum at the centre of the double propagator and extracting it at one of the other vertices. The result for $\Delta_{2,2}$ is well-known and given for instance in Ref.~\cite{klein}. But here for the purpose of displaying the $R^*$ procedure, we choose a different nullification as shown, so that there is an IR divergence requiring the use of the counterterm $\ncal_1$. The procedure is shown pictorially in Eq.~\eqref{exdiaga}.
\begin{align}
\Delta_{2,2}=-\Kcal\Rbar^*\left(\tikz[scale=0.6,baseline=(vert_cent.base),scale=1]{
	%\draw[help lines] (-3,-3) grid (3,3);
	\node (vert_cent) {\hspace{-13pt}$\phantom{-}$};
	\draw (0,0) circle[radius=1cm];
\draw (-1.25,0) -- (1.25,0);
\filldraw (0,-1) circle [radius=0.1];
}\right)=&\,-\Kcal\Bigl[\tikz[scale=0.6,baseline=(vert_cent.base),scale=1]{
	%\draw[help lines] (-3,-3) grid (3,3);
	\node (vert_cent) {\hspace{-13pt}$\phantom{-}$};
	\draw (0,0) circle[radius=1cm];
\draw (-1.25,0) -- (1.25,0);
\filldraw (0,-1) circle [radius=0.1];
}+\Delta\left(\tikz[scale=0.6,baseline=(vert_cent.base)]{
  \node (vert_cent) {\hspace{-13pt}$\phantom{-}$};
 \filldraw (1.3,0) circle [radius=0.1]; \filldraw (0.1,0) circle [radius=0.1];
   \draw     (0.7,0) ++(0:0.6cm) arc (0:360:0.6cm and 0.4cm);
}\right)\tikz[scale=0.6,baseline=(vert_cent.base)]{
  \node (vert_cent) {\hspace{-13pt}$\phantom{-}$};
  \draw (-0.1,0.1)--(0.1,0)
(-0.1,-0.1)--(0.1,0)
        (0.7,0) ++(0:0.6cm) arc (0:360:0.6cm and 0.4cm);
\filldraw (1.3,0) circle [radius=0.1];
}\nn
&+ \left(\,
\tikz[scale=0.6,baseline=(vert_cent.base)]{
  \node (vert_cent) {\hspace{-13pt}$\phantom{-}$};
  \draw (0,-.4)--(0,.4);
        \draw (0,.5) circle [radius=0.1];
\draw (0,-.5) circle [radius=0.1];
\filldraw (0,0) circle [radius=0.1];
}\,\right)_{IR}\left\{\tikz[scale=0.6,baseline=(vert_cent.base)]{
  \node (vert_cent) {\hspace{-13pt}$\phantom{-}$};
  \draw (-0.1,0)--(0.1,0)
        (0.7,0) ++(0:0.6cm) arc (0:360:0.6cm and 0.4cm)
        (1.3,0)--(1.5,0);
}+\Delta\left(\tikz[scale=0.6,baseline=(vert_cent.base)]{
  \node (vert_cent) {\hspace{-13pt}$\phantom{-}$};
  \filldraw (1.3,0) circle [radius=0.1]; \filldraw (0.1,0) circle [radius=0.1];
   \draw     (0.7,0) ++(0:0.6cm) arc (0:360:0.6cm and 0.4cm);
}\right)\right\}\Bigr].
\label{exdiaga}
\end{align}
Here the one-loop IR-divergent structure has been factored out. The RHS of Eq.~\eqref{exdiaga} may be succinctly described in terms of the $R$ and $\Rbar$ operations, in which only UV divergences are subtracted.
We denote by $I^{2,2}_0$ the terms in Eq.~\eqref{exdiaga} with no IR counterterm, and by $I^{2,2}_1$ the terms multiplied by the one-loop IR counterterm (in each case, before extracting the pole terms). We further denote the complete two-loop diagram on the LHS of Eq.~\eqref{exdiaga} by $G^{2,2}$ (this corresponds to the first term in $I_0^{2,2}$), and the one-loop diagram where the one-loop IR-divergent structure has been excised from $G^{2,2}$ by $G^{2,2}_1$ (this corresponds to the first term in $I_1^{2,2}$). We may then rewrite Eq.~\eqref{exdiaga} in the form
\be
\Delta_{2,2}=-\Kcal\left(I_0^{2,2}+\ncal_1I_1^{2,2}\right),
\label{del22exp}
\ee
where 
\be
I_0^{2,2}=\Rbar G^{2,2},\quad I_1^{2,2}=R G_1^{2,2}.
\ee
Note the appearance of $\Rbar$ rather than $R$ in the equation for $I_0^{2,2}$. We see in Eq.~\eqref{exdiaga} that the second term in $I_0^{2,2}$ is zero due to zero external momentum in the loop,  so that we have
\begin{subequations}\label{i22}
\begin{align}
I^{2,2}_0=&\,L(1,1)L(2,\tfrac12\epsilon)=-\tfrac{2}{\epsilon^2}-\tfrac{3}{\epsilon}+\ldots,\\
I^{2,2}_1=&\,L(1,1)-\tfrac{2}{\epsilon}=2+\ldots.\label{i22:b}
\end{align}
\end{subequations}
We note that here and elsewhere we have included an extra factor of $\Gamma\left(\tfrac{d}{2}-1\right)$ for each loop, which in any case has no effect on the UV pole structure after application of the $R^*$ procedure; this reduces the appearance of $\gamma$, $\zeta_2$ etc (and also products of powers of $\gamma$ and a single $\zeta_2$) in the intervening expressions.
Combining the results and using Eq.~\eqref{IR12:a}, we can expand Eq.~\eqref{del22exp} as
\be
\Delta_{2,2}=\tfrac{2}{\epsilon^2}-\tfrac{1}{\epsilon}.
\label{d22}
\ee
Of course, this is the well-known result as given in Ref.~\cite{klein} and elsewhere, and listed for completeness in our Eq.~\eqref{divs:b}. So far this is identical to the standard calculation.

We now consider the effect of omitting the factor $F_1$ from our new IR counterterm $\ncal_1'$. Since $F_1=1-\tfrac12\epsilon\gamma+\ldots$, and since $I^{2,2}_1$ is finite, it is immediately clear from Eq.~\eqref{del22exp} that this omission has no effect on the UV counterterm $\Delta_{2,2}$, as we claimed. We shall see later that the cancellation of leading poles in quantities such as $I^{2,2}_1$ is a generic feature.

We continue with a slightly less trivial example at three loops. We consider the UV counterterm $\Delta_{3,5}$. The diagram is depicted on the LHS of Eq.~\eqref{exdiagb}. Just as we saw at two loops, all the three-loop diagrams relevant for the $\beta$-function in $\phi^4$ theory may be evaluated using IR rearrangement by a judicious nullification of external momenta, without introducing IR divergences. In the current case, if a momentum enters at the centre of the double propagator and exits at any of the other vertices, there are no IR divergences and the counterterm may readily be evaluated as in Ref.~\cite{klein}. But as in the previous two-loop example, for pedagogic purposes and in order to display and check our alternative method, we choose a different nullification as shown, so that there is an IR divergence and two IR divergent substructures $\Ical_1$ and $\Ical_{2a}$, now requiring the use of both one and two loop counterterms, $\ncal_1$ and $\ncal_{2}$.
\begin{align}
\Delta_{3,5}=&\,-\Kcal\Rbar^*\left(\tikz[scale=0.6,baseline=(vert_cent.base),scale=1]{
\draw (180:1cm) arc (180:0:1cm) ;
\draw [bend right=40] (0,1) to (1,0);
\draw [bend left=40] (0,1) to (-1,0);
\filldraw (0,0) circle [radius=0.1];
\draw (-1.25,0) -- (1,0);
\draw (0,1.25) -- (0,1); 
}\right)=-\Kcal\Bigl[\tikz[scale=0.6,baseline=(vert_cent.base),scale=1]{
\draw (180:1cm) arc (180:0:1cm) ;
\draw [bend right=40] (0,1) to (1,0);
\draw [bend left=40] (0,1) to (-1,0);
\filldraw (0,0) circle [radius=0.1];
\draw (-1.25,0) -- (1,0);
\draw (0,1.25) -- (0,1); 
}\nn
&+\Delta\left(\tikz[scale=0.6,baseline=(vert_cent.base)]{
  \node (vert_cent) {\hspace{-13pt}$\phantom{-}$};
  \filldraw (1.3,0) circle [radius=0.1]; \filldraw (0.1,0) circle [radius=0.1];\draw 
        (0.7,0) ++(0:0.6cm) arc (0:360:0.6cm and 0.4cm);
}\right)\tikz[scale=0.6,baseline=(vert_cent.base),scale=1]{
	%\draw[help lines] (-3,-3) grid (3,3);
	\node (vert_cent) {\hspace{-13pt}$\phantom{-}$};
	\draw (0,0) circle[radius=1cm];
\draw (-1.25,0) -- (1.25,0);
\filldraw (0,-1) circle [radius=0.1];
}+\Delta\left(\tikz[scale=0.6,baseline=(vert_cent.base)]{
  \node (vert_cent) {\hspace{-13pt}$\phantom{-}$};
\filldraw (1.3,0) circle [radius=0.1]; \filldraw (0.1,0) circle [radius=0.1];\draw
        (0.7,0) ++(0:0.6cm) arc (0:360:0.6cm and 0.4cm);
}\right)\tikz[scale=0.6,baseline=(vert_cent.base),scale=1]{
	%\draw[help lines] (-3,-3) grid (3,3);
	\node (vert_cent) {\hspace{-13pt}$\phantom{-}$};
	\draw (0,0) circle[radius=1cm];
\draw (-1,0) -- (1,0);
\draw (-1.25,0.1) -- (-1,0);
\draw (-1.25,-0.1) -- (-1,0);
\filldraw (0,-1) circle [radius=0.1];
}\nn
&+\Delta\left(\tikz[scale=0.6,baseline=(vert_cent.base)]{
  \node (vert_cent) {\hspace{-13pt}$\phantom{-}$};
  \filldraw (1.3,0) circle [radius=0.1]; \filldraw (0.1,0) circle [radius=0.1];\draw
        (0.7,0) ++(0:0.6cm) arc (0:360:0.6cm and 0.4cm);
}\right)\Delta\left(\tikz[scale=0.6,baseline=(vert_cent.base)]{
  \node (vert_cent) {\hspace{-13pt}$\phantom{-}$};
  \filldraw (1.3,0) circle [radius=0.1]; \filldraw (0.1,0) circle [radius=0.1];\draw
        (0.7,0) ++(0:0.6cm) arc (0:360:0.6cm and 0.4cm);
}\right)\tikz[scale=0.6,baseline=(vert_cent.base)]{
  \node (vert_cent) {\hspace{-13pt}$\phantom{-}$};
  \draw (-0.1,0.1)--(0.1,0)
(-0.1,-0.1)--(0.1,0)
        (0.7,0) ++(0:0.6cm) arc (0:360:0.6cm and 0.4cm);
\filldraw (1.3,0) circle [radius=0.1];
}\nn
&+ \left(\,
\tikz[scale=0.6,baseline=(vert_cent.base)]{
  \node (vert_cent) {\hspace{-13pt}$\phantom{-}$};
  \draw (0,-.4)--(0,.4);
        \draw (0,.5) circle [radius=0.1];
\draw (0,-.5) circle [radius=0.1];
\filldraw (0,0) circle [radius=0.1];
}\,\right)_{IR}\Bigl\{\tikz[scale=0.6,baseline=(vert_cent.base)]{
  \node (vert_cent) {\hspace{-13pt}$\phantom{-}$};
  \draw [bend right=40] (-0.5,0) to (0.5,0);
\draw [bend left=40] (-0.5,0) to  (0.5,0);
\draw [bend right=30] (-0.5,0) to (-0.5,0.5);
\draw [bend left=30] (-0.5,0) to  (-0.5,0.5);
\draw (-0.75,0) -- (-0.5,0);
\draw (0.5,0) -- (0.75,0);
}+\Delta\left(\tikz[scale=0.6,baseline=(vert_cent.base)]{
  \node (vert_cent) {\hspace{-13pt}$\phantom{-}$};
  \filldraw (1.3,0) circle [radius=0.1]; \filldraw (0.1,0) circle [radius=0.1];\draw
        (0.7,0) ++(0:0.6cm) arc (0:360:0.6cm and 0.4cm);
}\right)\tikz[scale=0.6,baseline=(vert_cent.base)]{
  \node (vert_cent) {\hspace{-13pt}$\phantom{-}$};
  \draw (-0.1,0)--(0.1,0)
        (0.7,0) ++(0:0.6cm) arc (0:360:0.6cm and 0.4cm)
        (1.3,0)--(1.5,0);
}\nn
&+\Delta\left(\tikz[scale=0.6,baseline=(vert_cent.base)]{
  \node (vert_cent) {\hspace{-13pt}$\phantom{-}$};
  \filldraw (1.3,0) circle [radius=0.1]; \filldraw (0.1,0) circle [radius=0.1];\draw
        (0.7,0) ++(0:0.6cm) arc (0:360:0.6cm and 0.4cm);
}\right)\tikz[scale=0.6,baseline=(vert_cent.base)]{
  \node (vert_cent) {\hspace{-13pt}$\phantom{-}$};
\draw [bend right=30] (0,0) to (0,0.5);
\draw [bend left=30] (0,0) to  (0,0.5);
\draw (-0.25,0) -- (0.25,0);
}+\Delta\left(\tikz[scale=0.6,baseline=(vert_cent.base)]{
  \node (vert_cent) {\hspace{-13pt}$\phantom{-}$};
  \filldraw (1.3,0) circle [radius=0.1]; \filldraw (0.1,0) circle [radius=0.1];\draw
        (0.7,0) ++(0:0.6cm) arc (0:360:0.6cm and 0.4cm);
}\right)\Delta\left(\tikz[scale=0.6,baseline=(vert_cent.base)]{
  \node (vert_cent) {\hspace{-13pt}$\phantom{-}$};
  \filldraw (1.3,0) circle [radius=0.1]; \filldraw (0.1,0) circle [radius=0.1];\draw
        (0.7,0) ++(0:0.6cm) arc (0:360:0.6cm and 0.4cm);
}\right)\nn
&+\left(\,\tikz[scale=0.6,baseline=(vert_cent.base),scale=1]{
\filldraw  (135:1) circle [radius=0.1];
\draw (-1,0) circle [radius=0.1];
\draw (1,0) circle [radius=0.1];
\draw [bend right=40] (0,1) to (-1,0.1);
\draw [bend left=40] (0,1) to (1,0.1); 
\draw [bend right=40] (0,1) to (1,0.1);
}\,\right)_{IR}\Bigl\{\tikz[scale=0.6,baseline=(vert_cent.base)]{
  \node (vert_cent) {\hspace{-13pt}$\phantom{-}$};
  \draw (-0.1,0)--(0.1,0)
        (0.7,0) ++(0:0.6cm) arc (0:360:0.6cm and 0.4cm)
        (1.3,0)--(1.5,0);
}+\Delta\left(\tikz[scale=0.6,baseline=(vert_cent.base)]{
  \node (vert_cent) {\hspace{-13pt}$\phantom{-}$};
  \filldraw (1.3,0) circle [radius=0.1]; \filldraw (0.1,0) circle [radius=0.1];\draw
        (0.7,0) ++(0:0.6cm) arc (0:360:0.6cm and 0.4cm);
}\right)\Bigr\}\Bigl].
\label{exdiagb}
\end{align}
Here again the RHS of Eq.~\eqref{exdiagb} may be efficiently rewritten in terms of $R$ and $\Rbar$ operations. We denote by $I^{3,5}_0$ the terms with no IR counterterm, by $I^{3,5}_1$ the terms multiplied by the one-loop IR counterterm, and by $I^{3,5}_2$ the terms multiplied by the two-loop IR counterterm (in each case, before extracting the pole terms).  We further denote the complete three-loop diagram on the LHS of Eq.~\eqref{exdiagb} by $G^{3,5}$ (this is also the first term in $I_0^{3,5}$), the two-loop diagram where the one-loop IR-divergent structure has been excised by $G^{3,5}_1$ (this is the first term in $I_1^{3,5}$), and the one-loop diagram where the two-loop IR-divergent structure has been excised by $G^{3,5}_1$ (this is the first term in $I_2^{3,5}$). We may then rewrite Eq.~\eqref{exdiaga} in the form
\be
\Delta_{3,5}=-\Kcal\left(I^{3,5}_0+\ncal_1I^{3,5}_1+\ncal_{2}I^{3,5}_2\right),
\label{del35exp}
\ee
where
\be
I_0^{3,5}=\Rbar G^{3,5},\quad I_1^{3,5}=RG_1^{3,5}, \quad I_2^{3,5}=RG_2^{3,5}.
\ee
Here the third and fourth diagrams in $I_0^{3,5}$ and the first and third diagrams in $I_1^{3,5}$ are zero since they have internal loops with no external momentum circulating.
We then have
\begin{subequations}\label{oldexp}
\begin{align}
I^{3,5}_0=&\,L(1,1)^2L(2+\tfrac12\epsilon,\tfrac12\epsilon)-\tfrac{2}{\epsilon}L(1,1)L(2,\tfrac12\epsilon)=\tfrac83\tfrac{1}{\epsilon^3}+\tfrac83\tfrac{1}{\epsilon^2}+\tfrac43\tfrac{1}{\epsilon}+\ldots,\\
I^{3,5}_1=&\,-\tfrac{2}{\epsilon}L(1,1)+\tfrac{4}{\epsilon^2}=-\tfrac{2}{\epsilon}I^{2,2}_1=-\tfrac{4}{\epsilon}-4+\ldots,\label{oldexp:b}\\
I^{3,5}_2=&\,L(1,1)-\tfrac{2}{\epsilon}=I^{2,2}_1=2+2\epsilon+\ldots\label{oldexp:c}
\end{align}
\end{subequations}
Using the standard IR counterterms $\ncal_1$ and $\ncal_{2}$ from Eq.~\eqref{IR12}, we easily expand Eq.~\eqref{del35exp} as
\be
\Delta_{3,5}=-\tfrac83\tfrac{1}{\epsilon^3}+\tfrac43\tfrac{1}{\epsilon^2}+\tfrac23\tfrac{1}{\epsilon},
\label{d35}
\ee
which is the well-known result for the UV counterterm as derived in Ref.~\cite{kaz} and listed in Ref.~\cite{klein} and our Eq.~\eqref{divs:c}.

Now we turn to our alternative procedure, which presents significant differences in this case. We already know $\ncal_1'=\ncal_1$; and the result for $\ncal_{2}'$ was given in Eq.~\eqref{ir2p}.
As we remarked before, the poles in $\ncal_{2}'$ agree with the standard result in Eq.~\eqref{IR12}; but now there is a finite part, and of course $O(\epsilon)$ terms. It is trivial to see that the replacement of $\ncal_{2}$ by $\ncal'_{2}$ has no effect on the final result for $\Delta_{3,5}$; the finite term in $\ncal_{2}'$ has no effect due to the cancellation of leading poles in $I^{3,5}_2$ which is visible in the expansion given in Eq.~\eqref{oldexp:c}, together with a similar cancellation in $I^{3,5}_1$ - generic features which we saw already at two loops and which will also be observed at higher orders. Of course it is not yet apparent why we need to include the finite terms anyway; this will not be clear until we consider a four-loop example. If one is only concerned with comparing the two methods of computing counterterms, one can simply compute the difference 
\be
\delta_{3,5}=\Delta_{3,5}-\Delta'_{3,5}=\Kcal\left(\delta^{IR}_1I^{3,5}_1+\delta^{IR}_{2}I^{3,5}_2\right)=0
\ee
(using $\Delta'_{3,5}$ to denote the UV counterterm computed using our AIR counterterms), where $I^{3,5}_1$, $I^{3,5}_2$ are given in Eqs.~\eqref{oldexp:b}, \eqref{oldexp:c} and $\delta^{IR}_1$, $\delta^{IR}_2$ are given in Eqs.~\eqref{irdiffs:a}, \eqref{irdiffs:b}.
Once again we also want to show that the three-loop UV counterterm $\Delta_{3,5}$ computed in Eq.~\eqref{d35} is unaffected by the inclusion or otherwise of the factors $F_{1,2}$ in $C_{1,2}$. It is  easy to check that this follows from the leading behaviour of $I^{3,5}_1$, $I^{3,5}_2$ (as given in Eq.~\eqref{oldexp}), and $\ncal'_1=\ncal_1$, $\ncal'_2$ (as given in Eqs.~\eqref{IR12}, \eqref{ir2p}).

\section{Higher-loop examples}

We now turn to examine two four-loop examples. We consider first the UV counterterm $\Delta_{4,18}$. The diagram is depicted on the LHS of Eq.~\eqref{exdiagc}. Just as we saw at two and three loops, all the four-loop diagrams relevant for the $\beta$-function in $\phi^4$ theory may be evaluated using IR rearrangement without introducing IR divergences. In the case at hand, if a momentum enters at the centre of the double propagator and exits at any of the other vertices, there are no IR divergences and the counterterm may readily be evaluated as in Ref.~\cite{klein}. However, we are again instead choosing a momentum arrangement which introduces an IR divergence, in order to illustrate and check our method; the procedure being displayed in Eq.~\eqref{exdiagc}.
\begin{align}
\Delta_{4,18}=&\,-\Kcal\Rbar^*\left(\tikz[scale=0.6,baseline=(vert_cent.base),scale=1]{
\draw (180:1cm) arc (180:0:1cm) ;
\draw [bend right=40] (-1,0) to (120:1);
\draw [bend left=40] (60:1) to (120:1);
\draw [bend left=-40] (60:1) to (1,0);
\filldraw (0,0) circle [radius=0.1];
\draw (120:1) -- (120:1.2);
\draw (-1.25,0) -- (1,0); 
}\right)=-\Kcal\Bigl[\tikz[scale=0.6,baseline=(vert_cent.base),scale=1]{
\draw (180:1cm) arc (180:0:1cm) ;
\draw [bend right=40] (-1,0) to (120:1);
\draw [bend left=40] (60:1) to (120:1);
\draw [bend left=-40] (60:1) to (1,0);
\filldraw (0,0) circle [radius=0.1];
\draw (120:1) -- (120:1.2);
\draw (-1.25,0) -- (1,0); 
}+2\Delta\left(\tikz[scale=0.6,baseline=(vert_cent.base)]{
  \node (vert_cent) {\hspace{-13pt}$\phantom{-}$};
  \filldraw (1.3,0) circle [radius=0.1]; \filldraw (0.1,0) circle [radius=0.1];\draw
        (0.7,0) ++(0:0.6cm) arc (0:360:0.6cm and 0.4cm);
}\right)\tikz[scale=0.6,baseline=(vert_cent.base),scale=1]{
\draw (180:1cm) arc (180:0:1cm) ;
\draw [bend right=40] (0,1) to (1,0);
\draw [bend left=40] (0,1) to (-1,0);
\filldraw (0,0) circle [radius=0.1];
\draw (-1.25,0) -- (1,0);
\draw (0,1.25) -- (0,1); 
}\nn
&+\Delta\left(\tikz[scale=0.6,baseline=(vert_cent.base)]{
  \node (vert_cent) {\hspace{-13pt}$\phantom{-}$};
  \filldraw (1.3,0) circle [radius=0.1]; \filldraw (0.1,0) circle [radius=0.1];\draw
        (0.7,0) ++(0:0.6cm) arc (0:360:0.6cm and 0.4cm);
}\right)\Delta\left(\tikz[scale=0.6,baseline=(vert_cent.base)]{
  \node (vert_cent) {\hspace{-13pt}$\phantom{-}$};
  \filldraw (1.3,0) circle [radius=0.1]; \filldraw (0.1,0) circle [radius=0.1];\draw
        (0.7,0) ++(0:0.6cm) arc (0:360:0.6cm and 0.4cm);
}\right)\tikz[scale=0.6,baseline=(vert_cent.base),scale=1]{
	%\draw[help lines] (-3,-3) grid (3,3);
	\node (vert_cent) {\hspace{-13pt}$\phantom{-}$};
	\draw (0,0) circle[radius=1cm];
\draw (-1.25,0) -- (1.25,0);
\filldraw (0,-1) circle [radius=0.1];
}\nn
&+\left(\,\tikz[scale=0.6,baseline=(vert_cent.base)]{
  \node (vert_cent) {\hspace{-13pt}$\phantom{-}$};
  \draw (0,-.4)--(0,.4);
        \draw (0,.5) circle [radius=0.1];
\draw (0,-.5) circle [radius=0.1];
\filldraw (0,0) circle [radius=0.1];
}\,\right)_{IR}\Bigl\{\Delta\left(\tikz[scale=0.6,baseline=(vert_cent.base)]{
  \node (vert_cent) {\hspace{-13pt}$\phantom{-}$};
  \filldraw (1.3,0) circle [radius=0.1]; \filldraw (0.1,0) circle [radius=0.1];\draw
        (0.7,0) ++(0:0.6cm) arc (0:360:0.6cm and 0.4cm);
}\right)\Delta\left(\tikz[scale=0.6,baseline=(vert_cent.base)]{
  \node (vert_cent) {\hspace{-13pt}$\phantom{-}$};
  \filldraw (1.3,0) circle [radius=0.1]; \filldraw (0.1,0) circle [radius=0.1];\draw
        (0.7,0) ++(0:0.6cm) arc (0:360:0.6cm and 0.4cm);
}\right)\tikz[scale=0.6,baseline=(vert_cent.base)]{
  \node (vert_cent) {\hspace{-13pt}$\phantom{-}$};
  \draw (-0.1,0)--(0.1,0)
        (0.7,0) ++(0:0.6cm) arc (0:360:0.6cm and 0.4cm)
        (1.3,0)--(1.5,0);
}\nn
&+\Delta\left(\tikz[scale=0.6,baseline=(vert_cent.base)]{
  \node (vert_cent) {\hspace{-13pt}$\phantom{-}$};
  \filldraw (1.3,0) circle [radius=0.1]; \filldraw (0.1,0) circle [radius=0.1];\draw
        (0.7,0) ++(0:0.6cm) arc (0:360:0.6cm and 0.4cm);
}\right)\Delta\left(\tikz[scale=0.6,baseline=(vert_cent.base)]{
  \node (vert_cent) {\hspace{-13pt}$\phantom{-}$};
  \filldraw (1.3,0) circle [radius=0.1]; \filldraw (0.1,0) circle [radius=0.1];\draw
        (0.7,0) ++(0:0.6cm) arc (0:360:0.6cm and 0.4cm);
}\right)\Delta\left(\tikz[scale=0.6,baseline=(vert_cent.base)]{
  \node (vert_cent) {\hspace{-13pt}$\phantom{-}$};
  \filldraw (1.3,0) circle [radius=0.1]; \filldraw (0.1,0) circle [radius=0.1];\draw
        (0.7,0) ++(0:0.6cm) arc (0:360:0.6cm and 0.4cm);
}\right)\Bigr\}\nn
&+2\left(\,\tikz[scale=0.6,baseline=(vert_cent.base),scale=1]{
\filldraw  (135:1) circle [radius=0.1];
\draw (-1,0) circle [radius=0.1];
\draw (1,0) circle [radius=0.1];
\draw [bend right=40] (0,1) to (-1,0.1);
\draw [bend left=40] (0,1) to (1,0.1); 
\draw [bend right=40] (0,1) to (1,0.1);
}\,\right)_{IR}\Bigl\{\Delta\left(\tikz[scale=0.6,baseline=(vert_cent.base)]{
  \node (vert_cent) {\hspace{-13pt}$\phantom{-}$};
  \filldraw (1.3,0) circle [radius=0.1]; \filldraw (0.1,0) circle [radius=0.1];\draw
        (0.7,0) ++(0:0.6cm) arc (0:360:0.6cm and 0.4cm);
}\right)\tikz[scale=0.6,baseline=(vert_cent.base)]{
  \node (vert_cent) {\hspace{-13pt}$\phantom{-}$};
  \draw (-0.1,0)--(0.1,0)
        (0.7,0) ++(0:0.6cm) arc (0:360:0.6cm and 0.4cm)
        (1.3,0)--(1.5,0);
}\nn
&+\Delta\left(\tikz[scale=0.6,baseline=(vert_cent.base)]{
  \node (vert_cent) {\hspace{-13pt}$\phantom{-}$};
  \filldraw (1.3,0) circle [radius=0.1]; \filldraw (0.1,0) circle [radius=0.1];\draw
        (0.7,0) ++(0:0.6cm) arc (0:360:0.6cm and 0.4cm);
}\right)\Delta\left(\tikz[scale=0.6,baseline=(vert_cent.base)]{
  \node (vert_cent) {\hspace{-13pt}$\phantom{-}$};
  \filldraw (1.3,0) circle [radius=0.1]; \filldraw (0.1,0) circle [radius=0.1];\draw
        (0.7,0) ++(0:0.6cm) arc (0:360:0.6cm and 0.4cm);
}\right)\Bigr\}\nn
&+\left(\,\tikz[scale=0.6,baseline=(vert_cent.base),scale=1]{
\filldraw  (150:1) circle [radius=0.1];
\draw (-1,0) circle [radius=0.1];
\draw (1,0) circle [radius=0.1];
\draw [bend right=30] (120:1) to (-1,0.1);
\draw [bend left=30] (60:1) to (1,0.1); 
\draw [bend right=30] (60:1) to (1,0.1);
\draw [bend left=30] (120:1) to (60:1);
\draw [bend right=30] (120:1) to (60:1);
}\,\right)_{IR}\left\{\tikz[scale=0.6,baseline=(vert_cent.base)]{
  \node (vert_cent) {\hspace{-13pt}$\phantom{-}$};
  \draw (-0.1,0)--(0.1,0)
        (0.7,0) ++(0:0.6cm) arc (0:360:0.6cm and 0.4cm)
        (1.3,0)--(1.5,0);
}+\Delta\left(\tikz[scale=0.6,baseline=(vert_cent.base)]{
  \node (vert_cent) {\hspace{-13pt}$\phantom{-}$};
  \filldraw (1.3,0) circle [radius=0.1]; \filldraw (0.1,0) circle [radius=0.1];\draw
        (0.7,0) ++(0:0.6cm) arc (0:360:0.6cm and 0.4cm);
}\right)\right\}
\Bigr].
\label{exdiagc}
\end{align}
As we saw in the case of $\Delta_{3,5}$,  we denote the terms in Eq.~\eqref{exdiagc} with no IR divergence by $I_0^{4,18}$ and those multiplied by one, two and three-loop IR-divergent structures by $I_1^{4,18}$, $I_2^{4,18}$, $I_3^{4,18}$ respectively. The full four-loop diagram (the first term in $I_0^{4,18}$) is denoted $G^{4,18}$ and $G^{4,18}_1$, $G^{4,18}_2$ and $G^{4,18}_3$ are obtained by successively removing $\Ical_1$, $\Ical_{2a}$, $\Ical_{3,1d}$ (in other words, the IR-divergent substructures of $G^{4,18}$) from $G^{4,18}$; these diagrams are also the first terms in $I_1^{4,18}$, $I_2^{4,18}$, $I_3^{4,18}$ respectively. We then have, as usual (but noting the factor of 2 due to the two independent two-loop IR divergences)
\be
\Delta_{4,18}=-\Kcal\left(I^{4,18}_0+\ncal_1I^{4,18}_1+2\ncal_{2}I^{4,18}_2+\ncal_{3,1}I^{4,18}_3\right),
\label{del418exp}
\ee
where
\be
I_0^{4,18}=\Rbar G^{4,18},\quad I_1^{4,18}=RG_1^{4,18}, \quad I_2^{4,18}=RG_2^{4,18}, \quad I_3^{4,18}=RG_3^{4,18}.
\ee
In this case and subsequently, we refrain from depicting those diagrams which are manifestly zero due to a loop with no external momentum circulating. We find
\begin{subequations}\label{i418}
\begin{align}
I^{4,18}_0=&\,L(1,1)^3L(2+\epsilon,\tfrac12\epsilon)-2\tfrac{2}{\epsilon}L(1,1)^2L(2+\tfrac12\epsilon,\tfrac12\epsilon)+\tfrac{4}{\epsilon^2}L(1,1)L(2,\tfrac12\epsilon)\nn
=&\,-\tfrac{4}{\epsilon^4}-\tfrac{10}{3}\tfrac{1}{\epsilon^3}-\tfrac53\tfrac{1}{\epsilon^2}+\left(\tfrac13\zeta_3-\tfrac56\right)\tfrac{1}{\epsilon}+\ldots,\\
I^{4,18}_1=&\,\Delta_{1,1}^2I^{2,2}_1=\tfrac{4}{\epsilon^2}I^{2,2}_1,\\
I^{4,18}_2=&\,\Delta_{1,1}I^{2,2}_1=-\tfrac{2}{\epsilon}I^{2,2}_1,\\
I^{4,18}_3=&\,I^{2,2}_1.
\end{align}
\end{subequations}
In order to compare the results for the four-loop UV counterterm in Eq.~\eqref{exdiagc} obtained using firstly the standard and secondly the AIR counterterms, we need in particular expressions for the three-loop IR counterterm which we see in Eq.~\eqref{exdiagc}. We shall first explain our procedure in some generality for use in later examples in this Section and in Appendix B. One starts by locating the IR-divergent substructure in the second column of Tables~\ref{IRres12}, \ref{IRres3} or \ref{IRres4}. One can then read off our labelling for it from the first column. The structures with the same numerical labelling share the same CVG and the same IR counterterm. Our alternative form for this common IR counterterm may be read off from the fourth column in the Tables. The standard expressions for this IR counterterm derived in Ref.~\cite{klein} using the methods of Section 2 are listed in Eq.~\eqref{IRold}. We have computed the differences between the new and standard IR counterterms to the maximum order in $\epsilon$ required for our purposes, and listed them in Eq.~\eqref{irdiffs}. With these differences already computed and ready to hand in Eq.~\eqref{irdiffs}, the alternative results for the IR counterterms may most readily be obtained by combining the results from Eqs.~\eqref{IRold} and Eq.~\eqref{irdiffs} (rather than using the expressions in the Tables which need expanding in $\epsilon$ to be used in practice).

In the case of the three-loop IR-divergent structure in Eq.~\eqref{exdiagc}, we first identify it as $\Ical_{3,1d}$ in Table~\ref{IRres3}. The structures  $\Ical_{3,1a}$ - $\Ical_{3,1d}$ share the same CVG and the same IR counterterm. We see from Table~\ref{IRres3} that the AIR counterterm is $\ncal'_{3,1}=L(1,1)^2C_3$ and this may then be expanded in $\epsilon$ using Eqs.~\eqref{cdef}, \eqref{Lres} as
\be
\ncal_{3,1}'=L(1,1)^2C_3=\tfrac83\tfrac{1}{\epsilon^3}+\tfrac83\tfrac{1}{\epsilon^2}+\tfrac43\tfrac{1}{\epsilon}+\ldots
\label{ir5p}
\ee
We may obtain the standard expression for the counterterm, in this case $\ncal_{3,1}$, from Eq.~\eqref{IRold:c};
and the difference, $\delta^{IR}_{3,1}$, between the new and standard expressions has been computed in Eq.~\eqref{irdiffs:c}. It is easy to check by comparison with Eq.~\eqref{ir5p} that
\be
\ncal_{3,1}'=\ncal_{3,1}+\delta^{IR}_{3,1}
\label{ir5pnew}
\ee
up to finite terms; and as we said earlier, having available the differences in Eq.~\eqref{irdiffs} it will be easier for the reader to compute $\ncal_{3,1}'$ to the order we require by using Eq.~\eqref{ir5pnew} rather than by expanding the expression in Table~\ref{IRres3} quoted in Eq.~\eqref{ir5p} (of course the same will apply to other IR counterterms).

Starting with the standard IR counterterms, using the results for $\ncal_1$ and $\ncal_{2}$ from Eq.~\eqref{IR12}, and for $\ncal_{3,1}$ from Eq.~\eqref{IRold:c}, we easily expand Eq.~\eqref{del418exp} as
\be
\Delta_{4,18}=\tfrac{4}{\epsilon^4}-\tfrac{2}{\epsilon^3}-\tfrac{1}{\epsilon^2}
-\left(\tfrac12-\zeta_3\right)\tfrac{1}{\epsilon},
\ee
which is the result obtained in Ref.~\cite{kaz} and quoted in Ref.~\cite{klein} and in our Eq.~\eqref{divs:h}. Turning to our AIR counterterms, we know we have $\ncal_1'=\ncal_1$, and the AIR counterterm  $\ncal'_{2}$ was already given in Eq.~\eqref{ir2p}. 
We see that once again the new counterterms agree with the standard versions as to the leading and next-to-leading poles; but Eq.~\eqref{irdiffs:c} shows us that the simple poles in the three-loop IR counterterms differ. It is straightforward to check that the replacement of $\ncal_{2}$ and $\ncal_{3,1}$ by $\ncal'_{2}$ and $\ncal'_{3,1}$ nevertheless has no effect on the final result for $\Delta_{4,18}$. As usual, if one is only concerned with comparing the two methods of computing counterterms, one can simply compute the difference. Noting from Eq.~\eqref{irdiffs} that
\be
\delta^{IR}_1=0,\quad \delta^{IR}_{2}=\tfrac12+\ldots,\quad \delta^{IR}_{3,1}=\tfrac{2}{\epsilon}+\ldots
\label{ir125}
\ee
and then combining with Eq.~\eqref{i418} we have
\be
\delta_{4,18}=\Kcal\left(\delta^{IR}_1I^{4,18}_1+2\delta^{IR}_{2}I^{4,18}_2+\delta^{IR}_{3,1}I^{4,18}_3\right)=0.
\label{del418}
\ee
Here the finite term in $\delta^{IR}_{2}$ gives a simple pole which cancels against that produced by $\delta^{IR}_{3,1}$; so that as we have said before, we need to include finite terms (and later, $O(\epsilon)$ and higher powers) in our IR counterterms in order to compensate for the difference between the pole terms in our IR counterterms and the standard ones. One sees a pattern developing in the results for $I^{4,18}_1$, $I^{4,18}_2$ and $I^{4,18}_3$ and also those for $I^{3,5}_1$, $I^{3,5}_2$ in Eq.~\eqref{oldexp} and indeed $I^{2,2}_1$ in Eq.~\eqref{i22}; we shall have more to say about such patterns in due course.

As we have done before, we can check that the omission of $F_{1,2,3}$ from the one, two and three loop IR counterterms has no effect on the final result for the UV counterterm $\Delta_{4,18}$; at this level this requires a complex interplay between the results from different orders.

We now consider another four-loop example; the diagram is displayed on the LHS of Eq.~\eqref{exdiagd}. The UV counterterm may readily be evaluated by IR rearrangement without the need for IR counterterms. However, here we choose a somewhat perverse way of threading the momentum, as shown in the diagram on the LHS of Eq.~\eqref{exdiagd}. This choice introduces the two-loop IR-divergent structure $\Ical_{2b}$, which we have seen already in Eq.~\eqref{IRdiags}; but this is the first time we have encountered it in the evaluation of a UV counterterm.
As we have already explained in Section 2, the general results of Ref.~\cite{chet3,herz} imply that the IR counterterm may be deduced from the CVG, which is the same as for $\Ical_{2a}$, leading to $(\Ical_{2b})_{IR}=(\Ical_{2a})_{IR}=\ncal_2$. However, it is not clear that this  reasoning applies to our alternative counterterms. We have assumed it does and we have incorporated this assumption into our description of the alternative counterterms in Section 3 and applied it throughout Tables~\ref{IRres12}, \ref{IRres3} and \ref{IRres4}. Nevertheless we should check that this is a reasonable assumption and here we are performing this check in one of the simplest settings. We mention here that the IR-divergent structures encountered so far, namely $\Ical_1$, $\Ical_{2a}$ and $\Ical_{3,1d}$ are simply related to the severed graphs $G^{\rm{sev}}$ defined in Section 3, namely by switching the position of one of the external propagators in $G^{\rm{sev}}$. $\Ical_{2b}$ is not of this form and so it is here, if anywhere, that a discrepancy might reveal itself. We have
\begin{align}
\Delta_{4,25}=&\,-\Kcal\Rbar^*\left(\tikz[scale=0.6,baseline=(vert_cent.base),scale=1]{
	%\draw[help lines] (-3,-3) grid (3,3);
	\node (vert_cent) {\hspace{-13pt}$\phantom{-}$};
\draw [bend left=30] (-1,0) to (1,1);
         \draw [bend right=30] (-1,0) to (0,0);
           \draw [bend left=30] (-1,0) to (0,0);
\draw (0,0) -- (1,1);
\draw (0,0) -- (1,0);
\draw (1,1) -- (1,0);
\draw (-1.25,0) -- (-1,0);
\draw (1.25,0) -- (1,0);
\filldraw (0,0.9) circle [radius=0.1];
\draw (0:1cm) arc (0:-180:1cm) ;
}\right)=-\Kcal\Bigl[\tikz[scale=0.6,baseline=(vert_cent.base),scale=1]{
	%\draw[help lines] (-3,-3) grid (3,3);
	\node (vert_cent) {\hspace{-13pt}$\phantom{-}$};
\draw [bend left=30] (-1,0) to (1,1);
         \draw [bend right=30] (-1,0) to (0,0);
           \draw [bend left=30] (-1,0) to (0,0);
\draw (0,0) -- (1,1);
\draw (0,0) -- (1,0);
\draw (1,1) -- (1,0);
\draw (-1.25,0) -- (-1,0);
\draw (1.25,0) -- (1,0);
\filldraw (0,0.9) circle [radius=0.1];
\draw (0:1cm) arc (0:-180:1cm) ;
}+\Delta\left(\tikz[scale=0.6,baseline=(vert_cent.base)]{
  \node (vert_cent) {\hspace{-13pt}$\phantom{-}$};
  \filldraw (1.3,0) circle [radius=0.1]; \filldraw (0.1,0) circle [radius=0.1];\draw
        (0.7,0) ++(0:0.6cm) arc (0:360:0.6cm and 0.4cm);
}\right)\tikz[scale=0.6,baseline=(vert_cent.base),scale=1]{
	%\draw[help lines] (-3,-3) grid (3,3);
	\node (vert_cent) {\hspace{-13pt}$\phantom{-}$};
	\draw (0,0) circle[radius=1cm];
           \draw [bend right=40] (-1,0) to (0,1);
\draw (-1.25,0) -- (1.25,0);
\filldraw (135:1) circle [radius=0.1];
}\nn
&+\left(\,\tikz[scale=0.6,baseline=(vert_cent.base)]{
  \node (vert_cent) {\hspace{-13pt}$\phantom{-}$};
  \draw (0,-.4)--(0,.4);
        \draw (0,.5) circle [radius=0.1];
\draw (0,-.5) circle [radius=0.1];
\filldraw (0,0) circle [radius=0.1];
}\,\right)_{IR}\Bigl\{\tikz[scale=0.6,baseline=(vert_cent.base),scale=1]{
\draw (180:1cm) arc (180:0:1cm) ;
\draw [bend right=40] (0,1) to (1,0);
\draw [bend left=40] (0,1) to (-1,0);
\filldraw (45:1) circle [radius=0.1];
\draw (-1.25,0) -- (1.25,0);
}+\Delta\left(\tikz[scale=0.6,baseline=(vert_cent.base)]{
  \node (vert_cent) {\hspace{-13pt}$\phantom{-}$};
  \filldraw (1.3,0) circle [radius=0.1]; \filldraw (0.1,0) circle [radius=0.1];\draw
        (0.7,0) ++(0:0.6cm) arc (0:360:0.6cm and 0.4cm);
}\right)\tikz[scale=0.6,baseline=(vert_cent.base),scale=1]{
	%\draw[help lines] (-3,-3) grid (3,3);
	\node (vert_cent) {\hspace{-13pt}$\phantom{-}$};
	\draw (0,0) circle[radius=1cm];
\draw (-1.25,0) -- (1.25,0);
\filldraw (0,-1) circle [radius=0.1];
}+\Delta\left(\tikz[scale=0.6,baseline=(vert_cent.base),scale=1]{
\draw (180:1cm) arc (180:0:1cm) ;
\draw [bend right=40] (0,1) to (1,0);
\draw [bend left=40] (0,1) to (-1,0);
\filldraw (45:1) circle [radius=0.1];
\draw (-1,0) -- (1,0);
}\right)\Bigr\}\nn
&+\left(\tikz[scale=0.6,baseline=(vert_cent.base),scale=1]{
\filldraw  (0,0.5) circle [radius=0.1];
\draw (-1,0) circle [radius=0.1];
\draw (1,0) circle [radius=0.1];
\draw (0,0) circle [radius=0.1];
\draw (0,1) --(0,0.1);
\draw [bend right=40] (0,1) to (-1,0.1);
\draw [bend left=40] (0,1) to (1,0.1); 
}\right)_{IR}\Bigl\{\tikz[scale=0.6,baseline=(vert_cent.base),scale=1]{
	%\draw[help lines] (-3,-3) grid (3,3);
	\node (vert_cent) {\hspace{-13pt}$\phantom{-}$};
	\draw (0,0) circle[radius=1cm];
           \draw [bend left=40] (1,0) to (0,1);
\draw (-1.25,0) -- (-1,0);
\draw (1.25,0) -- (1,0);
}+\Delta\left(\tikz[scale=0.6,baseline=(vert_cent.base)]{
  \node (vert_cent) {\hspace{-13pt}$\phantom{-}$};
  \filldraw (1.3,0) circle [radius=0.1]; \filldraw (0.1,0) circle [radius=0.1];\draw
        (0.7,0) ++(0:0.6cm) arc (0:360:0.6cm and 0.4cm);
}\right)\tikz[scale=0.6,baseline=(vert_cent.base)]{
  \node (vert_cent) {\hspace{-13pt}$\phantom{-}$};
  \draw (-0.1,0)--(0.1,0)
        (0.7,0) ++(0:0.6cm) arc (0:360:0.6cm and 0.4cm)
        (1.3,0)--(1.5,0);
}+\Delta\left(\tikz[scale=0.6,baseline=(vert_cent.base),scale=1]{
	%\draw[help lines] (-3,-3) grid (3,3);
	\node (vert_cent) {\hspace{-13pt}$\phantom{-}$};
	\draw (0,0) circle[radius=1cm];
           \draw [bend left=40] (1,0) to (0,1);
\filldraw (-1,0) circle [radius=0.1];
}\right)\Bigr\}\Bigr].
\label{exdiagd}
\end{align}
As usual, we denote the terms in Eq.~\eqref{exdiagd} with no IR divergence by $I_0^{4,25}$ and those multiplied by one and two IR-divergent structures by $I_1^{4,25}$ and $I_2^{4,25}$ respectively. The full four-loop diagram (the first term in $I_0^{4,25}$) is denoted $G^{4,25}$ and $G^{4,25}_1$ and  $G^{4,25}_2$  are obtained by successively removing $\Ical_1$ and $\Ical_{2b}$ (in other words, the IR-divergent substructures of $G^{4,25}$) from $G_{4,25}$; these diagrams are the first terms in $I_1^{4,25}$ and $I_2^{4,25}$ respectively. We then have, as usual,
\be
\Delta_{4,25}=-\Kcal\left(I^{4,25}_0+\ncal_1I^{4,25}_1+\ncal_{2}I^{4,25}_2\right)
\label{del425exp}
\ee
where
\be
I_0^{4,25}=\Rbar G^{4,25},\quad I_1^{4,25}=RG_1^{4,25}, \quad I_2^{4,25}=RG_2^{4,25}.
\label{i425defs}
\ee
We find 
\begin{subequations}\label{i425}
\begin{align}
I^{4,25}_0=&\,L(1,1)L(2,\tfrac12\epsilon,1,1,1)L(1,1+\tfrac32\epsilon)-\tfrac{2}{\epsilon}L(1,1)L(1,2)L(2+\tfrac12\epsilon,\tfrac12\epsilon)\nn
=&\,-\tfrac23\tfrac{1}{\epsilon^4}+\tfrac{2}{\epsilon^3}+\tfrac{47}{6}\tfrac{1}{\epsilon^2}+\left(\tfrac{51}{2}-\tfrac{11}{3}\zeta_3\right)\tfrac{1}{\epsilon}+\ldots,\\
I^{4,25}_1=&\,L(1,1)L(1,2)L(1,1+\epsilon)-\tfrac{2}{\epsilon}L(1,2)L(1,1+\tfrac12\epsilon)+\Delta_{3,7}=-\frac{11}{\epsilon}+\ldots,\\
 I^{4,25}_2=&\,L(1,1)L(1,1+\tfrac12\epsilon)-\tfrac{2}{\epsilon}L(1,1)+\Delta_{2,2}=\frac{11}{2}+\ldots\label{i425:c}
\end{align}
\end{subequations}
 Using the results for $\ncal_1$ and $\ncal_{2}$ from Eq.~\eqref{IR12}, we easily expand Eq.~\eqref{del425exp} as 
\be
\Delta_{4,25}=\tfrac23\tfrac{1}{\epsilon^4}-\tfrac{2}{\epsilon^3}+\tfrac{19}{6}\tfrac{1}{\epsilon^2}
-\left(\tfrac52-2\zeta_3\right)\tfrac{1}{\epsilon},
\ee
which is the result obtained in Ref.~\cite{kaz} and quoted in Ref.~\cite{klein} and in our Eq.~\eqref{divs:l}. 
We also find, using Eq.~\eqref{ir125},
\be
\delta_{4,25}=\Kcal\left(\delta^{IR}_1I^{4,25}_1+\delta^{IR}_{2}I^{4,25}_2\right)=0,
\label{del425}
\ee
where as stated before, we have $(\Ical_{2b})_{IR}=(\Ical_{2a})_{IR}=\ncal_2$ and we have assumed $(\Ical'_{2b})_{IR}=(\Ical'_{2a})_{IR}=\ncal'_2$. It is noteworthy that both leading and non-leading poles cancel in $I^{4,25}_2$ (and also $I^{4,25}_1$), which is necessary for the finiteness of $\delta^{4,25}$. Together with the factor of $(-2)$ between the leading terms in $I^{4,25}_1$ and $I^{4,25}_2$, this also ensures that as usual omitting the factors of $F_{1,2}$ from $C_{1,2}$ has no effect. The fact that, as we see in Eq.~\eqref{del425}, our AIR counterterms once again produce the same UV counterterm as the standard ones is evidence that we are correct in assuming that our new IR counterterms are, like the standard ones, determined by the CVG. 

As we have mentioned, the three and four-loop UV divergences relevant for the $\beta$-function in $\phi^4$ theory may all be evaluated using IR rearrangement without the use of IR counterterms. Accordingly, in Sections 3 and 4 we have only covered a small number of examples where we deliberately 
chose a momentum nullification which introduced IR divergences, in order to show the ideas and issues involved in our alternative counterterms in simple realistic situations. In Appendix B,  however, we shall present a comprehensive discussion of the five-loop diagrams (again, focussing on those contributing to the $\beta$-function) in which IR divergences were found to be unavoidable in Ref.~\cite{klein}\footnote{Of course, subsequent advances might by now have provided alternative means to evaluate these diagrams.}; but concentrating on those involving double propagators. We shall complete the picture by discussing five-loop IR-divergent diagrams with no double propagators in Appendix C. 
The procedure in Appendix B follows that for $\Delta_{4,25}$ in Eq.~\eqref{exdiagd}; we simply expand quantities such as $I^{4,25}_1$, $I^{4,25}_2$ which appear as coefficients of IR counterterms to requisite orders in $\epsilon$ (as we did in Eq.~\eqref{i425}), and combine with explicit results for $\delta^{IR}_2$ etc in Eq.~\eqref{irdiffs} to check that differences such as $\delta_{4,25}$ in Eq.~\eqref{del425} are zero.

However, in the next Section we shall consider a different approach which reduces the required checking to a limited set of conditions. As we shall see, it will furthermore seem somewhat natural to expect that any viable set of alternative counterterms should satisfy these conditions.

\section{Towards a general proof?}

We start by pointing out that the validity of our AIR counterterms rests on two different features of the calculations, as may be observed in each of the checks we have performed so far. Firstly there are properties of combinations of diagrams which arise in applying the $R$ operation, such as $I_1^{4,25}$ and $I_2^{4,25}$ (as defined in Eq.~\eqref{i425defs}. These properties involve the absence of poles which might {\it a priori} be expected at a given loop order; and also unexpected relations between these combinations. We see both these features very clearly in the expansions of $I_1^{4,25}$ and $I_2^{4,25}$ given in Eq.~\eqref{i425}, as we pointed out towards the end of Section 4 (similar features may also be seen in $I_1^{4,18}$, $I_2^{4,18}$ and $I_3^{4,18}$ in Eq.~\eqref{i418}). These properties turn out to depend fairly straightforwardly on features of the $R$ operation, independently of our AIR counterterms. Secondly there are inherent properties of the AIR counterterms themselves, such as the fact that 
\be
\frac{\delta^{IR}_{3,1}}{\delta^{IR}_2}=-\frac{I_2^{4,18}}{I_3^{4,18}}+O(1),
\ee
 which we see in Eqs.~\eqref{ir125}, \eqref{i418} and which was necessary for Eq.~\eqref{del418} to work; this is less transparent, at least at present. (For simplicity of exposition we shall assume for the present that $\delta^{IR}_1=0$, as is the case for our AIR counterterms with the inclusion of the $F_1$ factor in $C_1$ in \eqref{cdef}; we shall return later to the possibility of relaxing this condition, in order to allow for dropping the $F_L$ factor in $C_L$.) We shall start by sketching an explanation of the relation apparent in Eq.~\eqref{i425} between the leading terms in $I^{4,25}_1$ and $I^{4,25}_2$. Recall the definition of  $I^{4,25}_1$ and $I^{4,25}_2$ in Eq.~\eqref{i425defs}, where for clarity the diagrams $G^{4,25}$, $G_1^{4,25}$ and $G_2^{4,25}$ are depicted here:
\be
G^{4,25}=\tikz[scale=0.6,baseline=(vert_cent.base),scale=1]{
	%\draw[help lines] (-3,-3) grid (3,3);
	\node (vert_cent) {\hspace{-13pt}$\phantom{-}$};
\draw [bend left=30] (-1,0) to (1,1);
         \draw [bend right=30] (-1,0) to (0,0);
           \draw [bend left=30] (-1,0) to (0,0);
\draw (0,0) -- (1,1);
\draw (0,0) -- (1,0);
\draw (1,1) -- (1,0);
\draw (-1.25,0) -- (-1,0);
\draw (1.25,0) -- (1,0);
\filldraw (0,0.9) circle [radius=0.1];
\draw (0:1cm) arc (0:-180:1cm) ;
},\quad
G^{4,25}_1=\tikz[scale=0.6,baseline=(vert_cent.base),scale=1]{
\draw (180:1cm) arc (180:0:1cm) ;
\draw [bend right=40] (0,1) to (1,0);
\draw [bend left=40] (0,1) to (-1,0);
\filldraw (45:1) circle [radius=0.1];
\draw (-1.25,0) -- (1.25,0);
},\quad G^{4,25}_2=\tikz[scale=0.6,baseline=(vert_cent.base),scale=1]{
	%\draw[help lines] (-3,-3) grid (3,3);
	\node (vert_cent) {\hspace{-13pt}$\phantom{-}$};
	\draw (0,0) circle[radius=1cm];
           \draw [bend left=40] (1,0) to (0,1);
\draw (-1.25,0) -- (-1,0);
\draw (1.25,0) -- (1,0);
}.
\label{G12def}
\ee
Since $G^{4,25}_2$ has no IR divergences, Eq.~\eqref{Rdef} implies immediately that $I^{4,25}_2$ is finite, despite being a two-loop quantity which might be expected to be $O(1/\epsilon^2)$; in other words, the leading and non-leading poles are absent (as we have already seen explicitly in Eq.~\eqref{i425:c}). However in the case of $I^{4,25}_1$, which does have IR divergences, Eq.~\eqref{Rdef} is replaced by Eq.~\eqref{rstar:a}.
In this instance, we may write Eq.~\eqref{rstar:a} explicitly as
\begin{align}
\Kcal R^*G^{4,25}_1=&\,\Kcal\Bigl[\tikz[scale=0.6,baseline=(vert_cent.base),scale=1]{
\draw (180:1cm) arc (180:0:1cm) ;
\draw [bend right=40] (0,1) to (1,0);
\draw [bend left=40] (0,1) to (-1,0);
\filldraw (45:1) circle [radius=0.1];
\draw (-1.25,0) -- (1.25,0);
}+\Delta\left(\tikz[scale=0.6,baseline=(vert_cent.base)]{
  \node (vert_cent) {\hspace{-13pt}$\phantom{-}$};
  \filldraw (1.3,0) circle [radius=0.1]; \filldraw (0.1,0) circle [radius=0.1];\draw
        (0.7,0) ++(0:0.6cm) arc (0:360:0.6cm and 0.4cm);
}\right)\tikz[scale=0.6,baseline=(vert_cent.base),scale=1]{
	%\draw[help lines] (-3,-3) grid (3,3);
	\node (vert_cent) {\hspace{-13pt}$\phantom{-}$};
	\draw (0,0) circle[radius=1cm];
\draw (-1.25,0) -- (1.25,0);
\filldraw (0,-1) circle [radius=0.1];
}+\Delta\left(\tikz[scale=0.6,baseline=(vert_cent.base),scale=1]{
\draw (180:1cm) arc (180:0:1cm) ;
\draw [bend right=40] (0,1) to (1,0);
\draw [bend left=40] (0,1) to (-1,0);
\filldraw (45:1) circle [radius=0.1];
\draw (-1,0) -- (1,0);
}\right)\nn
&+\left(\,\tikz[scale=0.6,baseline=(vert_cent.base)]{
  \node (vert_cent) {\hspace{-13pt}$\phantom{-}$};
  \draw (0,-.4)--(0,.4);
        \draw (0,.5) circle [radius=0.1];
\draw (0,-.5) circle [radius=0.1];
\filldraw (0,0) circle [radius=0.1];
}\,\right)_{IR}\Bigl\{\tikz[scale=0.6,baseline=(vert_cent.base),scale=1]{
	%\draw[help lines] (-3,-3) grid (3,3);
	\node (vert_cent) {\hspace{-13pt}$\phantom{-}$};
	\draw (0,0) circle[radius=1cm];
           \draw [bend left=40] (1,0) to (0,1);
\draw (-1.25,0) -- (-1,0);
\draw (1.25,0) -- (1,0);
}+\Delta\left(\tikz[scale=0.6,baseline=(vert_cent.base)]{
  \node (vert_cent) {\hspace{-13pt}$\phantom{-}$};
  \filldraw (1.3,0) circle [radius=0.1]; \filldraw (0.1,0) circle [radius=0.1];\draw
        (0.7,0) ++(0:0.6cm) arc (0:360:0.6cm and 0.4cm);
}\right)\tikz[scale=0.6,baseline=(vert_cent.base)]{
  \node (vert_cent) {\hspace{-13pt}$\phantom{-}$};
  \draw (-0.1,0)--(0.1,0)
        (0.7,0) ++(0:0.6cm) arc (0:360:0.6cm and 0.4cm)
        (1.3,0)--(1.5,0);
}+\Delta\left(\tikz[scale=0.6,baseline=(vert_cent.base),scale=1]{
	%\draw[help lines] (-3,-3) grid (3,3);
	\node (vert_cent) {\hspace{-13pt}$\phantom{-}$};
	\draw (0,0) circle[radius=1cm];
           \draw [bend left=40] (1,0) to (0,1);
\filldraw (-1,0) circle [radius=0.1];
}\right)\Bigr\}\Bigr]=0.
\label{i425a}
\end{align}
 Trivially rearranging Eq.~\eqref{i425a}, and again reading off $I^{4,25}_1$, $I^{4,25}_2$ from Eq.~\eqref{i425defs} we may write
\begin{align}
\Kcal I^{4,25}_1=\Kcal\left(RG_1^{4,25}\right)=&\,\Kcal\Bigl[\tikz[scale=0.6,baseline=(vert_cent.base),scale=1]{
\draw (180:1cm) arc (180:0:1cm) ;
\draw [bend right=40] (0,1) to (1,0);
\draw [bend left=40] (0,1) to (-1,0);
\filldraw (45:1) circle [radius=0.1];
\draw (-1.25,0) -- (1.25,0);
}+\Delta\left(\tikz[scale=0.6,baseline=(vert_cent.base)]{
  \node (vert_cent) {\hspace{-13pt}$\phantom{-}$};
  \filldraw (1.3,0) circle [radius=0.1]; \filldraw (0.1,0) circle [radius=0.1];\draw
        (0.7,0) ++(0:0.6cm) arc (0:360:0.6cm and 0.4cm);
}\right)\tikz[scale=0.6,baseline=(vert_cent.base),scale=1]{
	%\draw[help lines] (-3,-3) grid (3,3);
	\node (vert_cent) {\hspace{-13pt}$\phantom{-}$};
	\draw (0,0) circle[radius=1cm];
\draw (-1.25,0) -- (1.25,0);
\filldraw (0,-1) circle [radius=0.1];
}+\Delta\left(\tikz[scale=0.6,baseline=(vert_cent.base),scale=1]{
\draw (180:1cm) arc (180:0:1cm) ;
\draw [bend right=40] (0,1) to (1,0);
\draw [bend left=40] (0,1) to (-1,0);
\filldraw (45:1) circle [radius=0.1];
\draw (-1,0) -- (1,0);
}\right)\Bigr]\nn
=-&\Kcal\Bigl[\left(\,\tikz[scale=0.6,baseline=(vert_cent.base)]{
  \node (vert_cent) {\hspace{-13pt}$\phantom{-}$};
  \draw (0,-.4)--(0,.4);
        \draw (0,.5) circle [radius=0.1];
\draw (0,-.5) circle [radius=0.1];
\filldraw (0,0) circle [radius=0.1];
}\,\right)_{IR}\Bigl\{\tikz[scale=0.6,baseline=(vert_cent.base),scale=1]{
	%\draw[help lines] (-3,-3) grid (3,3);
	\node (vert_cent) {\hspace{-13pt}$\phantom{-}$};
	\draw (0,0) circle[radius=1cm];
           \draw [bend left=40] (1,0) to (0,1);
\draw (-1.25,0) -- (-1,0);
\draw (1.25,0) -- (1,0);
}+\Delta\left(\tikz[scale=0.6,baseline=(vert_cent.base)]{
  \node (vert_cent) {\hspace{-13pt}$\phantom{-}$};
  \filldraw (1.3,0) circle [radius=0.1]; \filldraw (0.1,0) circle [radius=0.1];\draw
        (0.7,0) ++(0:0.6cm) arc (0:360:0.6cm and 0.4cm);
}\right)\tikz[scale=0.6,baseline=(vert_cent.base)]{
  \node (vert_cent) {\hspace{-13pt}$\phantom{-}$};
  \draw (-0.1,0)--(0.1,0)
        (0.7,0) ++(0:0.6cm) arc (0:360:0.6cm and 0.4cm)
        (1.3,0)--(1.5,0);
}\nn
&+\Delta\left(\tikz[scale=0.6,baseline=(vert_cent.base),scale=1]{
	%\draw[help lines] (-3,-3) grid (3,3);
	\node (vert_cent) {\hspace{-13pt}$\phantom{-}$};
	\draw (0,0) circle[radius=1cm];
           \draw [bend left=40] (1,0) to (0,1);
\filldraw (-1,0) circle [radius=0.1];
}\right)\Bigr\}\Bigr]=-\Kcal\left(\ncal_1RG_2^{4,25}\right)=-\Kcal\left(\tfrac{2}{\epsilon}I^{4,25}_2\right).
\label{i425b}
\end{align}
We can summarise what we have found as
\begin{subequations}\label{i425c}
\begin{align}
I_1^{4,25}+\ncal_1I_2^{4,25}=&\,O(1),\label{i425c:a}\\
I_2^{4,25}=&\,O(1).\label{i425c:b}
\end{align}
\end{subequations}
We have therefore accounted for the relation between $I^{4,25}_1$ and $I^{4,25}_2$ in Eq.~\eqref{i425} and also for the cancellation of the leading and next-to-leading poles in these quantities. Note that due to the appearance of $\Rbar$ rather than $R$  in the expression for $I^{4,25}_0$ in Eq.~\eqref{i425defs}, we do not find any relation between $I^{4,25}_0$ and $I^{4,25}_1$, $I^{4,25}_2$. Furthermore it is apparent that (since $\delta_1^{IR}=0$) the only additional piece of information we require to ensure that $\delta_{4,25}=0$ in Eq.~\eqref{del425} is a (rather trivial) condition on $\delta^{IR}_2$, namely 
\be
\delta^{IR}_2=O(1),
\label{delrel1c}
\ee
which is clearly valid, as we see in Eq.~\eqref{irdiffs:b}.

 %We just pause here to remark that $we have already seen that I^{4,18}_1$ and $I^{4,18}_2$ share a similar relation 
We can provide a similar analysis for $\Delta_{4,18}$, though it is clear that this is a less trivial case since both the second and third terms on the right-hand side of Eq.~\eqref{del418} contain simple poles (from $I^{4,18}_2$ and $\delta_{3,1}^{IR}$) which must conspire to cancel. We observe from Eq.~\eqref{i418} that 
\begin{subequations}\label{i418rel}
\begin{align}
I_2^{4,18}-\Delta_{1,1}I_3^{4,18}=&\,0,\\
I_3^{4,18}=&\,O(1),
\end{align}
\end{subequations}
and since $\delta^{IR}_1=0$, we may rewrite Eq.~\eqref{del418} as
\begin{align}
\delta_{4,18}=\Kcal\left(2\delta^{IR}_{2}I^{4,18}_2+\delta^{IR}_{3,1}I^{4,18}_3\right)=&\,\Kcal\Bigl[2\delta^{IR}_{2}(I^{4,18}_2-\Delta_{1,1}I^{4,18}_3)\nn
&+(\delta^{IR}_{3,1}+2\Delta_{1,1}\delta^{IR}_{2})I^{4,18}_3\Bigr],
\label{newdel418}
\end{align}
so that, in the light of Eq.~\eqref{i418rel}, Eq.~\eqref{del418} is valid provided Eq.~\eqref{delrel1c} is valid, together with
\be
\delta^{IR}_{3,1}+2\Delta_{1,1}\delta^{IR}_{2}=O(1).
\label{delrel1}
\ee
Eq.~\eqref{delrel1} may straightforwardly be verified using Eqs.~\eqref{irdiffs:b}, \eqref{irdiffs:c} and \eqref{divs:a}. As in the case of $\Delta_{4,25}$, the validity of our AIR counterterms rests on an RG property (Eq.~\eqref{i418rel}) coupled with conditions on the IR counterterms themselves, namely Eqs.~\eqref{delrel1c} as before, now accompanied by Eq.~\eqref{delrel1}.

We shall soon see how a similar combination of properties derived from RG arguments and properties depending on the definitions of IR counterterms plays a crucial role at higher orders. Firstly one can see how to generalise the process applied to  the graphs $G_1^{4,25}$ and $G_2^{4,25}$ which resulted in Eq.~\eqref{i425c}. For any $L$-loop graph $G^{L,m}$ there is a sequence $G^{L,m}_n$, $n=1,2,\ldots N$ formed by removing one, two, three... $N$-loop IR divergences from $G^{L,m}$, and a corresponding sequence $I^{L,m}_n=RG^{L,m}_n$, $n=1,2,\ldots N$. The sequence terminates in a graph $G^{L,m}_N$ which has no IR divergences and for which $I^{L,m}_N=RG^{L,m}_N=R^*G^{L,m}_N$ is finite. Applying the $R^*$ procedure to $G_1^{L,m}$ generates a linear relation $R_1(I_1^{L,m},I_2^{L,m},\ldots I_N^{L,m})=O(1)$; then applying the $R^*$ procedure to $G_2^{L,m}$ generates a linear relation $R_2(I_2^{L,m},I_3^{L,m}\ldots I_N^{L,m})=O(1)$, and so on, up to $R_{N}=I_{N}^{L,m}=O(1)$.This is the generalisation of Eqs.~\eqref{i425c}, \eqref{i418rel}. We can then derive an expression for $\delta_{L,m}$, the difference between the UV divergences $\Delta_{L,m}$, $\Delta'_{L,m}$ computed using the standard and AIR counterterms, which is expressed in terms of $R_1,R_2,\ldots R_{N-1}$ with coefficients which are combinations of the $\delta^{IR}_{p,q}$ (the differences between the alternative  and standard IR counterterms). This is the generalisation of Eq.~\eqref{newdel418}. This will manifestly be zero provided certain conditions on the $\delta^{IR}_{p,q}$  are satisfied - the generalisation of Eqs.~\eqref{delrel1c}, \eqref{delrel1}. We shall perform this exercise for each of the diagrams considered in Appendix B, showing that the resulting conditions on the $\delta^{IR}_{p,q}$ are satisfied by the expressions as computed in Eq.~\eqref{irdiffs}.
%Just as in our simple example, we find relations between $I^{L,m}_n$ and $I^{L,m}_N$; but since $I^{L,m}_n$ may have IR divergences of several %loop orders, such relations may be more complicated than we have seen here. We shall refrain from showing the details in other cases. We should %emphasise that in any case one should beware of using these relations, since they only relates the pole terms in $I^{L,m}_n$ to those in  $I^{L,m}%_{n'}$; but in the computation of $\Delta^{L,m}$, $I^{L,m}_n$ is multiplied by an IR counterterm and therefore the finite terms are important too.

We start by gathering together for reference the expressions for the five-loop analogues of $\delta_{4,18}$ and $\delta_{4,25}$ which were considered in Appendix B. As derived in Eqs.~\eqref{g67res}, \eqref{g101res}, \eqref{del537}, \eqref{del5107},  \eqref{del553}, \eqref{del575}, \eqref{del5111}, \eqref{del573}, \eqref{del593}, they are
\begin{subequations}\label{delsum}
\begin{align}
\delta_{5,67}=&\,\Kcal\left(\delta^{IR}_1I^{5,67}_1+\delta^{IR}_{2}I^{5,67}_2\right),\label{delsum:a}\\
\delta_{5,101}=&\,\Kcal\left(\delta^{IR}_1I^{5,101}_1+\delta^{IR}_{2}I^{5,101}_2\right),\label{delsum:b}\\
\delta_{5,37}=&\,\Kcal\left(\delta_1^{IR}I_1^{5,37}+\delta_{4,3}^{IR}I_2^{5,37}\right),\label{delsum:c}\\
\delta_{5,107}=&\,\Kcal\left(\delta^{IR}_1I^{5,107}_1+2\delta^{IR}_{2}I^{5,107}_2+\delta^{IR}_{3,1}I^{5,107}_3\right),\label{delsum:d}\\
\delta_{5,53}=&\,\Kcal\left(\delta^{IR}_1I^{5,53}_1+\delta^{IR}_{2}I^{5,53}_2+\delta^{IR}_{3,2}I^{5,53}_3+\delta^{IR}_{4,2}I^{5,53}_4\right),\label{delsum:e}\\
\delta_{5,75}=&\,\Kcal\left(\delta^{IR}_1I^{5,75}_1+\delta^{IR}_{2}I^{5,75}_2+\delta^{IR}_{3,2}I^{5,75}_3+\delta^{IR}_{4,2}I^{5,75}_4\right),\label{delsum:f}\\
\delta_{5,111}=&\,\Kcal\left(\delta^{IR}_1I^{5,111}_1+2\delta^{IR}_{2}I^{5,111}_2+\delta^{IR}_{3,1}I^{5,111}_3+\delta^{IR}_{4,1}I^{5,111}_4\right),\label{delsum:g}\\
\delta_{5,73}=&\,\Kcal\left(2\delta^{IR}_1I^{5,73}_1+(\ncal_1^{\prime2}-\ncal_1^2)I^{5,73}_2+\delta^{IR}_{3,3}I^{5,73}_3+\delta^{IR}_{4,4}I^{5,73}_4\right),\label{delsum:h}\\
\delta_{5,93}=&\,\Kcal\left(2\delta^{IR}_1I^{5,93}_1+(\ncal_1^{\prime2}-\ncal_1^2)I^{5,93}_2+\delta^{IR}_{3,3}I^{5,93}_3+\delta^{IR}_{4,4}I^{5,93}_4\right).\label{delsum:i}
\end{align}
\end{subequations}
The detailed derivations of these equations and the definitions of quantities such as $I^{5,67}_1$ may be found in Appendix B.

The first two diagrams are fairly straightforward to deal with, and we only require the finiteness of the final $I^{L,m}_N$ (the analogue of Eq.~\eqref{i425c:b}) rather than relations such as Eq.~\eqref{i425c:a}.
For instance, looking at $\delta_{5,67}$ as defined in Eq.~\eqref{delsum:a}, firstly we have $\delta^{IR}_1=0$. $I^{5,67}_2$ is the last in the chain of $I^{5,67}_n$ and hence is $O(1)$. We see from Eq.~\eqref{delsum:a} that $\delta_{5,67}$ will vanish provided $\delta^{IR}_{2}$ is $O(1)$ too. Of course we have already met this condition in Eq.~\eqref{delrel1c}, so there is nothing new here.  A similar argument applies to $\delta_{5,101}$. 

Next we consider $\delta_{5,37}$ in Eq.~\eqref{delsum:c}. As we pointed out in Appendix B, since $\delta_1^{IR}=0$ and $I^{5,37}_2$ is finite due to being the last in the chain of $I^{5,37}_n$, we simply require 
\be
\delta^{IR}_{4,3}=O(1),
\label{delrel1a}
\ee
which we can check in Eq.~\eqref{irdiffs:i}.
Turning now to Eq.~\eqref{delsum:d}, we see that $\delta_{5,107}$ involves $\delta^{IR}_{3,1}$, and in fact this is a similar case to $\delta_{4,18}$ which we discussed a short while ago. We have, following the argument which led to Eq.~\eqref{i425c} and was generalised earlier in this Section,
\begin{subequations}\label{i5107rels}
\begin{align}
I^{5,107}_1+2\ncal_1I^{5,107}_2+(\ncal_1)^2I^{5,107}_3=&\,O(1),\\
I^{5,107}_2+\ncal_1I^{5,107}_3=&\,O(1),\\
I^{5,107}_3=&\,O(1),
\end{align}
\end{subequations}
and we can write
\begin{align}
\delta_{5,107}=\Kcal\left(2\delta^{IR}_{2}I^{5,107}_2+\delta^{IR}_{3,1}I^{5,107}_3\right)=&\,\Kcal\Bigl[2\delta^{IR}_{2}(I^{5,107}_2+\ncal_1I^{5,107}_3)\nn
&+(\delta^{IR}_{3,1}-2\ncal_1\delta^{IR}_{2})I^{5,107}_3\Bigr],
\end{align}
so that, in the light of Eqs.~\eqref{i5107rels} and \eqref{iruv}, Eq.~\eqref{delsum:d} is again valid provided Eq.~\eqref{delrel1} is satisfied.

Now turning to $\delta_{5,53}$ in Eq.~\eqref{delsum:e}, since $\delta^{IR}_{4,2}$ (as we see in Eq.~\eqref{irdiffs:h}) is divergent, we again need to use the relations between the $I^{5,53}_n$. We have, again following the argument which led to Eq.~\eqref{i425c},
\begin{subequations}
\begin{align}
I^{5,53}_1+\ncal_1I^{5,53}_2+\ncal_{2}I^{5,53}_3+\ncal_{3,2}I^{5,53}_4=&\,O(1),\\
I^{5,53}_2+\ncal_1I^{5,53}_3+\ncal_{2}I^{5,53}_4=&\,O(1),\\
I^{5,53}_3+\ncal_1I^{5,53}_4=&\,O(1),\\
I^{5,53}_4=&\,O(1),
\end{align}
\end{subequations}
We then see that
\begin{align}
\delta_{5,53}=\Kcal\left(\delta^{IR}_{2}I^{5,53}_2+\delta^{IR}_{3,2}I^{5,53}_3+\delta^{IR}_{4,2}I^{5,53}_4\right)=&\,\Kcal\Bigl[\delta^{IR}_{2}(I^{5,53}_2+\ncal_1I^{5,53}_3+\ncal_{2}I^{5,53}_4)\nn
&+(\delta^{IR}_{3,2}-\ncal_1\delta^{IR}_{2})(I^{5,53}_3+\ncal_1I^{5,53}_4)\nn
&+(\delta^{IR}_{4,2}+\Delta_{1,1}\delta^{IR}_{3,2}+\Delta_{2,2}\delta^{IR}_{2})I^{5,53}_4\Bigr],
\end{align}
and so vanishes provided Eq.~\eqref{delrel1c} is valid, together with
\begin{subequations}\label{delrel}
\begin{align}
\delta^{IR}_{3,2}+\Delta_{1,1}\delta^{IR}_{2}=&\,O(1),\label{delrel:a}\\
\delta^{IR}_{4,2}+\Delta_{1,1}\delta^{IR}_{3,2}+\Delta_{2,2}\delta^{IR}_{2}=&\,O(1),\label{delrel:b}
\end{align}
\end{subequations}
where we have used Eq.~\eqref{iruv} to rewrite $\ncal_1$ and $\ncal_{2}$. Eqs.~\eqref{delrel} may also readily be verified using Eqs.~\eqref{irdiffs:b}, \eqref{irdiffs:d}, \eqref{irdiffs:h}, \eqref{divs:a} and \eqref{divs:b}. It is clear that $\delta_{5,75}$ in Eq.~\eqref{delsum:f} vanishes under the same condition.

Now examining $\delta_{5,111}$ in Eq.~\eqref{delsum:g}, we can show that 
\begin{subequations}
\begin{align}
I^{5,111}_1+2\ncal_1I^{5,111}_2+(\ncal_1)^2I^{5,111}_3+\ncal_{3,3}I^{5,111}_4=&\,O(1),\\
I^{5,111}_2+\ncal_1I^{5,111}_3+\ncal_{2}I^{5,111}_4=&\,O(1),\\
I^{5,111}_3+\ncal_1I^{5,111}_4=&\,O(1),\\
I^{5,111}_4=&\,O(1).
\end{align}
\end{subequations}
In a similar argument to that for $\delta_{5,53}$, Eq.~\eqref{delsum:g} then follows provided Eqs.~\eqref{delrel1c}, \eqref{delrel1} are satisfied, together with
\be
\delta^{IR}_{4,1}+\Delta_{1,1}\delta^{IR}_{3,1}+2\Delta_{2,2}\delta^{IR}_{2}=O(1),
\label{delrel2}
\ee
which again may be checked empirically  using Eqs.~\eqref{irdiffs:b}, \eqref{irdiffs:c} and \eqref{irdiffs:g}.

Furthermore we have, again following the argument which led to Eq.~\eqref{i425c},
\begin{subequations}
\begin{align}
I^{5,73}_1+\ncal_1I^{5,73}_2+\ncal_{2}I^{5,73}_3+\ncal_{3,2}I^{5,73}_4=&\,O(1),\\
I^{5,73}_2+\ncal_1I^{5,73}_3+\ncal_{2}I^{5,73}_4=&\,O(1),\\
I^{5,73}_3+\ncal_1I^{5,73}_4=&\,O(1),\\
I^{5,73}_4=&\,O(1).
\end{align}
\end{subequations}
Once again, in a similar argument to that for $\delta_{5,53}$, Eq.~\eqref{delsum:h} then follows provided
\begin{subequations}\label{delrel3}
\begin{align}
\delta^{IR}_{3,3}=&\,O(1),\label{delrel3:a}\\
\delta^{IR}_{4,4}+\Delta_{1,1}\delta^{IR}_{3,3}=&\,O(1),\label{delrel3:b}
\end{align}
\end{subequations}
which again may be checked empirically using Eqs.~\eqref{irdiffs:e} and \eqref{irdiffs:j}. $\delta_{5,93}$ in Eq.~\eqref{delsum:i} also vanishes by a similar argument using the same conditions.

We have shown that the AIR counterterms (for IR-divergent structures with double propagators) result in the same UV counterterms as their standard counterparts up to five loops provided the conditions Eqs.~\eqref{delrel1c}, \eqref{delrel1}, \eqref{delrel1a}, \eqref{delrel}, \eqref{delrel2} and \eqref{delrel3} are satisfied. This seems to be a significant step forward compared with the explicit checks of Sections 3, 4 and Appendix B, since these conditions are intrinsic to the IR counterterms themselves and are independent of the renormalisation process for a particular five-loop diagram; indeed, we have seen that the same condition arises in different diagrams. However, we should reiterate here that we are assuming throughout that the UV counterterms are defined by minimal subtraction, and indeed, as we stated at the beginning of Section 3, are the same whether we use the standard IR counterterms or our alternative versions.

The  next realisation is that these conditions all follow if each AIR counterterm is related to the UV counterterm for the corresponding CVG by a similar relation to that for the standard one, described in Section 2. 
For instance, neglecting manifestly zero contributions, Eq.~\eqref{IRtwo:b} may be written
\be
-\Delta_{2,2}=\ncal_{2}+\Delta_{1,1}\ncal_1.
\ee
If the new counterterms obeyed a similar relation we would have
\be
-\Delta_{2,2}=\ncal'_{2}+\Delta_{1,1}\ncal_1+O(1),
\ee
where the $O(1)$ terms are necessary since as we have argued, we need to include the finite terms in our new counterterms; and we know $\ncal_1=\ncal'_1$. Subtracting, we have
\be
\delta^{IR}_{2}=O(1),
\label{del2a}
\ee
which of course is Eq.~\eqref{delrel1c}. In order to proceed to the next order, recall that the relation between the IR and UV counterterms follows from the scale invariance of the CVG which has no masses and no external momenta. The two corresponding relations for the CVG corresponding to the 3-loop Figure 5 in Ref.~\cite{klein} are
\begin{subequations}
\begin{align}
-\Delta_{3,5}=&\,\ncal_{3,1}+2\Delta_{1,1}\ncal_{2}+(\Delta_{1,1})^2\ncal_1,\\
-\Delta_{3,5}=&\,\ncal'_{3,1}+2\Delta_{1,1}\ncal'_{2}+(\Delta_{1,1})^2\ncal_1+O(1),
\end{align}
\end{subequations}
and subtracting we immediately obtain Eq.~\eqref{delrel1}.  
Similarly, we have 
\begin{subequations}
\begin{align}
-\Delta_{4,21}=&\,\ncal_{4,3}+\Delta_{3,4}\ncal_1,\\
-\Delta_{4,21}=&\,\ncal'_{4,3}+\Delta_{3,4}\ncal_1+O(1),
\end{align}
\end{subequations}
and subtracting we  immediately obtain Eq.~\eqref{delrel1a}; and furthermore
\begin{subequations}
\begin{align}
-\Delta_{4,25}=&\,\ncal_{4,2}+\Delta_{1,1}\ncal_{3,2}+\Delta_{2,2}\ncal_{2}+\Delta_{3,7}\ncal_1,\\
-\Delta_{4,25}=&\,\ncal'_{4,2}+\Delta_{1,1}\ncal'_{3,2}+\Delta_{2,2}\ncal'_{2}+\Delta_{3,7}\ncal_1+O(1),
\end{align}
\end{subequations}
and subtracting we obtain Eq.~\eqref{delrel:b}.
Likewise, we have 
\begin{subequations}
\begin{align}
-\Delta_{4,17}=&\,\ncal_{4,1}+\Delta_{1,1}\ncal_{3,1}+2\Delta_{2,2}\ncal_{2}+\Delta_{3,8}\ncal_1,\\
-\Delta_{4,17}=&\,\ncal'_{4,1}+\Delta_{1,1}\ncal'_{3,1}+2\Delta_{2,2}\ncal'_{2}+\Delta_{3,8}\ncal_1+O(1),
\end{align}
\end{subequations}
and subtracting we obtain Eq.~\eqref{delrel2}. Finally, we have
\begin{subequations}
\begin{align}
-\Delta_{4,8}=&\,\ncal_{4,4}+\Delta_{1,1}\ncal_{3,3}+\Delta_{2,2}(\ncal_1)^2+\Delta_{3,7}\ncal_1,\nn
-\Delta_{4,8}=&\,\ncal'_{4,4}+\Delta_{1,1}\ncal'_{3,3}+\Delta_{2,2}(\ncal_1)^2+\Delta_{3,7}\ncal_1+O(1),
\end{align}
\end{subequations}
and subtracting we obtain Eq.~\eqref{delrel3:b}.

Eq.~\eqref{delrel:a} corresponds in a similar way to $\Delta_{3,7}$, and Eq.~\eqref{delrel3:a} to $\Delta_{3,8}$; in fact clearly there is a similar identity for every diagram with at least one double propagator contributing to the $\beta$-function. For instance,  
\be
\delta^{IR}_{4,7}+3\Delta_{1,1}\delta^{IR}_{3,1}+3(\Delta_{1,1})^2\delta^{IR}_{2}=O(1),
\ee
corresponding to $\Delta_{4,18}$. This identity may also readily be verified. (We have considered $\Delta_{4,18}$ previously in Section 4, but there we considered the arrangement of external momenta shown on the LHS of Eq.~\eqref{exdiagc}, rather than currently where we consider the corresponding CVG which is the same graph but with no external momenta.) 

We have shown that our new IR counterterms are related to the standard (i.e. $\msbar$) UV counterterms by the general relation which was derived in Refs.~\cite{chet3,herz}, at least as far as required to derive all the standard results for the five-loop UV counterterms. One might wonder if it would have been sufficient to check this without our laborious individual calculations for the individual diagrams; but we have seen that in practice more is required, namely relations such as Eq.~\eqref{i5107rels} between various combinations of diagrams generated by the $R$ operation. This might be a clue as to how to proceed towards a general proof, as we shall discuss in the Conclusions.

\section{Conclusions} As we have said, it was shown in Refs.~\cite{chet3,herz} that any viable set of IR counterterms are related to the UV counterterms by certain relations which follow from the scale-invariance of the CVG. One may ask whether this necessary condition is also sufficient; to be more precise, if we have a proposed set of IR counterterms which satisfies these relations (using the usual $\msbar$ UV counterterms), is it guaranteed that the entire renormalisation process will be compatible with these usual $\msbar$ UV counterterms? Our practical calculations seem to sound a note of caution; we have shown that our AIR counterterms are related as required to the $\msbar$ UV counterterms up to four loops, but nevertheless 
an additional property of the renormalisation procedure is required to guarantee that the correct UV counterterms are obtained at the next order, i.e. five loops - for instance as in Eq.~\eqref{i5107rels}. This additional property is satisfied, as far as we have seen; but nevertheless it appears that there will be two aspects to a general proof of the validity of our AIR counterterms. Our detailed arguments in Section 5 may give an indication of how to proceed in general.

The existence of an alternative form for the IR counterterms seems to us quite remarkable, especially as it provides a relatively simple closed form description for them (though of course there may be a non-trivial momentum integral still to perform). However, in practical terms it is not clear that this provides any saving of effort in computing the UV counterterms; at each loop order, it is not a huge effort to compute the standard IR counterterms from the UV counterterms using relations such as Eq.~\eqref{IR12}. On the other hand, if one imagined a scenario where one had not yet computed any UV or IR counterterms, one could consider inverting the procedure and using the IR counterterms as a basis from which to compute the UV counterterms. This really would represent a saving in overall labour. Another nice feature of our method is that one only has to perform loop integrals one loop lower than the loop order one is working at; though in practice the order of difficulty may not depend too much on the loop order in any case. In the main body of the text we have only considered IR counterterms corresponding to CVGs with at least one double propagator, but in Appendix B, we extend our results to cases with no double propagators; the triple (and higher multiple) propagator case is still an open question, unfortunately. We have also concentrated on IR divergences arising in scalar theories where the numerators of the Feynman integrals are trivial; a next step is to extend to tensor numerators as required for fermions and gauge particles; this was thoroughly addressed for the case of standard IR counterterms in Ref.~\cite{herz}.

Finally, as we have admitted before, we have obtained our results by heuristic means and it would be more satisfying (and maybe simpler) if there was an alternative direct derivation.

\section*{Acknowledgments}
 We are very grateful for many useful conversations and correspondence with John Gracey, Tim Jones, and Hugh Osborn. This research was supported by the University of Liverpool.

\appendix

\section{Alternative IR counterterms}

\begin{figure}
\center\begin{tikzpicture}
{
  \node (vert_cent) {\hspace{-13pt}$\phantom{-}$};
\draw (0,0) circle [radius=0.5];
\draw (-.75,0) to (-.5,0);
\draw (.5,0) to  (.75,0);
\draw (0,-.5) node[below]{$\small{a}$};       
\draw (0,.5) node[above]{$\small{b}$};   
        }
\end{tikzpicture}
\caption{One-loop bubble} \label{diag1}
\end{figure}
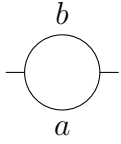

\begin{figure}
\center\begin{tikzpicture}
{
  \node (vert_cent) {\hspace{-13pt}$\phantom{-}$};
\draw (-3,0)--(-2,0);
 \draw(-2,0)--(0,2);
\draw(-2,0)--(0,-2);
\draw(2,0)--(0,2);
\draw(2,0)--(0,-2);
    \draw (0,2)--(0,-2);
    \draw (2,0)--(3,0);    
\draw (-1,-1) node[below]{$\small{a_4}$};       
\draw (-1,1) node[above]{$a_1$};  
\draw (1,1) node[above]{$a_2$};  
\draw (1,-1) node[below]{$a_3$};  
\draw (0.2,0.2) node[below]{$a_5$};  
        }
\end{tikzpicture}
\caption{Two-loop diagram} \label{diag2}
\end{figure}
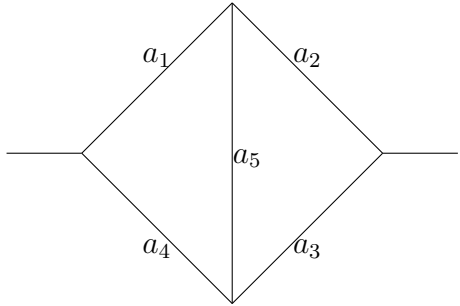

We start by defining a couple of quantities which often arise in evaluating our counterterms. We write
\be
\tfrac{1}{(p^2)^{a+b-\tfrac{d}{2}}}L(a,b)
\label{Ldef}
\ee
for the result of the general one-loop bubble shown in Fig.~\ref{diag1}, where $a$, $b$ are the weights of the propagators and $p$ is the incoming/outgoing momentum.
Then we have
\be
 L(a,b)=(4\pi)^{\tfrac{\epsilon}{2}}\tfrac{\Gamma(\tfrac{d}{2}-a)\Gamma(\tfrac{d}{2}-b)\Gamma(a+b-\tfrac{d}{2})}{\Gamma(a)\Gamma(b)\Gamma(d-a-b)}.
\label{Lres}
\ee
We also write
\be
\tfrac{1}{(p^2)^{a_1+a_2+a_3+a_4+a_5-d}}L(a_1,a_2,a_3,a_4,a_5),
\label{L4def}
\ee
for the result of the general two-loop diagram shown in Fig.~\ref{diag2}, where $a_1-a_5$ are the weights of the propagators and $p$ is the incoming/outgoing momentum.

Our alternative results for the IR counterterms are then given in Tables~\ref{IRres12}, \ref{IRres3} and \ref{IRres4}, together with the associated CVG,  the severed graph $G^{\rm{sev}}$ and the IR counterterm constructed  according to Eq.~\eqref{Xdef}. Note however the extra terms in the expressions for $\ncal_{3,3}$ and $\ncal_{4,4}$, where the CVG has two double propagators and therefore $G^{\rm{sev}}$ itself has an IR divergence which requires an IR counterterm. We should also mention here that in these simple cases, it makes no difference which of the two double propagators are severed, but this may not always be the case; this deserves further investigation.
\begin{table}
\begin{tabular}{|c|c|c|c|c|}
\hline
Label&IR divergence&CVG&$G^{\rm{sev}}$&IR counterterm\nn
\hline
$\Ical_1$&\tikz[scale=0.6,baseline=(vert_cent.base)]{
  \node (vert_cent) {\hspace{-13pt}$\phantom{-}$};
\node at (0,1.1) () {};
\node at (0,-1.1) () {};
\draw (0,.5) circle [radius=0.1];
\draw (0,-.5) circle [radius=0.1];
  \draw (0,-.4)--(0,.4);
\filldraw (0,0) circle [radius=0.1];
}&\tikz[scale=0.6,baseline=(vert_cent.base)]{
	%\draw[help lines] (-3,-3) grid (3,3);
	\node (vert_cent) {\hspace{-13pt}$\phantom{-}$};
	\draw (0,0) circle[radius=1cm];
\filldraw (0,-1) circle [radius=0.1];
\filldraw (0,1) circle [radius=0.1];
}&\tikz[scale=0.6,baseline=(vert_cent.base)]{
\draw (0,0) --(1,0);
\filldraw (0.5,0) circle [radius=0.1];
}&$\ncal'_1=C_1$\nn
\hline
$\Ical_{2a}$&\tikz[scale=0.6,baseline=(vert_cent.base),scale=1]{
\filldraw  (135:1) circle [radius=0.1];
\draw (-1,0) circle [radius=0.1];
\draw (1,0) circle [radius=0.1];
\draw [bend right=40] (0,1) to (-1,0.1);
\draw [bend left=40] (0,1) to (1,0.1); 
\draw [bend right=40] (0,1) to (1,0.1);
}&
\tikz[scale=0.6,baseline=(vert_cent.base),scale=1]{
	%\draw[help lines] (-3,-3) grid (3,3);
	\node (vert_cent) {\hspace{-13pt}$\phantom{-}$};
\node at (0,1.1) () {};
	\draw (0,0) circle[radius=1cm];
\draw (0,-1) -- (0,1);
\filldraw (0,0) circle [radius=0.1];
}&\tikz[scale=0.6,baseline=(vert_cent.base),scale=1]{
\draw (0,0) circle [radius=0.5];
\draw (-.75,0) -- (-0.5,0);
\draw (0.75,0) -- (0.5,0);
}&$\ncal'_2=L(1,1)C_2$\nn
$\Ical_{2b}$&\tikz[scale=0.6,baseline=(vert_cent.base),scale=1]{
\filldraw  (0,0.5) circle [radius=0.1];
\draw (-1,0) circle [radius=0.1];
\draw (1,0) circle [radius=0.1];
\draw (0,0) circle [radius=0.1];
\draw (0,1) --(0,0.1);
\draw [bend right=40] (0,1) to (-1,0.1);
\draw [bend left=40] (0,1) to (1,0.1); 
}&
&&\nn
$\Ical_{2c}$&\tikz[scale=0.6,baseline=(vert_cent.base),scale=1]{
\filldraw  (45:1) circle [radius=0.1];
\draw (-1,0) circle [radius=0.1];
\draw (1,0) circle [radius=0.1];
\draw [bend right=40] (0,1) to (-1,0.1);
\draw [bend left=40] (0,1) to (1,0.1); 
\draw [bend right=40] (0,1) to (1,0.1);
}&
&&\\
\hline
\end{tabular}
\caption{\label{IRres12}Alternative results for one- and two-loop infra-red counterterms}
\end{table}

\begin{table}
\begin{tabular}{|c|c|c|c|c|}
\hline
Label&IR divergence&CVG&$G^{\rm{sev}}$&IR counterterm\nn
\hline
\rule{0pt}{20pt}
$\Ical_{3,1a}$&\tikz[scale=0.6,baseline=(vert_cent.base),scale=1]{
\filldraw  (0,0.5) circle [radius=0.1];
\draw (-1,0) circle [radius=0.1];
\draw (1,0) circle [radius=0.1];
\draw (0,1) --(0,0);
\draw [bend right=40] (0,1) to (-1,0.1);
\draw [bend left=40] (0,1) to (1,0.1);
\draw (-.9,0) -- (0.9,0) 
}&
\tikz[scale=0.6,baseline=(vert_cent.base),scale=1]{
	%\draw[help lines] (-3,-3) grid (3,3);
	\node (vert_cent) {\hspace{-13pt}$\phantom{-}$};
\node at (0,1.1) () {};
	\draw (0,0) circle[radius=1cm];
	\draw [bend right=40] (-1,0) to (0,1);
           \draw [bend left=40] (1,0) to (0,1);
\filldraw (0,-1) circle [radius=0.1];
}&\tikz[scale=0.6,baseline=(vert_cent.base),scale=1]{
\draw (-0.5,0) circle [radius=0.5];
\draw (0.5,0) circle [radius=0.5];
\draw (-1.25,0) -- (-1,0);
\draw (1.25,0) -- (1,0);
}&$\ncal'_{3,1}=L(1,1)^2C_3$\nn
$\Ical_{3,1b}$&\tikz[scale=0.6,baseline=(vert_cent.base)]{
  \node (vert_cent) {\hspace{-13pt}$\phantom{-}$};
  \draw
        (5:1cm) arc (5:175:1cm) ;
\draw (0,0.1)--(45:1);
\draw (0,0.1)--(135:1);
\draw (-1,0) circle [radius=0.1];
\draw (1,0) circle [radius=0.1];
\draw (0,0) circle [radius=0.1];
\filldraw (0,1) circle [radius=0.1];
}&&&\nn 
$\Ical_{3,1c}$&\tikz[scale=0.6,baseline=(vert_cent.base),scale=1]{
\filldraw  (30:1) circle [radius=0.1];
\draw (-1,0) circle [radius=0.1];
\draw (1,0) circle [radius=0.1];
\draw [bend right=30] (120:1) to (-1,0.1);
\draw [bend left=30] (60:1) to (1,0.1); 
\draw [bend right=30] (120:1) to (1,0.1);
\draw [bend left=30] (120:1) to (60:1);
\draw [bend right=30] (120:1) to (60:1);
}&&&\nn 
$\Ical_{3,1d}$&\tikz[scale=0.6,baseline=(vert_cent.base),scale=1]{
\filldraw  (150:1) circle [radius=0.1];
\draw (-1,0) circle [radius=0.1];
\draw (1,0) circle [radius=0.1];
\draw [bend right=30] (120:1) to (-1,0.1);
\draw [bend left=30] (60:1) to (1,0.1); 
\draw [bend right=30] (60:1) to (1,0.1);
\draw [bend left=30] (120:1) to (60:1);
\draw [bend right=30] (120:1) to (60:1);
}&&&\nn
\hline
$\Ical_{3,2a}$&\tikz[scale=0.6,baseline=(vert_cent.base)]{
  \node (vert_cent) {\hspace{-13pt}$\phantom{-}$};
  \draw
        (5:1cm) arc (5:175:1cm) ;
\draw (0,0.1)--(45:1);
\draw (0,0.1)--(135:1);
\draw (-1,0) circle [radius=0.1];
\draw (1,0) circle [radius=0.1];
\draw (0,0) circle [radius=0.1];
\filldraw (135:0.5) circle [radius=0.1];
}&\tikz[scale=0.6,baseline=(vert_cent.base),scale=1]{
	%\draw[help lines] (-3,-3) grid (3,3);
	\node (vert_cent) {\hspace{-13pt}$\phantom{-}$};
\node at (0,1.1) () {};
	\draw (0,0) circle[radius=1cm];
	\draw [bend right=40] (-1,0) to (0,1);
           \draw [bend left=40] (1,0) to (0,1);
\filldraw (135:1) circle [radius=0.1];
}&\tikz[scale=0.6,baseline=(vert_cent.base),scale=1]{
	%\draw[help lines] (-3,-3) grid (3,3);
	\node (vert_cent) {\hspace{-13pt}$\phantom{-}$};
	\draw (0,0) circle[radius=1cm];
           \draw [bend left=40] (1,0) to (0,1);
\draw (-1.25,0) -- (-1,0);
\draw (1.25,0) -- (1,0);
}&$\ncal'_{3,2}=L(1,1)L(1,1+\tfrac12\epsilon)C_3$\nn
$\Ical_{3,2b}$&\tikz[scale=0.6,baseline=(vert_cent.base)]{
  \node (vert_cent) {\hspace{-13pt}$\phantom{-}$};
  \draw
        (5:1cm) arc (5:175:1cm) ;
\draw (0,0.1)--(45:1);
\draw (0,0.1)--(135:1);
\draw (-1,0) circle [radius=0.1];
\draw (1,0) circle [radius=0.1];
\draw (0,0) circle [radius=0.1];
\filldraw (157.5:1) circle [radius=0.1];
}&&&\nn 
$\Ical_{3,2c}$&\tikz[scale=0.6,baseline=(vert_cent.base)]{
  \node (vert_cent) {\hspace{-13pt}$\phantom{-}$};
  \draw
        (5:1cm) arc (5:175:1cm) ;
\draw (0,0.1)--(0,1);
\draw (-1,0) circle [radius=0.1];
\draw (1,0) circle [radius=0.1];
\draw (0,0) circle [radius=0.1];
\draw [bend right=40] (135:1) to (0,1);
\filldraw (0,0.5) circle [radius=0.1];
}&&&\nn 
$\Ical_{3,2d}$&\tikz[scale=0.6,baseline=(vert_cent.base)]{
  \node (vert_cent) {\hspace{-13pt}$\phantom{-}$};
  \draw
        (5:1cm) arc (5:175:1cm) ;
\draw (-1,0) circle [radius=0.1];
\draw (1,0) circle [radius=0.1];
\filldraw (135:1) circle [radius=0.1];
\draw (0,0) to (0,1);
\draw (-0.9,0) -- (0.9,0);
\draw (0.9,0) -- (0.1,0);
}
&&&\nn
\hline
$\Ical_{3,3a}$&\tikz[scale=0.6,baseline=(vert_cent.base),scale=1]{
\filldraw  (0.5,0.5) circle [radius=0.1];
\filldraw  (0.5,-0.5) circle [radius=0.1];
\draw (-1,0) circle [radius=0.1];
\draw [bend right=40] (0,0) to (-0.5,0.5);
\draw [bend left=40] (-1,0.07) to (-0.5,0.5);
\draw [bend left=40] (0,0) to (-0.5,-0.5);
\draw [bend right=40] (-1,-0.07) to (-0.5,-0.5);
\draw (1,0) circle [radius=0.1];
\draw [bend left=40] (0,0) to (0.5,0.5);
\draw [bend right=40] (1,0.07) to (0.5,0.5);
\draw [bend right=40] (0,0) to (0.5,-0.5);
\draw [bend left=40] (1,-0.07) to (0.5,-0.5);
}&\tikz[scale=0.6,baseline=(vert_cent.base),scale=1]{
	%\draw[help lines] (-3,-3) grid (3,3);
	\node (vert_cent) {\hspace{-13pt}$\phantom{-}$};
\node at (0,1.1) () {};
	\draw (0,0) circle[radius=1cm];
         \draw [bend right=40] (0,1) to (0,-1);
           \draw [bend left=40] (0,1) to (0,-1);
\filldraw (180:1) circle [radius=0.1];
\filldraw (0:1) circle [radius=0.1];
}&\tikz[scale=0.6,baseline=(vert_cent.base),scale=1]{
	%\draw[help lines] (-3,-3) grid (3,3);
	\node (vert_cent) {\hspace{-13pt}$\phantom{-}$};
	\draw (0,0) circle[radius=1cm];
\filldraw (0,0) circle [radius=0.1];
\draw (-1.25,0) -- (1.25,0);
}&$\ncal'_{3,3}=L(1,1)L(2,\tfrac12\epsilon)C_3+L(1,1)C_1C_2$\nn
$\Ical_{3,3b}$&\tikz[scale=0.6,baseline=(vert_cent.base),scale=1]{
\filldraw  (0.5,0.5) circle [radius=0.1];
\filldraw  (-0.5,-0.5) circle [radius=0.1];
\draw (-1,0) circle [radius=0.1];
\draw [bend right=40] (0,0) to (-0.5,0.5);
\draw [bend left=40] (-1,0.07) to (-0.5,0.5);
\draw [bend left=40] (0,0) to (-0.5,-0.5);
\draw [bend right=40] (-1,-0.07) to (-0.5,-0.5);
\draw (1,0) circle [radius=0.1];
\draw [bend left=40] (0,0) to (0.5,0.5);
\draw [bend right=40] (1,0.07) to (0.5,0.5);
\draw [bend right=40] (0,0) to (0.5,-0.5);
\draw [bend left=40] (1,-0.07) to (0.5,-0.5);
}&&&\\
\hline
$\Ical_{3,4}$&\tikz[scale=0.6,baseline=(vert_cent.base)]{
  \node (vert_cent) {\hspace{-13pt}$\phantom{-}$};
  \draw
        (5:1cm) arc (5:175:1cm) ;
\draw (0,0.5)--(45:1);
\draw (0,0.5)--(135:1);
\draw (0,0.1) -- (0,0.5);
\draw (-1,0) circle [radius=0.1];
\draw (1,0) circle [radius=0.1];
\draw (0,0) circle [radius=0.1];
}&
\tikz[scale=0.6,baseline=(vert_cent.base),scale=1]{
	%\draw[help lines] (-3,-3) grid (3,3);
	\node (vert_cent) {\hspace{-13pt}$\phantom{-}$};
\node at (0,1.1) () {};
\node at (0,-1.1) () {};
	\draw (0,0) circle[radius=1cm];
           \draw (0,0) -- (45:1);
           \draw (0,0) -- (135:1);
	\draw (0,0) to (0,-1);
}&
\tikz[scale=0.6,baseline=(vert_cent.base),scale=1]{
	%\draw[help lines] (-3,-3) grid (3,3);
	\node (vert_cent) {\hspace{-13pt}$\phantom{-}$};
	\draw (0,0) circle[radius=1cm];
\draw (0,-1) -- (0,1);
\draw (-1.5,0) -- (-1,0);
\vsq at (1,0) {};
}&$\ncal'_{3,4}=L(1,1,1,1,1)C_3$\nn 
\hline
\end{tabular}
\caption{\label{IRres3}Alternative results for three-loop infra-red counterterms}
\end{table}

\begin{table}
\begin{tabular}{|c|c|c|c|c|}
\hline
Label&IR divergence&CVG&$G^{\rm{sev}}$&IR counterterm\nn
\hline
$\Ical_{4,1a}$&\tikz[scale=0.6,baseline=(vert_cent.base)]{
  \node (vert_cent) {\hspace{-13pt}$\phantom{-}$};
  \draw
        (5:1cm) arc (5:175:1cm) ;
\draw (0,0)--(45:1);
\draw (0,0)--(135:1);
\draw (-1,0) circle [radius=0.1];
\draw (1,0) circle [radius=0.1];
\filldraw (0,1) circle [radius=0.1];
\draw (-0.9,0) -- (0.9,0);
\draw (0.9,0) -- (0.1,0);
}&\tikz[scale=0.6,baseline=(vert_cent.base),scale=1]{
	%\draw[help lines] (-3,-3) grid (3,3);
	\node (vert_cent) {\hspace{-13pt}$\phantom{-}$};
\node at (0,1.1) () {};
	\draw (0,0) circle[radius=1cm];
           \draw (0,0) -- (45:1);
           \draw (0,0) -- (135:1);
	\draw [bend right=40] (0,0) to (0,-1);
           \draw [bend left=40] (0,0) to (0,-1);
\filldraw (0,1) circle [radius=0.1];
}&\tikz[scale=0.6,baseline=(vert_cent.base),scale=1]{
	%\draw[help lines] (-3,-3) grid (3,3);
	\node (vert_cent) {\hspace{-13pt}$\phantom{-}$};
	\draw (0,0) circle[radius=1cm];
         \draw [bend right=40] (0,1) to (0,-1);
           \draw [bend left=40] (0,1) to (0,-1);
\draw (-1.25,0) -- (-1,0);
\draw (1.25,0) -- (1,0);
}&$\ncal'_{4,1}=L(1,1)L(1,1,1,1,\tfrac12\epsilon)C_4$\nn
$\Ical_{4,1b}$&\tikz[scale=0.6,baseline=(vert_cent.base)]{
  \node (vert_cent) {\hspace{-13pt}$\phantom{-}$};
  \draw
        (5:1cm) arc (5:180:1cm) ;
\draw (-2,0) circle [radius=0.1];
\draw (1,0) circle [radius=0.1];
\filldraw (-1.5,0) circle [radius=0.1];
\draw [bend left=40] (0,0) to (0,1);
\draw [bend right=40] (0,0) to (0,1);
\draw (-1,0) -- (0.9,0);
\draw (-1.9,0) -- (-1,0);
}&&&\\
\hline
$\Ical_{4,2a}$&\tikz[scale=0.6,baseline=(vert_cent.base)]{
  \node (vert_cent) {\hspace{-13pt}$\phantom{-}$};
  \draw
        (5:1cm) arc (5:175:1cm) ;
\draw (0,0)--(45:1);
\draw (0,0)--(135:1);
\draw (-1,0) circle [radius=0.1];
\draw (1,0) circle [radius=0.1];
\filldraw (135:0.5) circle [radius=0.1];
\draw (-0.9,0) -- (0.9,0);
\draw (0.9,0) -- (0.1,0);
}&\tikz[scale=0.6,baseline=(vert_cent.base),scale=1]{
	%\draw[help lines] (-3,-3) grid (3,3);
	\node (vert_cent) {\hspace{-13pt}$\phantom{-}$};
\node at (0,1.1) () {};
	\draw (0,0) circle[radius=1cm];
           \draw (0,0) -- (45:1);
           \draw (0,0) -- (135:1);
	\draw [bend right=40] (0,0) to (0,-1);
           \draw [bend left=40] (0,0) to (0,-1);
\filldraw (135:0.5) circle [radius=0.1];
}&\tikz[scale=0.6,baseline=(vert_cent.base),scale=1]{
	%\draw[help lines] (-3,-3) grid (3,3);
	\node (vert_cent) {\hspace{-13pt}$\phantom{-}$};
	\draw (0,0) circle[radius=1cm];
           \draw [bend left=40] (1,0) to (0,1);
\draw (0,-1) -- (0,1);
\draw (-1.25,0) -- (-1,0);
\draw (1.25,0) -- (1,0);
}&$\ncal'_{4,2}=L(1,1)L(\tfrac12\epsilon,1,1,1,1)C_4$\nn
$\Ical_{4,2b}$&\tikz[scale=0.6,baseline=(vert_cent.base)]{
  \node (vert_cent) {\hspace{-13pt}$\phantom{-}$};
  \draw
        (5:1cm) arc (5:175:1cm) ;
\draw (0,0)--(45:1);
\draw (0,0)--(135:1);
\draw (-1,0) circle [radius=0.1];
\draw (1,0) circle [radius=0.1];
\filldraw (157.5:1) circle [radius=0.1];
\draw (-0.9,0) -- (0.9,0);
\draw (0.9,0) -- (0.1,0);
}&&&\nn
$\Ical_{4,2c}$&\tikz[scale=0.6,baseline=(vert_cent.base)]{
  \node (vert_cent) {\hspace{-13pt}$\phantom{-}$};
  \draw
        (5:1cm) arc (5:180:1cm) ;
\draw (-2,0) circle [radius=0.1];
\draw (1,0) circle [radius=0.1];
\filldraw (-1.5,0) circle [radius=0.1];
\draw [bend right=40] (-1,0) to (0,1);
\draw  (0,0) to (0,1);
\draw (-1,0) -- (0.9,0);
\draw (-1.9,0) -- (-1,0);
}&&&\\
\hline
$\Ical_{4,3}$&\tikz[scale=0.6,baseline=(vert_cent.base)]{
  \node (vert_cent) {\hspace{-13pt}$\phantom{-}$};
  \draw
        (5:1cm) arc (5:175:1cm) ;
\draw (0,0)--(45:1);
\draw (0,0)--(135:1);
\draw (-1,0) circle [radius=0.1];
\draw (1,0) circle [radius=0.1];
\filldraw (-0.5,0) circle [radius=0.1];
\draw (-0.9,0) -- (0.9,0);
\draw (0.9,0) -- (0.1,0);
}&\tikz[scale=0.6,baseline=(vert_cent.base),scale=1]{
	%\draw[help lines] (-3,-3) grid (3,3);
	\node (vert_cent) {\hspace{-13pt}$\phantom{-}$};
\node at (0,1.1) () {};
\node at (0,-1.1) () {};
	\draw (0,0) circle[radius=1cm];
           \draw (0,0) -- (45:1);
           \draw (0,0) -- (135:1);
	\draw [bend right=40] (0,0) to (0,-1);
           \draw [bend left=40] (0,0) to (0,-1);
\filldraw (-0.2,-0.5) circle [radius=0.1];
}&\tikz[scale=0.6,baseline=(vert_cent.base),scale=1]{
	%\draw[help lines] (-3,-3) grid (3,3);
	\node (vert_cent) {\hspace{-13pt}$\phantom{-}$};
	\draw (0,0) circle[radius=1cm];
\draw (-1.25,0) -- (1.25,0);
\draw (0,0) -- (0,-1);
}&$\ncal'_{4,3}=L(1,1+\epsilon)L(1,1,1,1,1)C_4$\nn
\hline
$\Ical_{4,4a}$&\tikz[scale=0.6,baseline=(vert_cent.base)]{
  \node (vert_cent) {\hspace{-13pt}$\phantom{-}$};
  \draw
        (5:1cm) arc (5:175:1cm) ;
\draw (-1,0) circle [radius=0.1];
\draw (1,0) circle [radius=0.1];
\filldraw (45:1) circle [radius=0.1];
\filldraw (0.5,0) circle [radius=0.1];
\draw [bend left=40] (0,0) to (0,1);
\draw [bend right=40] (0,0) to (0,1);
\draw (-0.9,0) -- (0.9,0);
\draw (0.9,0) -- (0.1,0);
}&\tikz[scale=0.6,baseline=(vert_cent.base),scale=1]{
	%\draw[help lines] (-3,-3) grid (3,3);
	\node (vert_cent) {\hspace{-13pt}$\phantom{-}$};
\node at (0,1.1) () {};
	\draw (0,0) circle[radius=1cm];
           \draw [bend left=40] (1,0) to (0,1);
\draw [bend right=40] (-1,0) to (0,1);
\draw (-1,0) -- (1,0);
\filldraw (135:1) circle [radius=0.1];
\filldraw (45:1) circle [radius=0.1];
}&\tikz[scale=0.6,baseline=(vert_cent.base),scale=1]{
\draw (180:1cm) arc (180:0:1cm) ;
\draw [bend right=40] (0,1) to (1,0);
\draw [bend left=40] (0,1) to (-1,0);
\filldraw (135:1) circle [radius=0.1];
\draw (-1.25,0) -- (1.25,0);
}&$\ncal'_{4,4}=L(1,1)L(1,2)L(1,1+\epsilon)C_4$\nn
$\Ical_{4,4b}$&\tikz[scale=0.6,baseline=(vert_cent.base)]{
  \node (vert_cent) {\hspace{-13pt}$\phantom{-}$};
  \draw
        (5:1cm) arc (5:175:1cm) ;
\draw (-1,0) circle [radius=0.1];
\draw (1,0) circle [radius=0.1];
\filldraw (135:1) circle [radius=0.1];
\filldraw (0.5,0) circle [radius=0.1];
\draw [bend left=40] (0,0) to (0,1);
\draw [bend right=40] (0,0) to (0,1);
\draw (-0.9,0) -- (0.9,0);
}&&&$+L(1,1)L(1,1+\tfrac12\epsilon)C_1C_3$\\
$\Ical_{4,4c}$&\tikz[scale=0.6,baseline=(vert_cent.base)]{
  \node (vert_cent) {\hspace{-13pt}$\phantom{-}$};
  \draw
        (5:1cm) arc (5:180:1cm) ;
\draw (-2,0) circle [radius=0.1];
\draw (1,0) circle [radius=0.1];
\filldraw (-1.5,0) circle [radius=0.1];
\filldraw (135:1) circle [radius=0.1];
\draw [bend right=40] (-1,0) to (0,1);
\draw [bend left=40] (1,0) to (0,1);
\draw (-1,0) -- (0.9,0);
\draw (-1.9,0) -- (-1,0);
}&&&\\
\hline
$\Ical_{4,5}$&\tikz[scale=0.6,baseline=(vert_cent.base)]{
  \node (vert_cent) {\hspace{-13pt}$\phantom{-}$};
  \draw
        (5:1cm) arc (5:175:1cm) ;
\draw (0,0.5)--(45:1);
\draw (0,0.5)--(135:1);
\draw (0,0) -- (0,0.5);
\draw (-1,0) circle [radius=0.1];
\draw (1,0) circle [radius=0.1];
\draw (-0.9,0) -- (0.9,0);
}&
\tikz[scale=0.6,baseline=(vert_cent.base),scale=1]{
	%\draw[help lines] (-3,-3) grid (3,3);
	\node (vert_cent) {\hspace{-13pt}$\phantom{-}$};
\node at (0,1.1) () {};
\node at (0,-1.1) () {};
	\draw (0,0) circle[radius=1cm];
           \draw (0,0.2) -- (45:1);
           \draw (0,0.2) -- (135:1);
	\draw (0,0.2) to (0,-0.3);
\draw [bend right=40] (0,-0.3) to (0,-1);
           \draw [bend left=40] (0,-0.3) to (0,-1);
}&\tikz[scale=0.6,baseline=(vert_cent.base),scale=1]{
	%\draw[help lines] (-3,-3) grid (3,3);
	\node (vert_cent) {\hspace{-13pt}$\phantom{-}$};
	\draw (0,0) circle[radius=1cm];
\draw (0,-1) -- (0,1);
\draw (-1.5,0) circle [radius=0.5];
\draw (-2.5,0) -- (-2,0);
\vsq at (1,0) {};
}&$\ncal'_{4,5}=L(1,1)L(1,1,1,1,1)C_4$\nn 
\hline
$\Ical_{4,6}$&\tikz[scale=0.6,baseline=(vert_cent.base)]{
  \node (vert_cent) {\hspace{-13pt}$\phantom{-}$};
\node at (0,0.6) () {};
\node at (0,-0.6) () {};
\draw (-1,0) circle [radius=0.1];
\draw (1,0) circle [radius=0.1];
\draw (-0.5,0.5) -- (0.5,0.5) (-0.5,0.5) -- (-0.5,-0.5)  (-0.5,-0.5) -- (0.5,-0.5) (0.5,-0.5) -- (0.5,0.5)
(-0.92,0.08) -- (-0.5,0.5) (-0.92,-0.08) -- (-0.5,-0.5) (0.92,0.08) -- (0.5,0.5) (0.92,-0.08) -- (0.5,-0.5);
}&
\tikz[scale=0.6,baseline=(vert_cent.base)]{
  \node (vert_cent) {\hspace{-13pt}$\phantom{-}$};
\draw (-0.5,0.5) -- (0.5,0.5) (-0.5,0.5) -- (-0.5,-0.5)  (-0.5,-0.5) -- (0.5,-0.5) (0.5,-0.5) -- (0.5,0.5)
(-0.5,-0.5) -- (0.5,0.5) (-0.5,0.5) -- (0.5,-0.5);
}&
\tikz[scale=0.6,baseline=(vert_cent.base)]{
  \node (vert_cent) {\hspace{-13pt}$\phantom{-}$};
\draw (-1,-0.5) -- (0,-0.5) (-1,-0.5) -- (-0.5,0.5)  (-0.5,0.5) -- (0,-0.5) (-0.5,0.5) -- (0.5,0.5)
(0.5,0.5) -- (0,-0.5) (0.5,0.5) -- (1,-0.5) (0,-0.5) -- (1,-0.5) (-1.5,-0.5) -- (-1,-0.5);
\vsq at (1,-0.5) {};
}&$\ncal'_{4,6}=\left(\tfrac{20\zeta_5}{\epsilon}+\ldots\right)C_4$\\
\hline
$\Ical_{4,7a}$&\tikz[scale=0.6,baseline=(vert_cent.base),scale=1]{
\filldraw  (157:1) circle [radius=0.1];
\draw (-1,0) circle [radius=0.1];
\draw (1,0) circle [radius=0.1];
\draw [bend right=30] (135:1) to (-1,0.1);
\draw [bend left=30] (45:1) to (1,0.1); 
\draw [bend right=30] (45:1) to (1,0.1);
\draw [bend left=30] (90:1) to (45:1);
\draw [bend right=30] (90:1) to (45:1);
\draw [bend left=30] (135:1) to (90:1);
\draw [bend right=30] (135:1) to (90:1);
}
&\tikz[scale=0.6,baseline=(vert_cent.base),scale=1]{
	%\draw[help lines] (-3,-3) grid (3,3);
	\node (vert_cent) {\hspace{-13pt}$\phantom{-}$};
\node at (0,1.1) () {};
	\draw (0,0) circle[radius=1cm];
	\draw [bend right=40] (-1,0) to (120:1);
\draw [bend right=40] (120:1) to (60:1);
           \draw [bend left=40] (1,0) to (60:1);
\filldraw (0,-1) circle [radius=0.1];
}&\tikz[scale=0.6,baseline=(vert_cent.base),scale=1]{
\draw (-1,0) circle [radius=0.25];
\draw (-0.5,0) circle [radius=0.25];
\draw (0,0) circle [radius=0.25];
\draw (-1.25,0) -- (-1.5,0);
\draw (.25,0) -- (0.5,0);
}&$\ncal'_{4,7}=L(1,1)^3C_4$\nn 
$\Ical_{4,7b}$&\tikz[scale=0.6,baseline=(vert_cent.base),scale=1]{
\filldraw  (23:1) circle [radius=0.1];
\draw (-1,0) circle [radius=0.1];
\draw (1,0) circle [radius=0.1];
\draw [bend right=30] (135:1) to (-1,0.1);
\draw [bend right=30] (135:1) to (1,0.1); 
\draw [bend left=30] (45:1) to (1,0.1);
\draw [bend left=30] (90:1) to (45:1);
\draw [bend right=30] (90:1) to (45:1);
\draw [bend left=30] (135:1) to (90:1);
\draw [bend right=30] (135:1) to (90:1);
}&&&\\
\hline
\end{tabular}
\caption{\label{IRres4}Alternative results for four-loop infra-red counterterms}
\end{table}
The standard results for the IR counterterms as computed and listed in Appendix A.3 of Ref.~\cite{klein} are given for completeness in Eq.~\eqref{IRold} (though for the reasons given in the main text, we quote here our expression for $\ncal_{2a}$ which differs from that given in Ref.~\cite{klein}) 
\begin{subequations}\label{IRold}
\begin{align}
\ncal_1=&\,\tfrac{2}{\epsilon},\\
\ncal_{2}=&\,\tfrac{1}{\epsilon^2}(2+\epsilon),\\
\ncal_{3,1}=&\,\tfrac{1}{\epsilon^3}\left(\tfrac83+\tfrac83\epsilon-\tfrac23\epsilon^2\right),\label{IRold:c}\\
\ncal_{3,2}=&\,\tfrac{1}{\epsilon^3}\left(\tfrac43+2\epsilon+\tfrac43\epsilon^2\right),\\
\ncal_{3,3}=&\,\tfrac{1}{\epsilon^3}\left(\tfrac83+\tfrac43\epsilon+\tfrac23\epsilon^2\right),\\
\ncal_{3,4}=&\,\tfrac{4\zeta_3}{\epsilon},\\
\ncal_{4,1}=&\,\tfrac{1}{\epsilon^4}\left(\tfrac43+\tfrac83\epsilon+\tfrac13\epsilon^2+[2\zeta_3-1]\epsilon^3\right),\\
\ncal_{4,2}=&\,\tfrac{1}{\epsilon^4}\left(\tfrac23+2\epsilon+\tfrac{19}{6}\epsilon^2+\left[\tfrac52-2\zeta_3\right]\epsilon^3\right),\\
\ncal_{4,3}=&\,\tfrac{1}{\epsilon^2}\left(2\zeta_3+\left[3\zeta_3+\tfrac32\zeta_4\right]\epsilon\right),\\
\ncal_{4,4}=&\,\tfrac{1}{\epsilon^4}\left(\tfrac43+2\epsilon+\tfrac{7}{3}\epsilon^2+\left[\tfrac{11}{6}-\zeta_3\right]\epsilon^3\right),\\
\ncal_{4,5}=&\,\tfrac{1}{\epsilon^2}\left(6\zeta_3+\left[3\zeta_3-\tfrac32\zeta_4\right]\epsilon\right),\\
\ncal_{4,6}=&\,\tfrac{20\zeta_5}{\epsilon},\\
\ncal_{4,7}=&\,\tfrac{1}{\epsilon^4}\left(4+6\epsilon-3\epsilon^2+\left[\tfrac12-\zeta_3\right]\epsilon^3\right).
\end{align}
\end{subequations}
Finally we give here explicit results for the differences between our AIR counterterms and the standard ones, assuming that AIR counterterms with the same CVG have the same expansions, just as in the standard case; so that $\ncal'_{2b}=\ncal'_{2a}$, $\ncal'_{3,1b}=\ncal'_{3,1a}$, etc.
\begin{subequations}\label{irdiffs}
\begin{align}
\delta^{IR}_1=&\,0,\label{irdiffs:a}\\
\delta^{IR}_{2}=&\,\tfrac12+\left(\tfrac14-\tfrac12\zeta_3\right)\epsilon+\ldots,\label{irdiffs:b}\\
\delta^{IR}_{3,1}=&\,\tfrac{2}{\epsilon}-\left(\tfrac23+\tfrac43\zeta_3\right)+\ldots,\label{irdiffs:c}\\
\delta^{IR}_{3,2}=&\,\tfrac{1}{\epsilon}+\tfrac52-\tfrac53\zeta_3+\ldots,\label{irdiffs:d}\\
\delta^{IR}_{3,3}=&\,\tfrac13+\tfrac23\zeta_3+\ldots,\label{irdiffs:e}\\
\delta^{IR}_{3,4}=&\,3\zeta_4,\label{irdiffs:f}\\
\delta^{IR}_{4,1}=&\,\tfrac{2}{\epsilon^2}-\left(\tfrac43+\tfrac23\zeta_3\right)\tfrac{1}{\epsilon}+\ldots,\label{irdiffs:g}\\
\delta^{IR}_{4,2}=&\,\tfrac{1}{\epsilon^2}+\left(5-\tfrac73\zeta_3\right)\tfrac{1}{\epsilon}+\ldots,\label{irdiffs:h}\\
\delta^{IR}_{4,3}=&\,O(1),\label{irdiffs:i}\\
\delta^{IR}_{4,4}=&\,\tfrac{2}{\epsilon}\delta^{IR}_{3,3}+O(1),\label{irdiffs:j}\\
\delta^{IR}_{4,5}=&\,\tfrac{6\zeta_4}{\epsilon},\label{irdiffs:k}\\
\delta^{IR}_{4,6}=&\,O(1),\label{irdiffs:l}\\
\delta^{IR}_{4,7}=&\,\tfrac{6}{\epsilon^2}-(7+2\zeta_3)\tfrac{1}{\epsilon}.
\end{align}
\end{subequations}
\section{Example calculations}
In this Appendix we perform numerous comparisons to show that our alternative counterterms give the same results for UV counterterms as the standard ones, for all the five-loop diagrams contributing to the scalar $\phi^4$ $\beta$-function where IR divergences were found in Ref.~\cite{klein} to be unavoidable even with IR rearrangement. We focus on IR divergences with double propagators, covering the case of IR divergences with no double propagators in Appendix C. We use the diagrammatic results for the counterterms given in Appendix A.1 of Ref.~\cite{klein}, which are reproduced in equations such as Eqs.~\eqref{g67graph}, denoting the counterterms $\Delta_{L,n}$ where $L$ records the loop number and $n$ is the label used in Ref.~\cite{klein}. The ellipses in our expressions represent terms not involving IR divergences, which of course are the same using either method of computing the IR counterterms and therefore do not concern us. We use the notation introduced in Sections 3 and 4. For instance in Eq.~\eqref{g67graph}, we denote the full five-loop graph by $G^{5,67}$. The graphs resulting from excising from $G^{5,67}$ the IR-divergent substructures at one, two,... loops are then denoted by $G_1^{5,67}$, $G_2^{5,67}$ etc. The expressions multiplying the successive IR counterterms in the equations such as Eq.~\eqref{g67graph} for $\Delta_{5,67}$ etc are similarly denoted $I_1^{5,67}$, $I_2^{5,67}$ etc and are given by $I_1^{5,67}=RG_1^{5,67}$ etc. Explicit expressions are given  immediately afterwards, in equations such as Eq.~\eqref{i567}.    We also need here results for the counterterms for various lower loop graphs which are recorded in Eq.~\eqref{divs}. For each diagram, the differences $\delta_{5,67}$ etc between the results for the UV counterterms of using our AIR counterterms as described in Tables~\ref{IRres12}, \ref{IRres3} and \ref{IRres4} and the results of using the standard ones as given in Eq.~\eqref{IRold} are then evaluated and also given in Eqs.~\eqref{g67res}, etc. In each case the difference turns out to be zero upon using Eq.~\eqref{irdiffs}, verifying that our AIR counterterms are equally valid. 

We group the diagrams according to the order and similarity of the IR singularities; thus the first two diagrams, No. 67 and No. 101, have only one and two-loop IR singularities, and the next one, No. 37, has one and four-loop IR singularities (there is no need to consider in detail diagrams such as No. 47 which only contain one-loop IR divergences, and are manifestly unchanged by the switch to our AIR counterterms). The following No. 107 has one, two and three-loop IR singularities. Then diagrams Nos. 53, 75 and 111 have one, two, three and four-loop IR singularities of similar kind, as do the following two diagrams Nos. 73 and 93. 

Starting with Diagram 67, we first observe that there is an error in the depiction of the five-loop diagram in Ref.~\cite{klein}. Correcting this, we have
\begin{align}
\Delta_{5,67}=&\,-\Kcal\Rbar^*\left(\tikz[scale=0.6,baseline=(vert_cent.base),scale=1]{
	%\draw[help lines] (-3,-3) grid (3,3);
	\node (vert_cent) {\hspace{-13pt}$\phantom{-}$};
	\draw (0,0) circle[radius=1cm];
           \draw [bend left=40] (1,0) to (0,1);
      \draw [bend left=40] (-1,0) to (0,0);
\draw (-1.25,0) -- (1.25,0);
\draw (0,0) -- (0,-1);
\filldraw (0,-0.5) circle [radius=0.1];
}\right)=-\Kcal\Bigl[\ldots + \left(\,
\tikz[scale=0.6,baseline=(vert_cent.base)]{
  \node (vert_cent) {\hspace{-13pt}$\phantom{-}$};
  \draw (0,-.4)--(0,.4);
        \draw (0,.5) circle [radius=0.1];
\draw (0,-.5) circle [radius=0.1];
\filldraw (0,0) circle [radius=0.1];
}\,\right)_{IR}\Bigl\{\tikz[scale=0.6,baseline=(vert_cent.base),scale=1]{
	%\draw[help lines] (-3,-3) grid (3,3);
	\node (vert_cent) {\hspace{-13pt}$\phantom{-}$};
	\draw (0,0) circle[radius=1cm];
           \draw [bend left=40] (1,0) to (0,1);
      \draw [bend left=40] (-1,0) to (0,0);
\draw (-1.25,0) -- (1.25,0);
\filldraw (0,-1) circle [radius=0.1];
}+2\Delta\left(\tikz[scale=0.6,baseline=(vert_cent.base)]{
  \node (vert_cent) {\hspace{-13pt}$\phantom{-}$};
  \filldraw (1.3,0) circle [radius=0.1]; \filldraw (0.1,0) circle [radius=0.1];\draw
        (0.7,0) ++(0:0.6cm) arc (0:360:0.6cm and 0.4cm);
}\right)\tikz[scale=0.6,baseline=(vert_cent.base),scale=1]{
	%\draw[help lines] (-3,-3) grid (3,3);
	\node (vert_cent) {\hspace{-13pt}$\phantom{-}$};
	\draw (0,0) circle[radius=1cm];
           \draw [bend left=40] (1,0) to (0,1);
\draw (-1.25,0) -- (1.25,0);
\filldraw (0,-1) circle [radius=0.1];
}\nn
&+\Delta\left(\tikz[scale=0.6,baseline=(vert_cent.base)]{
  \node (vert_cent) {\hspace{-13pt}$\phantom{-}$};
  \filldraw (1.3,0) circle [radius=0.1]; \filldraw (0.1,0) circle [radius=0.1];\draw
        (0.7,0) ++(0:0.6cm) arc (0:360:0.6cm and 0.4cm);
}\right)\Delta\left(\tikz[scale=0.6,baseline=(vert_cent.base)]{
  \node (vert_cent) {\hspace{-13pt}$\phantom{-}$};
  \filldraw (1.3,0) circle [radius=0.1]; \filldraw (0.1,0) circle [radius=0.1];\draw
        (0.7,0) ++(0:0.6cm) arc (0:360:0.6cm and 0.4cm);
}\right)\tikz[scale=0.6,baseline=(vert_cent.base),scale=1]{
	%\draw[help lines] (-3,-3) grid (3,3);
	\node (vert_cent) {\hspace{-13pt}$\phantom{-}$};
	\draw (0,0) circle[radius=1cm];
\draw (-1.25,0) -- (1.25,0);
\filldraw (0,-1) circle [radius=0.1];
}+\Delta\left(\tikz[scale=0.6,baseline=(vert_cent.base),scale=1]{
	%\draw[help lines] (-3,-3) grid (3,3);
	\node (vert_cent) {\hspace{-13pt}$\phantom{-}$};
	\draw (0,0) circle[radius=1cm];
           \draw [bend left=40] (1,0) to (0,1);
      \draw [bend left=40] (-1,0) to (0,0);
\draw (-1,0) -- (1,0);
\filldraw (0,-1) circle [radius=0.1];
}\right)\Bigr\}\nn
&+\left(\,\tikz[scale=0.6,baseline=(vert_cent.base)]{
  \node (vert_cent) {\hspace{-13pt}$\phantom{-}$};
  \draw
        (10:0.5cm) arc (10:170:0.5cm) ;
\draw (0,0.1)--(0,.5);
\draw (-0.5,0) circle [radius=0.1];
\draw (0.5,0) circle [radius=0.1];
\draw (0,0) circle [radius=0.1];
\filldraw (0,0.25) circle [radius=0.1];
}\,\right)_{IR}\Bigl\{\tikz[scale=0.6,baseline=(vert_cent.base),scale=1]{
	%\draw[help lines] (-3,-3) grid (3,3);
	\node (vert_cent) {\hspace{-13pt}$\phantom{-}$};
	\draw (0,0) circle[radius=1cm];
           \draw [bend left=40] (1,0) to (0,1);
\draw [bend left=40] (-1,0) to (0,-1);
\draw (-1.25,0) -- (-1,0);
\draw (1.25,0) -- (1,0);
}+2\Delta\left(\tikz[scale=0.6,baseline=(vert_cent.base)]{
  \node (vert_cent) {\hspace{-13pt}$\phantom{-}$};
  \filldraw (1.3,0) circle [radius=0.1]; \filldraw (0.1,0) circle [radius=0.1];\draw
        (0.7,0) ++(0:0.6cm) arc (0:360:0.6cm and 0.4cm);
}\right)\tikz[scale=0.6,baseline=(vert_cent.base),scale=1]{
	%\draw[help lines] (-3,-3) grid (3,3);
	\node (vert_cent) {\hspace{-13pt}$\phantom{-}$};
	\draw (0,0) circle[radius=1cm];
           \draw [bend left=40] (1,0) to (0,1);
\draw (-1.25,0) -- (-1,0);
\draw (1.25,0) -- (1,0);
}\nn
&+\Delta\left(\tikz[scale=0.6,baseline=(vert_cent.base)]{
  \node (vert_cent) {\hspace{-13pt}$\phantom{-}$};
  \filldraw (1.3,0) circle [radius=0.1]; \filldraw (0.1,0) circle [radius=0.1];\draw
        (0.7,0) ++(0:0.6cm) arc (0:360:0.6cm and 0.4cm);
}\right)\Delta\left(\tikz[scale=0.6,baseline=(vert_cent.base)]{
  \node (vert_cent) {\hspace{-13pt}$\phantom{-}$};
  \filldraw (1.3,0) circle [radius=0.1]; \filldraw (0.1,0) circle [radius=0.1];\draw
        (0.7,0) ++(0:0.6cm) arc (0:360:0.6cm and 0.4cm);
}\right)\tikz[scale=0.6,baseline=(vert_cent.base)]{
  \node (vert_cent) {\hspace{-13pt}$\phantom{-}$};
  \draw (0.05,0)--(-0.2,0) (1.3,0) --(1.55,0) 
        (0.7,0) ++(0:0.6cm) arc (0:360:0.6cm and 0.4cm);
}+\Delta\left(\tikz[scale=0.6,baseline=(vert_cent.base),scale=1]{
	%\draw[help lines] (-3,-3) grid (3,3);
	\node (vert_cent) {\hspace{-13pt}$\phantom{-}$};
	\draw (0,0) circle[radius=1cm];
           \draw [bend left=40] (1,0) to (0,1);
\draw [bend left=40] (-1,0) to (0,-1);
}\right)\Bigr\}\Bigr].
\label{g67graph}
\end{align}
We have
\begin{subequations}\label{i567}
\begin{align}
I^{5,67}_1=&\,L(1,1)^2L(1+\tfrac12\epsilon,1+\tfrac12\epsilon)L(2,\tfrac32\epsilon)-2\tfrac{2}{\epsilon}L(1,1)L(1,1+\tfrac12\epsilon)L(2,\epsilon)\nn
&+\left(\tfrac{2}{\epsilon}\right)^2L(1,1)L(2,\tfrac12\epsilon)+\Delta_{4,22}=\left(\tfrac83\zeta_3-\tfrac{74}{3}\right)\tfrac{1}{\epsilon}+\ldots,\\
I^{5,67}_2=&\,L(1,1)^2L(1+\tfrac12\epsilon,1+\tfrac12\epsilon)-2\tfrac{2}{\epsilon}L(1,1)L(1,1+\tfrac12\epsilon)+\left(\tfrac{2}{\epsilon}\right)^2L(1,1)+\Delta_{3,6}\nn
=&\,\tfrac{37}{3}-\tfrac43\zeta_3+\ldots,
\end{align}
\end{subequations}
so that using Eq.~\eqref{irdiffs}, we find
\be
\delta_{5,67}=\Kcal\left(\delta^{IR}_1I^{5,67}_1+\delta^{IR}_{2}I^{5,67}_2\right)=0.
\label{g67res}
\ee
In view of the vanishing of $\delta^{IR}_1$ and the finiteness of $\delta^{IR}_{2}$ which we see in Eq.~\eqref{irdiffs}, the finiteness of $I^{5,67}_2$ is all we need for our current purposes; but we observe that $I^{5,67}_1$ and $I^{5,67}_2$ are related by a similar equation to that for $I^{4,25}_1$ and $I^{4,25}_2$ in Eq.~\eqref{i425c}, which is again responsible for factors of $F_{1,2}$ in $C_{1,2}$ having no overall effect.

Continuing with Diagram 101, we have
\begin{align}
\Delta_{5,101}=&\,-\Kcal\Rbar^*\left(\tikz[scale=0.6,baseline=(vert_cent.base),scale=1]{
	%\draw[help lines] (-3,-3) grid (3,3);
	\node (vert_cent) {\hspace{-13pt}$\phantom{-}$};
	\draw (0,0) circle[radius=1cm];
           \draw [bend left=40] (1,0) to (0,1);
      \draw [bend left=40] (-1,0) to (0,0);
\draw (-1.25,0) -- (1.25,0);
\draw (0,0) -- (0,-1);
\filldraw (-45:1) circle [radius=0.1];
}\right)=-\Kcal\Bigl[\ldots + \left(\,
\tikz[scale=0.6,baseline=(vert_cent.base)]{
  \node (vert_cent) {\hspace{-13pt}$\phantom{-}$};
  \draw (0,-.4)--(0,.4);
        \draw (0,.5) circle [radius=0.1];
\draw (0,-.5) circle [radius=0.1];
\filldraw (0,0) circle [radius=0.1];
}\,\right)_{IR}\Bigl\{\tikz[scale=0.6,baseline=(vert_cent.base),scale=1]{
	%\draw[help lines] (-3,-3) grid (3,3);
	\node (vert_cent) {\hspace{-13pt}$\phantom{-}$};
	\draw (0,0) circle[radius=1cm];
           \draw [bend left=40] (1,0) to (0,1);
\draw [bend left=40] (-1,0) to (0,-1);
\draw (-1,0) -- (0,-1);
\filldraw (-135:1) circle [radius=0.1];
\draw (-1.25,0) -- (-1,0);
\draw (1.25,0) -- (1,0);
}+\Delta\left(\tikz[scale=0.6,baseline=(vert_cent.base)]{
  \node (vert_cent) {\hspace{-13pt}$\phantom{-}$};
  \filldraw (1.3,0) circle [radius=0.1]; \filldraw (0.1,0) circle [radius=0.1];\draw
        (0.7,0) ++(0:0.6cm) arc (0:360:0.6cm and 0.4cm);
}\right)\tikz[scale=0.6,baseline=(vert_cent.base),scale=1]{
	%\draw[help lines] (-3,-3) grid (3,3);
	\node (vert_cent) {\hspace{-13pt}$\phantom{-}$};
	\draw (0,0) circle[radius=1cm];
\draw [bend left=40] (-1,0) to (0,-1);
\draw (-1,0) -- (0,-1);
\filldraw (-135:1) circle [radius=0.1];
\draw (-1.25,0) -- (-1,0);
\draw (1.25,0) -- (1,0);
}\nn
&+\Delta\left(\tikz[scale=0.6,baseline=(vert_cent.base),scale=1]{
	%\draw[help lines] (-3,-3) grid (3,3);
	\node (vert_cent) {\hspace{-13pt}$\phantom{-}$};
	\draw (0,0) circle[radius=1cm];
           \draw [bend left=40] (1,0) to (0,1);
\draw (-1.25,0) -- (-1,0);
\draw (1.25,0) -- (1,0);
}\right)\tikz[scale=0.6,baseline=(vert_cent.base),scale=1]{
	%\draw[help lines] (-3,-3) grid (3,3);
	\node (vert_cent) {\hspace{-13pt}$\phantom{-}$};
	\draw (0,0) circle[radius=1cm];
           \draw [bend left=40] (1,0) to (0,1);
\draw (-1.25,0) -- (-1,0);
\draw (1.25,0) -- (1,0);
}+\Delta\left(\tikz[scale=0.6,baseline=(vert_cent.base)]{
  \node (vert_cent) {\hspace{-13pt}$\phantom{-}$};
  \filldraw (1.3,0) circle [radius=0.1]; \filldraw (0.1,0) circle [radius=0.1];\draw
        (0.7,0) ++(0:0.6cm) arc (0:360:0.6cm and 0.4cm);
}\right)\Delta\left(\tikz[scale=0.6,baseline=(vert_cent.base),scale=1]{
	%\draw[help lines] (-3,-3) grid (3,3);
	\node (vert_cent) {\hspace{-13pt}$\phantom{-}$};
	\draw (0,0) circle[radius=1cm];
           \draw [bend left=40] (1,0) to (0,1);
\draw (-1.25,0) -- (-1,0);
\draw (1.25,0) -- (1,0);
}\right)\tikz[scale=0.6,baseline=(vert_cent.base)]{
  \node (vert_cent) {\hspace{-13pt}$\phantom{-}$};
 \draw (0.05,0)--(-0.2,0) (1.3,0) --(1.55,0);
 \draw (0.7,0) ++(0:0.6cm) arc (0:360:0.6cm and 0.4cm);
}\nn
&+\Delta\left(\tikz[scale=0.6,baseline=(vert_cent.base),scale=1]{
	%\draw[help lines] (-3,-3) grid (3,3);
	\node (vert_cent) {\hspace{-13pt}$\phantom{-}$};
	\draw (0,0) circle[radius=1cm];
           \draw [bend left=40] (1,0) to (0,1);
\draw [bend left=40] (-1,0) to (0,-1);
\draw (-1,0) -- (0,-1);
\filldraw (-135:1) circle [radius=0.1];
}\right)\Bigr\}\nn
&+\left(\,\tikz[scale=0.6,baseline=(vert_cent.base)]{
  \node (vert_cent) {\hspace{-13pt}$\phantom{-}$};
  \draw
        (10:0.5cm) arc (10:170:0.5cm) ;
\draw (0,0.1)--(0,.5);
\draw (-0.5,0) circle [radius=0.1];
\draw (0.5,0) circle [radius=0.1];
\draw (0,0) circle [radius=0.1];
\filldraw (0,0.25) circle [radius=0.1];
}\,\right)_{IR}\Bigl\{\tikz[scale=0.6,baseline=(vert_cent.base),scale=1]{
	%\draw[help lines] (-3,-3) grid (3,3);
	\node (vert_cent) {\hspace{-13pt}$\phantom{-}$};
	\draw (0,0) circle[radius=1cm];
           \draw [bend left=40] (1,0) to (0,1);
\draw [bend left=40] (-1,0) to (0,-1);
\draw (-1.25,0) -- (-1,0);
\draw (1.25,0) -- (1,0);
}+2\Delta\left(\tikz[scale=0.6,baseline=(vert_cent.base)]{
  \node (vert_cent) {\hspace{-13pt}$\phantom{-}$};
  \filldraw (1.3,0) circle [radius=0.1]; \filldraw (0.1,0) circle [radius=0.1];\draw
        (0.7,0) ++(0:0.6cm) arc (0:360:0.6cm and 0.4cm);
}\right)\tikz[scale=0.6,baseline=(vert_cent.base),scale=1]{
	%\draw[help lines] (-3,-3) grid (3,3);
	\node (vert_cent) {\hspace{-13pt}$\phantom{-}$};
	\draw (0,0) circle[radius=1cm];
           \draw [bend left=40] (1,0) to (0,1);
\draw (-1.25,0) -- (-1,0);
\draw (1.25,0) -- (1,0);
}\nn
&+\Delta\left(\tikz[scale=0.6,baseline=(vert_cent.base)]{
  \node (vert_cent) {\hspace{-13pt}$\phantom{-}$};
  \filldraw (1.3,0) circle [radius=0.1]; \filldraw (0.1,0) circle [radius=0.1];\draw
        (0.7,0) ++(0:0.6cm) arc (0:360:0.6cm and 0.4cm);
}\right)\Delta\left(\tikz[scale=0.6,baseline=(vert_cent.base)]{
  \node (vert_cent) {\hspace{-13pt}$\phantom{-}$};
  \filldraw (1.3,0) circle [radius=0.1]; \filldraw (0.1,0) circle [radius=0.1];\draw
        (0.7,0) ++(0:0.6cm) arc (0:360:0.6cm and 0.4cm);
}\right)\tikz[scale=0.6,baseline=(vert_cent.base)]{
  \node (vert_cent) {\hspace{-13pt}$\phantom{-}$};
 \draw (0.05,0)--(-0.2,0) (1.3,0) --(1.55,0);
  \draw (0.7,0) ++(0:0.6cm) arc (0:360:0.6cm and 0.4cm);
}+\Delta\left(\tikz[scale=0.6,baseline=(vert_cent.base),scale=1]{
	%\draw[help lines] (-3,-3) grid (3,3);
	\node (vert_cent) {\hspace{-13pt}$\phantom{-}$};
	\draw (0,0) circle[radius=1cm];
           \draw [bend left=40] (1,0) to (0,1);
\draw [bend left=40] (-1,0) to (0,-1);
}\right)\Bigr\}\Bigr].
\end{align}
We have
\begin{subequations}\label{i5101}
\begin{align}
I^{5,101}_1=&\,L(1,1)^2L(2,\tfrac12\epsilon)L(1+\tfrac12\epsilon,1+\epsilon)-\tfrac{2}{\epsilon}L(1,1)L(2,\tfrac12\epsilon)L(1,1+\epsilon)
\nn
&+\left(\tfrac{2}{\epsilon^2}-\tfrac{1}{\epsilon}\right)L(1,1)L(1,1+\tfrac12\epsilon)-\tfrac{2}{\epsilon}\left(\tfrac{2}{\epsilon^2}-\tfrac{1}{\epsilon}\right)
L(1,1)+\Delta_{4,9}\nn
=&\,I^{2,67}_2+O(1)=\left(\tfrac83\zeta_3-\tfrac{74}{3}\right)\tfrac{1}{\epsilon}+\ldots,\\
I^{5,101}_2=&\,I^{2,67}_2=\tfrac{37}{3}-\tfrac43\zeta_3+\ldots,
\end{align}
\end{subequations}
so that using Eq.~\eqref{irdiffs} we find
\be
\delta_{5,101}=\Kcal\left(\delta^{IR}_1I^{5,101}_1+\delta^{IR}_{2}I^{5,101}_2\right)=0.
\label{g101res}
\ee
Similar remarks regarding the leading behaviour of $I^{5,101}_1$, $I^{5,101}_2$ apply as in the case of the previous diagram, Diagram 67.

Diagram 37 gives us
\begin{align}
\Delta_{5,37}=&\,-\Kcal\Rbar^*\left(\tikz[scale=0.6,baseline=(vert_cent.base),scale=1]{
	%\draw[help lines] (-3,-3) grid (3,3);
	\node (vert_cent) {\hspace{-13pt}$\phantom{-}$};
	\draw (0,0) circle[radius=1cm];
\draw [bend left=40] (-1,0) to (1,0);
\draw (-1.25,0) -- (1.25,0);
\draw (0,0) -- (-120:1);
\draw (0,0) -- (-60:1);
\filldraw (-0.5,0) circle [radius=0.1];
}\right)=-\Kcal\Bigl[\ldots +\left(\,
\tikz[scale=0.6,baseline=(vert_cent.base)]{
  \node (vert_cent) {\hspace{-13pt}$\phantom{-}$};
  \draw (0,-.4)--(0,.4);
        \draw (0,.5) circle [radius=0.1];
\draw (0,-.5) circle [radius=0.1];
\filldraw (0,0) circle [radius=0.1];
}\,\right)_{IR}\Bigl\{\tikz[scale=0.6,baseline=(vert_cent.base),scale=1]{
	%\draw[help lines] (-3,-3) grid (3,3);
	\node (vert_cent) {\hspace{-13pt}$\phantom{-}$};
	\draw (0,0) circle[radius=1cm];
\draw [bend left=40] (-1,0) to (1,0);
\draw (0,0) to (0,-1);
\draw  (0,0) to (-45:1);
\draw (-1.25,0) --(-1,0) (0,0) -- (1.25,0);
}+\Delta\left(\tikz[scale=0.6,baseline=(vert_cent.base),scale=1]{
	%\draw[help lines] (-3,-3) grid (3,3);
	\node (vert_cent) {\hspace{-13pt}$\phantom{-}$};
	\draw (0,0) circle[radius=1cm];
\draw [bend left=40] (-1,0) to (1,0);
\draw (0,0) to (0,-1);
\draw  (0,0) to (-45:1);
\draw  (0,0) -- (1,0);
}\right)\Bigr\}\nn
&+\left(\tikz[scale=0.6,baseline=(vert_cent.base)]{
  \node (vert_cent) {\hspace{-13pt}$\phantom{-}$};
  \draw
        (5:1cm) arc (5:175:1cm) ;
\draw (0,0)--(45:1);
\draw (0,0)--(135:1);
\draw (-1,0) circle [radius=0.1];
\draw (1,0) circle [radius=0.1];
\filldraw (-0.5,0) circle [radius=0.1];
\draw (-0.9,0) -- (0.9,0);
\draw (0.9,0) -- (0.1,0);
}\right)_{IR}\Bigl\{\tikz[scale=0.6,baseline=(vert_cent.base)]{
  \node (vert_cent) {\hspace{-13pt}$\phantom{-}$};
  \draw (-0.1,0)--(0.1,0)
        (0.7,0) ++(0:0.6cm) arc (0:360:0.6cm and 0.4cm)
        (1.3,0)--(1.5,0);
}+\Delta\left(\tikz[scale=0.6,baseline=(vert_cent.base)]{
  \node (vert_cent) {\hspace{-13pt}$\phantom{-}$};
  \filldraw (1.3,0) circle [radius=0.1]; \filldraw (0.1,0) circle [radius=0.1];\draw
        (0.7,0) ++(0:0.6cm) arc (0:360:0.6cm and 0.4cm);
}\right)\Bigr\}\Bigr].
\end{align}
Here $I_2^{5,37}=I_1^{2,2}$ (in Eq.~\eqref{i22:b}) is finite. Since $\delta_1^{IR}=0$ and $\delta_{4,3}^{IR}$ is finite, we immediately see
\be
\delta_{5,37}=\Kcal\left(\delta_1^{IR}I_1^{5,37}+\delta_{4,3}^{IR}I_2^{5,37}\right)=0.
\label{del537}
\ee
($I_1^{5,37}$ {\it does} have a 3-loop IR divergent structure with no double propagators, which is not listed in Table~\ref{IRres3}, since it does not occur elsewhere; its CVG is that for $\ncal_{3,4}$.)

Diagram 107 has one, two and three-loop IR divergences, we have
\begin{align}
\Delta_{5,107}=&\,-\Kcal\Rbar^*\left(\tikz[scale=0.6,baseline=(vert_cent.base),scale=1]{
	%\draw[help lines] (-3,-3) grid (3,3);
	\node (vert_cent) {\hspace{-13pt}$\phantom{-}$};
	\draw (0,0) circle[radius=1cm];
           \draw [bend left=40] (1,0) to (0,1);
\draw [bend right=40] (-1,0) to (0,1);
\draw (-1.25,0) -- (1.25,0);
\draw (0,0) -- (0,-1);
\filldraw (0,-0.5) circle [radius=0.1];
}\right)=-\Kcal\Bigl[\ldots +\left(\,
\tikz[scale=0.6,baseline=(vert_cent.base)]{
  \node (vert_cent) {\hspace{-13pt}$\phantom{-}$};
  \draw (0,-.4)--(0,.4);
        \draw (0,.5) circle [radius=0.1];
\draw (0,-.5) circle [radius=0.1];
\filldraw (0,0) circle [radius=0.1];
}\,\right)_{IR}\Bigl\{
\tikz[scale=0.6,baseline=(vert_cent.base),scale=1]{
	%\draw[help lines] (-3,-3) grid (3,3);
	\node (vert_cent) {\hspace{-13pt}$\phantom{-}$};
	\draw (0,0) circle[radius=1cm];
           \draw [bend left=40] (1,0) to (0,1);
\draw [bend right=40] (-1,0) to (0,1);
\draw (-1.25,0) -- (1.25,0);
\filldraw (0,0) circle [radius=0.1];
\filldraw (0,-1) circle [radius=0.1];
}+2\Delta\left(\tikz[scale=0.6,baseline=(vert_cent.base)]{
  \node (vert_cent) {\hspace{-13pt}$\phantom{-}$};
  \filldraw (1.3,0) circle [radius=0.1]; \filldraw (0.1,0) circle [radius=0.1];\draw
        (0.7,0) ++(0:0.6cm) arc (0:360:0.6cm and 0.4cm);
}\right)\tikz[scale=0.6,baseline=(vert_cent.base),scale=1]{
	%\draw[help lines] (-3,-3) grid (3,3);
	\node (vert_cent) {\hspace{-13pt}$\phantom{-}$};
	\draw (0,0) circle[radius=1cm];
\draw [bend left=40] (-1,0) to (1,0);
\draw (-1.25,0) -- (1.25,0);
\filldraw (0,0) circle [radius=0.1];
\filldraw (0,-1) circle [radius=0.1];
}\nn
&
+\Delta\left(\tikz[scale=0.6,baseline=(vert_cent.base),scale=1]{
	%\draw[help lines] (-3,-3) grid (3,3);
	\node (vert_cent) {\hspace{-13pt}$\phantom{-}$};
	\draw (0,0) circle[radius=1cm];
           \draw [bend left=40] (1,0) to (0,1);
\draw [bend right=40] (-1,0) to (0,1);
\draw (-1,0) -- (1,0);
\filldraw (0,0) circle [radius=0.1];
\filldraw (0,-1) circle [radius=0.1];
}\right)
\Bigr\}\nn
&+2\left(\,\tikz[scale=0.6,baseline=(vert_cent.base)]{
  \node (vert_cent) {\hspace{-13pt}$\phantom{-}$};
  \draw
        (10:0.5cm) arc (10:170:0.5cm) ;
\draw (0,0.1)--(0,.5);
\draw (-0.5,0) circle [radius=0.1];
\draw (0.5,0) circle [radius=0.1];
\draw (0,0) circle [radius=0.1];
\filldraw (0,0.25) circle [radius=0.1];
}\,\right)_{IR}\Bigl\{\tikz[scale=0.6,baseline=(vert_cent.base),scale=1]{
\draw (180:1cm) arc (180:0:1cm) ;
\draw [bend right=40] (0,1) to (1,0);
\draw [bend left=40] (0,1) to (-1,0);
\filldraw (0,0) circle [radius=0.1];
\draw (-1.25,0) -- (1.25,0);
}+2\Delta\left(\tikz[scale=0.6,baseline=(vert_cent.base)]{
  \node (vert_cent) {\hspace{-13pt}$\phantom{-}$};
  \filldraw (1.3,0) circle [radius=0.1]; \filldraw (0.1,0) circle [radius=0.1];\draw
        (0.7,0) ++(0:0.6cm) arc (0:360:0.6cm and 0.4cm);
}\right)\tikz[scale=0.6,baseline=(vert_cent.base),scale=1]{
\draw (180:1cm) arc (180:0:1cm) ;
\draw [bend right=60] (1,0) to (-1,0);
\filldraw (0,0) circle [radius=0.1];
\draw (-1.25,0) -- (1.25,0);
}
+\Delta\left(\tikz[scale=0.6,baseline=(vert_cent.base),scale=1]{
\draw (180:1cm) arc (180:0:1cm) ;
\draw [bend right=40] (0,1) to (1,0);
\draw [bend left=40] (0,1) to (-1,0);
\filldraw (0,0) circle [radius=0.1];
\draw (-1,0) -- (1,0);
}\right)\Bigr\}\nn
&+\left(\tikz[scale=0.6,baseline=(vert_cent.base),scale=1]{
\filldraw  (0,0.5) circle [radius=0.1];
\draw (-1,0) circle [radius=0.1];
\draw (1,0) circle [radius=0.1];
\draw (0,1) --(0,0);
\draw [bend right=40] (0,1) to (-1,0.1);
\draw [bend left=40] (0,1) to (1,0.1);
\draw (-.9,0) -- (0.9,0) 
}\right)_{IR}\Bigl\{\tikz[scale=0.6,baseline=(vert_cent.base),scale=1]{
\draw (-0.5,0) circle [radius=0.5];
\draw (0.5,0) circle [radius=0.5];
\draw (-1.25,0) -- (-1,0) (1,0) -- (1.25,0);
}+2\Delta\left(\tikz[scale=0.6,baseline=(vert_cent.base)]{
  \node (vert_cent) {\hspace{-13pt}$\phantom{-}$};
  \filldraw (1.3,0) circle [radius=0.1]; \filldraw (0.1,0) circle [radius=0.1];\draw
        (0.7,0) ++(0:0.6cm) arc (0:360:0.6cm and 0.4cm);
}\right)\tikz[scale=0.6,baseline=(vert_cent.base),scale=1]{
\draw (0,0) circle [radius=0.5];
\draw (-0.75,0) -- (-0.5,0) (0.5,0) -- (0.75,0);
}
+\Delta\left(\tikz[scale=0.6,baseline=(vert_cent.base),scale=1]{
\draw (-0.5,0) circle [radius=0.5];
\draw (0.5,0) circle [radius=0.5];
\filldraw (-1,0) circle [radius=0.1];
\filldraw (1,0) circle [radius=0.1];
}\right)
\Bigr\}\Bigr].
\label{diag5107}
\end{align}
Note there is a mistake in the diagrams contributing to $I^{5,107}_1$ in Ref.~\cite{klein}.
\begin{subequations}\label{i5107}
\begin{align}
I^{5,107}_1=&\,L(1,1)^2L(2,\epsilon)L(2,\tfrac32\epsilon)-2\tfrac{2}{\epsilon}L(1,1)L(2,\tfrac12\epsilon)L(2,\epsilon)+\Delta_{4,14}=\tfrac{16}{\epsilon^2}+\ldots,\\
I^{5,107}_2=&\,L(1,1)^2L(2,\epsilon)-2\tfrac{2}{\epsilon}L(1,1)L(2,\tfrac12\epsilon)+\Delta_{3,5}=-\tfrac{8}{\epsilon}+\ldots,\\
I^{5,107}_3=&\,L(1,1)^2-2\tfrac{2}{\epsilon}L(1,1)+\left(\tfrac{2}{\epsilon}\right)^2=4+\ldots,
\end{align}
\end{subequations}
so that using Eq.~\eqref{irdiffs}, we find
\be
\delta_{5,107}=\Kcal\left(\delta^{IR}_1I^{5,107}_1+2\delta^{IR}_{2}I^{5,107}_2+\delta^{IR}_{3,1}I^{5,107}_3\right)=0.
\label{del5107}
\ee

The next three diagrams, 53, 75 and 111, have one, two, three and four-loop IR divergences of a similar kind (in fact their CVGs are identical in the case of Diagrams 53 and 75); firstly we have
\begin{align}
\Delta_{5,53}=&\,-\Kcal\Rbar^*\left(\tikz[scale=0.6,baseline=(vert_cent.base),scale=1]{
	%\draw[help lines] (-3,-3) grid (3,3);
	\node (vert_cent) {\hspace{-13pt}$\phantom{-}$};
	\draw (0,0) circle[radius=1cm];
\draw [bend left=40] (-1,0) to (1,0);
\draw (-1.25,0) -- (1.25,0);
\draw (0,0) -- (-120:1);
\draw (0,0) -- (-60:1);
\filldraw (-150:1) circle [radius=0.1];
}\right)=-\Kcal\Bigl[\ldots +\left(\,
\tikz[scale=0.6,baseline=(vert_cent.base)]{
  \node (vert_cent) {\hspace{-13pt}$\phantom{-}$};
  \draw (0,-.4)--(0,.4);
        \draw (0,.5) circle [radius=0.1];
\draw (0,-.5) circle [radius=0.1];
\filldraw (0,0) circle [radius=0.1];
}\,\right)_{IR}\Bigl\{\tikz[scale=0.6,baseline=(vert_cent.base),scale=1]{
	%\draw[help lines] (-3,-3) grid (3,3);
	\node (vert_cent) {\hspace{-13pt}$\phantom{-}$};
	\draw (0,0) circle[radius=1cm];
\draw [bend left=40] (-1,0) to (1,0);
\draw [bend left=40] (0,0) to (0,-1);
\draw [bend right=40] (0,0) to (0,-1);
\draw (-1.25,0) --(-1,0) (0,0) -- (1.25,0);
\filldraw (-.2,-0.5) circle [radius=0.1];
}+\Delta\left(\tikz[scale=0.6,baseline=(vert_cent.base),scale=1]{
	%\draw[help lines] (-3,-3) grid (3,3);
	\node (vert_cent) {\hspace{-13pt}$\phantom{-}$};
	\draw (0,0) circle[radius=1cm];
\draw [bend left=40] (-1,0) to (1,0);
\draw [bend left=40] (0,0) to (0,-1);
\draw [bend right=40] (0,0) to (0,-1);
\draw (0,0) -- (1,0);
\filldraw (-.2,-0.5) circle [radius=0.1];
}\right)\Bigr\}\nn
&+\left(\,\tikz[scale=0.6,baseline=(vert_cent.base)]{
  \node (vert_cent) {\hspace{-13pt}$\phantom{-}$};
  \draw
        (10:0.5cm) arc (10:170:0.5cm) ;
\draw (0,0.1)--(0,.5);
\draw (-0.5,0) circle [radius=0.1];
\draw (0.5,0) circle [radius=0.1];
\draw (0,0) circle [radius=0.1];
\filldraw (0,0.25) circle [radius=0.1];
}\,\right)_{IR}\Bigl\{\tikz[scale=0.6,baseline=(vert_cent.base),scale=1]{
	%\draw[help lines] (-3,-3) grid (3,3);
	\node (vert_cent) {\hspace{-13pt}$\phantom{-}$};
	\draw (0,0) circle[radius=1cm];
           \draw [bend left=40] (1,0) to (0,1);
\draw (-1.25,0) -- (1.25,0);
\filldraw (45:1) circle [radius=0.1];
}+\Delta\left(\tikz[scale=0.6,baseline=(vert_cent.base),scale=1]{
	%\draw[help lines] (-3,-3) grid (3,3);
	\node (vert_cent) {\hspace{-13pt}$\phantom{-}$};
	\draw (0,0) circle[radius=1cm];
           \draw [bend left=40] (1,0) to (0,1);
\draw (-1,0) -- (1,0);
\filldraw (45:1) circle [radius=0.1];
}\right)\Bigr\}\nn
&+\left(\tikz[scale=0.6,baseline=(vert_cent.base)]{
  \node (vert_cent) {\hspace{-13pt}$\phantom{-}$};
  \draw
        (5:1cm) arc (5:175:1cm) ;
\draw (0,0.1)--(45:1);
\draw (0,0.1)--(135:1);
\draw (-1,0) circle [radius=0.1];
\draw (1,0) circle [radius=0.1];
\draw (0,0) circle [radius=0.1];
\filldraw (157.5:1) circle [radius=0.1];
}\right)_{IR}\Bigl\{\tikz[scale=0.6,baseline=(vert_cent.base),scale=1]{
	%\draw[help lines] (-3,-3) grid (3,3);
	\node (vert_cent) {\hspace{-13pt}$\phantom{-}$};
	\draw (0,0) circle[radius=1cm];
\draw (-1.25,0) -- (1.25,0);
\filldraw (0,-1) circle [radius=0.1];
}+\Delta\left(\tikz[scale=0.6,baseline=(vert_cent.base),scale=1]{
	%\draw[help lines] (-3,-3) grid (3,3);
	\node (vert_cent) {\hspace{-13pt}$\phantom{-}$};
	\draw (0,0) circle[radius=1cm];
\draw (-1,0) -- (1,0);
\filldraw (0,-1) circle [radius=0.1];
}\right)\Bigr\}\nn
&+\left(\tikz[scale=0.6,baseline=(vert_cent.base)]{
  \node (vert_cent) {\hspace{-13pt}$\phantom{-}$};
  \draw
        (5:1cm) arc (5:175:1cm) ;
\draw (0,0)--(45:1);
\draw (0,0)--(135:1);
\draw (-1,0) circle [radius=0.1];
\draw (1,0) circle [radius=0.1];
\filldraw (157.5:1) circle [radius=0.1];
\draw (-0.9,0) -- (0.9,0);
\draw (0.9,0) -- (0.1,0);
}\right)_{IR}\Bigl\{\tikz[scale=0.6,baseline=(vert_cent.base)]{
  \node (vert_cent) {\hspace{-13pt}$\phantom{-}$};
  \draw (-0.1,0)--(0.1,0)
        (0.7,0) ++(0:0.6cm) arc (0:360:0.6cm and 0.4cm)
        (1.3,0)--(1.5,0);
}+\Delta\left(\tikz[scale=0.6,baseline=(vert_cent.base)]{
  \node (vert_cent) {\hspace{-13pt}$\phantom{-}$};
  \filldraw (1.3,0) circle [radius=0.1]; \filldraw (0.1,0) circle [radius=0.1];\draw
        (0.7,0) ++(0:0.6cm) arc (0:360:0.6cm and 0.4cm);
}\right)\Bigr\}\Bigr].
\end{align}
We have
\begin{subequations}\label{i553}
\begin{align}
I^{5,53}_1=&\,L(1,1)L(1,2)L(1,2+\tfrac12\epsilon)  L(2+\epsilon,\tfrac12\epsilon)+\Delta_{4,26}=-\tfrac83\tfrac{1}{\epsilon^3}+\tfrac13\tfrac{1}{\epsilon^2}+\ldots\\
I^{5,53}_2=&\,L(1,1)L(1,2) L(2+\tfrac12\epsilon,\tfrac12\epsilon)+\Delta_{3,7}=\tfrac{4}{\epsilon^2}+\tfrac{3}{\epsilon}+\ldots\\
I^{5,53}_3=&\,L(1,1)L(2,\tfrac12\epsilon)+\Delta_{2,2}=-\tfrac{4}{\epsilon}-\tfrac92+\ldots,\\
I^{5,53}_4=&\,L(1,1)+\Delta_{1,1}=2+2\epsilon+\ldots,
\end{align}
\end{subequations}
and then using Eq.~\eqref{irdiffs} we find
\be
\delta_{5,53}=\Kcal\left(\delta^{IR}_1I^{5,53}_1+\delta^{IR}_{2}I^{5,53}_2+\delta^{IR}_{3,2}I^{5,53}_3+\delta^{IR}_{4,2}I^{5,53}_4\right)=0.
\label{del553}
\ee
 
Diagram 75 is very similar; we find
\begin{align}
\Delta_{5,75}=&\,-\Kcal\Rbar^*\left(\tikz[scale=0.6,baseline=(vert_cent.base),scale=1]{
	%\draw[help lines] (-3,-3) grid (3,3);
	\node (vert_cent) {\hspace{-13pt}$\phantom{-}$};
	\draw (0,0) circle[radius=1cm];
\draw [bend left=40] (-1,0) to (1,0);
\draw (-1.25,0) -- (1.25,0);
\draw (0,0) -- (-120:1);
\draw (0,0) -- (-60:1);
\filldraw (-120:0.5) circle [radius=0.1];
}\right)=-\Kcal\Bigl[\ldots +\left(\,
\tikz[scale=0.6,baseline=(vert_cent.base)]{
  \node (vert_cent) {\hspace{-13pt}$\phantom{-}$};
  \draw (0,-.4)--(0,.4);
        \draw (0,.5) circle [radius=0.1];
\draw (0,-.5) circle [radius=0.1];
\filldraw (0,0) circle [radius=0.1];
}\,\right)_{IR}\Bigl\{\tikz[scale=0.6,baseline=(vert_cent.base),scale=1]{
	%\draw[help lines] (-3,-3) grid (3,3);
	\node (vert_cent) {\hspace{-13pt}$\phantom{-}$};
	\draw (0,0) circle[radius=1cm];
\draw [bend left=40] (-1,0) to (1,0);
\draw (-1.25,0) -- (1.25,0);
\draw (0,0) -- (-0,-1);
\filldraw (-135:1) circle [radius=0.1];
}+\Delta\left(\tikz[scale=0.6,baseline=(vert_cent.base),scale=1]{
	%\draw[help lines] (-3,-3) grid (3,3);
	\node (vert_cent) {\hspace{-13pt}$\phantom{-}$};
	\draw (0,0) circle[radius=1cm];
\draw [bend left=40] (-1,0) to (1,0);
\draw (-1,0) -- (1,0);
\draw (0,0) -- (-0,-1);
\filldraw (-135:1) circle [radius=0.1];
}\right)\Bigr\}\nn
&+\left(\,\tikz[scale=0.6,baseline=(vert_cent.base)]{
  \node (vert_cent) {\hspace{-13pt}$\phantom{-}$};
  \draw
        (10:0.5cm) arc (10:170:0.5cm) ;
\draw (0,0.1)--(0,.5);
\draw (-0.5,0) circle [radius=0.1];
\draw (0.5,0) circle [radius=0.1];
\draw (0,0) circle [radius=0.1];
\filldraw (0,0.25) circle [radius=0.1];
}\,\right)_{IR}\Bigl\{\tikz[scale=0.6,baseline=(vert_cent.base),scale=1]{
	%\draw[help lines] (-3,-3) grid (3,3);
	\node (vert_cent) {\hspace{-13pt}$\phantom{-}$};
	\draw (0,0) circle[radius=1cm];
           \draw [bend left=40] (1,0) to (0,1);
\draw (-1.25,0) -- (1.25,0);
\filldraw (45:1) circle [radius=0.1];
}+\Delta\left(\tikz[scale=0.6,baseline=(vert_cent.base),scale=1]{
	%\draw[help lines] (-3,-3) grid (3,3);
	\node (vert_cent) {\hspace{-13pt}$\phantom{-}$};
	\draw (0,0) circle[radius=1cm];
           \draw [bend left=40] (1,0) to (0,1);
\draw (-1,0) -- (1,0);
\filldraw (45:1) circle [radius=0.1];
}\right)\Bigr\}\nn
&+\left(\tikz[scale=0.6,baseline=(vert_cent.base)]{
  \node (vert_cent) {\hspace{-13pt}$\phantom{-}$};
  \draw
        (5:1cm) arc (5:175:1cm) ;
\draw (0,0.1)--(45:1);
\draw (0,0.1)--(135:1);
\draw (-1,0) circle [radius=0.1];
\draw (1,0) circle [radius=0.1];
\draw (0,0) circle [radius=0.1];
\filldraw (135:0.5) circle [radius=0.1];
}\right)_{IR}\Bigl\{\tikz[scale=0.6,baseline=(vert_cent.base),scale=1]{
	%\draw[help lines] (-3,-3) grid (3,3);
	\node (vert_cent) {\hspace{-13pt}$\phantom{-}$};
	\draw (0,0) circle[radius=1cm];
\draw (-1.25,0) -- (1.25,0);
\filldraw (0,-1) circle [radius=0.1];
}+\Delta\left(\tikz[scale=0.6,baseline=(vert_cent.base),scale=1]{
	%\draw[help lines] (-3,-3) grid (3,3);
	\node (vert_cent) {\hspace{-13pt}$\phantom{-}$};
	\draw (0,0) circle[radius=1cm];
\draw (-1,0) -- (1,0);
\filldraw (0,-1) circle [radius=0.1];
}\right)\Bigr\}\nn
&+\left(\tikz[scale=0.6,baseline=(vert_cent.base)]{
  \node (vert_cent) {\hspace{-13pt}$\phantom{-}$};
  \draw
        (5:1cm) arc (5:175:1cm) ;
\draw (0,0)--(45:1);
\draw (0,0)--(135:1);
\draw (-1,0) circle [radius=0.1];
\draw (1,0) circle [radius=0.1];
\filldraw (135:0.5) circle [radius=0.1];
\draw (-0.9,0) -- (0.9,0);
\draw (0.9,0) -- (0.1,0);
}\right)_{IR}\Bigl\{\tikz[scale=0.6,baseline=(vert_cent.base)]{
  \node (vert_cent) {\hspace{-13pt}$\phantom{-}$};
  \draw (-0.1,0)--(0.1,0)
        (0.7,0) ++(0:0.6cm) arc (0:360:0.6cm and 0.4cm)
        (1.3,0)--(1.5,0);
}+\Delta\left(\tikz[scale=0.6,baseline=(vert_cent.base)]{
  \node (vert_cent) {\hspace{-13pt}$\phantom{-}$};
  \filldraw (1.3,0) circle [radius=0.1]; \filldraw (0.1,0) circle [radius=0.1];\draw
        (0.7,0) ++(0:0.6cm) arc (0:360:0.6cm and 0.4cm);
}\right)\Bigr\}\Bigr].
\end{align}
We have
\begin{subequations}
\begin{align}
I^{5,75}_1=&\,L(1,1)L(1,2,1,1,1)L(2+\epsilon,\tfrac12\epsilon)+\Delta_{4,25}=I^{5,53}_1+O(\tfrac{1}{\epsilon}),\\
I^{5,75}_2=&\,I^{5,53}_2,\\
I^{5,75}_3=&\,I^{5,53}_3,\\
I^{5,75}_4=&\,I^{5,53}_4,
\end{align}
\end{subequations}
so that using Eq.~\eqref{irdiffs}, we find
\be
\delta_{5,75}=\Kcal\left(\delta^{IR}_1I^{5,75}_1+\delta^{IR}_{2}I^{5,75}_2+\delta^{IR}_{3,2}I^{5,75}_3+\delta^{IR}_{4,2}I^{5,75}_4\right)=0.
\label{del575}
\ee

In the case of Diagram 111 we have
\begin{align}
\Delta_{5,111}=&\,-\Kcal\Rbar^*\left(\tikz[scale=0.6,baseline=(vert_cent.base),scale=1]{
	%\draw[help lines] (-3,-3) grid (3,3);
	\node (vert_cent) {\hspace{-13pt}$\phantom{-}$};
	\draw (0,0) circle [radius=1];
\draw [bend left=40] (-1,0) to (1,0);
\draw (-1.25,0) -- (1.25,0);
\draw (0,0) -- (-120:1);
\draw (0,0) -- (-60:1);
\filldraw (-90:1) circle [radius=0.1];
}\right)=-\Kcal\Bigl[\ldots +\left(\,
\tikz[scale=0.6,baseline=(vert_cent.base)]{
  \node (vert_cent) {\hspace{-13pt}$\phantom{-}$};
  \draw (0,-.4)--(0,.4);
        \draw (0,.5) circle [radius=0.1];
\draw (0,-.5) circle [radius=0.1];
\filldraw (0,0) circle [radius=0.1];
}\,\right)_{IR}\Bigl\{\tikz[scale=0.6,baseline=(vert_cent.base),scale=1]{
	%\draw[help lines] (-3,-3) grid (3,3);
	\node (vert_cent) {\hspace{-13pt}$\phantom{-}$};
	\draw (0:1) arc (0:180:1);
\draw (0,0) arc (0:-180:0.5);
\draw (0,0) arc (-180:0:0.5);
\draw [bend left=40] (-1,0) to (1,0);
\draw (-1.25,0) -- (1.25,0);
\filldraw (-0.5,0) circle [radius=0.1];
\filldraw (0.5,0) circle [radius=0.1];
}+\Delta\left(\tikz[scale=0.6,baseline=(vert_cent.base),scale=1]{
	%\draw[help lines] (-3,-3) grid (3,3);
	\node (vert_cent) {\hspace{-13pt}$\phantom{-}$};
	\draw (0:1) arc (0:180:1);
\draw (0,0) arc (0:-180:0.5);
\draw (0,0) arc (-180:0:0.5);
\draw [bend left=40] (-1,0) to (1,0);
\draw (-1,0) -- (1,0);
\filldraw (-0.5,0) circle [radius=0.1];
\filldraw (0.5,0) circle [radius=0.1];
}\right)\Bigr\}\nn
&+2\left(\,\tikz[scale=0.6,baseline=(vert_cent.base)]{
  \node (vert_cent) {\hspace{-13pt}$\phantom{-}$};
  \draw
        (10:0.5cm) arc (10:170:0.5cm) ;
\draw (0,0.1)--(0,.5);
\draw (-0.5,0) circle [radius=0.1];
\draw (0.5,0) circle [radius=0.1];
\draw (0,0) circle [radius=0.1];
\filldraw (0,0.25) circle [radius=0.1];
}\,\right)_{IR}\Bigl\{\tikz[scale=0.6,baseline=(vert_cent.base),scale=1]{
	%\draw[help lines] (-3,-3) grid (3,3);
	\node (vert_cent) {\hspace{-13pt}$\phantom{-}$};
	\draw (0,0) circle[radius=1cm];
           \draw [bend left=40] (1,0) to (0,1);
\draw (-1.25,0) -- (1.25,0);
\filldraw (45:1) circle [radius=0.1];
}+\Delta\left(\tikz[scale=0.6,baseline=(vert_cent.base),scale=1]{
	%\draw[help lines] (-3,-3) grid (3,3);
	\node (vert_cent) {\hspace{-13pt}$\phantom{-}$};
	\draw (0,0) circle[radius=1cm];
           \draw [bend left=40] (1,0) to (0,1);
\draw (-1,0) -- (1,0);
\filldraw (45:1) circle [radius=0.1];
}\right)\Bigr\}\nn
&+\left(\tikz[scale=0.6,baseline=(vert_cent.base)]{
  \node (vert_cent) {\hspace{-13pt}$\phantom{-}$};
  \draw
        (5:1cm) arc (5:175:1cm) ;
\draw (0,0.1)--(45:1);
\draw (0,0.1)--(135:1);
\draw (-1,0) circle [radius=0.1];
\draw (1,0) circle [radius=0.1];
\draw (0,0) circle [radius=0.1];
\filldraw (90:1) circle [radius=0.1];
}\right)_{IR}\Bigl\{\tikz[scale=0.6,baseline=(vert_cent.base),scale=1]{
	%\draw[help lines] (-3,-3) grid (3,3);
	\node (vert_cent) {\hspace{-13pt}$\phantom{-}$};
	\draw (0,0) circle[radius=1cm];
\draw (-1.25,0) -- (1.25,0);
\filldraw (0,-1) circle [radius=0.1];
}+\Delta\left(\tikz[scale=0.6,baseline=(vert_cent.base),scale=1]{
	%\draw[help lines] (-3,-3) grid (3,3);
	\node (vert_cent) {\hspace{-13pt}$\phantom{-}$};
	\draw (0,0) circle[radius=1cm];
\draw (-1,0) -- (1,0);
\filldraw (0,-1) circle [radius=0.1];
}\right)\Bigr\}\nn
&+\left(\tikz[scale=0.6,baseline=(vert_cent.base)]{
  \node (vert_cent) {\hspace{-13pt}$\phantom{-}$};
  \draw
        (5:1cm) arc (5:175:1cm) ;
\draw (0,0)--(45:1);
\draw (0,0)--(135:1);
\draw (-1,0) circle [radius=0.1];
\draw (1,0) circle [radius=0.1];
\filldraw (90:1) circle [radius=0.1];
\draw (-0.9,0) -- (0.9,0);
\draw (0.9,0) -- (0.1,0);
}\right)_{IR}\Bigl\{\tikz[scale=0.6,baseline=(vert_cent.base)]{
  \node (vert_cent) {\hspace{-13pt}$\phantom{-}$};
  \draw (-0.1,0)--(0.1,0)
        (0.7,0) ++(0:0.6cm) arc (0:360:0.6cm and 0.4cm)
        (1.3,0)--(1.5,0);
}+\Delta\left(\tikz[scale=0.6,baseline=(vert_cent.base)]{
  \node (vert_cent) {\hspace{-13pt}$\phantom{-}$};
  \filldraw (1.3,0) circle [radius=0.1]; \filldraw (0.1,0) circle [radius=0.1];\draw
        (0.7,0) ++(0:0.6cm) arc (0:360:0.6cm and 0.4cm);
}\right)\Bigr\}\Bigr].
\end{align}
We have
\begin{subequations}
\begin{align}
I^{5,111}_1=&\,L(1,1)L(1,2)^2L(2+\epsilon,\tfrac12\epsilon)+\Delta_{4,8}=-\tfrac{16}{3}\tfrac{1}{\epsilon^3}-\tfrac{2}{\epsilon^2}+\ldots,\\
I^{5,111}_2=&\,I^{5,53}_2,\\
I^{5,111}_3=&\,I^{5,53}_3,\\
I^{5,111}_4=&\,I^{5,53}_4,
\end{align}
\end{subequations}
and then using Eq.~\eqref{irdiffs} we find
\be
\delta_{5,111}=\Kcal\left(\delta^{IR}_1I^{5,111}_1+2\delta^{IR}_{2}I^{5,111}_2+\delta^{IR}_{3,1}I^{5,111}_3+\delta^{IR}_{4,1}I^{5,111}_4\right)=0.
\label{del5111}
\ee

The next two diagrams also have one, two, three and four-loop IR divergences of a similar kind; firstly we have
\begin{align}
\Delta_{5,73}=&\,-\Kcal\Rbar^*\left(\tikz[scale=0.6,baseline=(vert_cent.base),scale=1]{
	%\draw[help lines] (-3,-3) grid (3,3);
	\node (vert_cent) {\hspace{-13pt}$\phantom{-}$};
	\draw (0,0) circle[radius=1cm];
\draw [bend left=40] (-1,0) to (1,0);
\draw [bend left=40] (0,0) to (0,-1);
\draw [bend right=40] (0,0) to (0,-1);
\draw (-1.25,0) -- (1.25,0);
\filldraw (-150:1) circle [radius=0.1];
\filldraw (0.5,0) circle [radius=0.1];
}\right)
=-\Kcal\Bigl[\ldots +2\left(\tikz[scale=0.6,baseline=(vert_cent.base)]{
  \node (vert_cent) {\hspace{-13pt}$\phantom{-}$};
  \draw (0,-.4)--(0,.4);
        \draw (0,.5) circle [radius=0.1];
\draw (0,-.5) circle [radius=0.1];
\filldraw (0,0) circle [radius=0.1];
}\,\right)_{IR}
\Bigl\{\tikz[scale=0.6,baseline=(vert_cent.base),scale=1]{
	%\draw[help lines] (-3,-3) grid (3,3);
	\node (vert_cent){\hspace{-13pt}$\phantom{-}$};
	\draw (0,0) circle [radius=1cm];
\draw [bend left=40] (-1,0) to (1,0);
\draw [bend left=40] (0,0) to (0,-1);
\draw [bend right=40] (0,0) to (0,-1);
\draw (-1.25,0) --(-1,0) (0,0) -- (1.25,0);
\filldraw (-45:1) circle [radius=0.1];
}
+\Delta\left(\tikz[scale=0.6,baseline=(vert_cent.base)]{
  \node (vert_cent) {\hspace{-13pt}$\phantom{-}$};
  \filldraw (1.3,0) circle [radius=0.1]; \filldraw (0.1,0) circle [radius=0.1];\draw
        (0.7,0) ++(0:0.6cm) arc (0:360:0.6cm and 0.4cm);
}\right)\tikz[scale=0.6,baseline=(vert_cent.base),scale=1]{
	%\draw[help lines] (-3,-3) grid (3,3);
	\node (vert_cent) {\hspace{-13pt}$\phantom{-}$};
	\draw (0,0) circle[radius=1cm];
\draw [bend right=40] (1,0) to (0,-1);
\draw [bend left=40] (-1,0) to (1,0);
\draw (-1.25,0) --(-1,0) (1,0) -- (1.25,0);
\filldraw (-45:1) circle [radius=0.1];
}\nn
&+\Delta\left(\tikz[scale=0.6,baseline=(vert_cent.base),scale=1]{
	%\draw[help lines] (-3,-3) grid (3,3);
	\node (vert_cent){\hspace{-13pt}$\phantom{-}$};
	\draw (0,0) circle [radius=1cm];
\draw [bend left=40] (-1,0) to (1,0);
\draw [bend left=40] (0,0) to (0,-1);
\draw [bend right=40] (0,0) to (0,-1);
\draw  (0,0) -- (1,0);
\filldraw (-45:1) circle [radius=0.1];
}\right)\Bigl\}\nn
&+\left(\tikz[scale=0.6,baseline=(vert_cent.base)]{
  \node (vert_cent) {\hspace{-13pt}$\phantom{-}$};
  \draw (0,-.4)--(0,.4);
        \draw (0,.5) circle [radius=0.1];
\draw (0,-.5) circle [radius=0.1];
\filldraw (0,0) circle [radius=0.1];
}\,\right)_{IR}\left(\tikz[scale=0.6,baseline=(vert_cent.base)]{
  \node (vert_cent) {\hspace{-13pt}$\phantom{-}$};
  \draw (0,-.4)--(0,.4);
        \draw (0,.5) circle [radius=0.1];
\draw (0,-.5) circle [radius=0.1];
\filldraw (0,0) circle [radius=0.1];
}\,\right)_{IR}\Bigl\{\tikz[scale=0.6,baseline=(vert_cent.base),scale=1]{
	%\draw[help lines] (-3,-3) grid (3,3);
	\node (vert_cent){\hspace{-13pt}$\phantom{-}$};
	\draw (0,0) circle[radius=1cm];
\draw [bend left=40] (-1,0) to (1,0);
\draw [bend left=40] (-120:1) to (-60:1);
\draw (-1.25,0) --(-1,0) (1,0) -- (1.25,0);
}+\Delta\left(\tikz[scale=0.6,baseline=(vert_cent.base)]{
  \node (vert_cent) {\hspace{-13pt}$\phantom{-}$};
  \filldraw (1.3,0) circle [radius=0.1]; \filldraw (0.1,0) circle [radius=0.1];\draw
        (0.7,0) ++(0:0.6cm) arc (0:360:0.6cm and 0.4cm);
}\right)\tikz[scale=0.6,baseline=(vert_cent.base),scale=1]{
	%\draw[help lines] (-3,-3) grid (3,3);
	\node (vert_cent) {\hspace{-13pt}$\phantom{-}$};
	\draw (0,0) circle[radius=1cm];
\draw (-1.25,0) -- (1.25,0);
\filldraw (0,-1) circle [radius=0.1];
}+\Delta\left(\tikz[scale=0.6,baseline=(vert_cent.base),scale=1]{
	%\draw[help lines] (-3,-3) grid (3,3);
	\node (vert_cent){\hspace{-13pt}$\phantom{-}$};
	\draw (0,0) circle[radius=1cm];
\draw [bend left=40] (-1,0) to (1,0);
\draw [bend left=40] (-120:1) to (-60:1);
}\right)\Bigr\}\nn
&+\left(\tikz[scale=0.6,baseline=(vert_cent.base),scale=1]{
\filldraw  (0.5,0.5) circle [radius=0.1];
\filldraw  (-0.5,-0.5) circle [radius=0.1];
\draw (-1,0) circle [radius=0.1];
\draw [bend right=40] (0,0) to (-0.5,0.5);
\draw [bend left=40] (-1,0.07) to (-0.5,0.5);
\draw [bend left=40] (0,0) to (-0.5,-0.5);
\draw [bend right=40] (-1,-0.07) to (-0.5,-0.5);
\draw (1,0) circle [radius=0.1];
\draw [bend left=40] (0,0) to (0.5,0.5);
\draw [bend right=40] (1,0.07) to (0.5,0.5);
\draw [bend right=40] (0,0) to (0.5,-0.5);
\draw [bend left=40] (1,-0.07) to (0.5,-0.5);
}\right)_{IR}\Bigl\{\Delta\left(\tikz[scale=0.6,baseline=(vert_cent.base)]{
  \node (vert_cent) {\hspace{-13pt}$\phantom{-}$};
  \filldraw (1.3,0) circle [radius=0.1]; \filldraw (0.1,0) circle [radius=0.1];\draw
        (0.7,0) ++(0:0.6cm) arc (0:360:0.6cm and 0.4cm);
}\right)\tikz[scale=0.6,baseline=(vert_cent.base)]{
  \node (vert_cent) {\hspace{-13pt}$\phantom{-}$};
  \draw (-0.1,0)--(0.1,0)
        (0.7,0) ++(0:0.6cm) arc (0:360:0.6cm and 0.4cm)
        (1.3,0)--(1.5,0);
}+\Delta\left(\tikz[scale=0.6,baseline=(vert_cent.base)]{
  \node (vert_cent) {\hspace{-13pt}$\phantom{-}$};
  \filldraw (1.3,0) circle [radius=0.1]; \filldraw (0.1,0) circle [radius=0.1];\draw
        (0.7,0) ++(0:0.6cm) arc (0:360:0.6cm and 0.4cm);
}\right)\Delta\left(\tikz[scale=0.6,baseline=(vert_cent.base)]{
  \node (vert_cent) {\hspace{-13pt}$\phantom{-}$};
  \filldraw (1.3,0) circle [radius=0.1]; \filldraw (0.1,0) circle [radius=0.1];\draw
        (0.7,0) ++(0:0.6cm) arc (0:360:0.6cm and 0.4cm);
}\right)\Bigl\}\nn
&+\left(\tikz[scale=0.6,baseline=(vert_cent.base)]{
  \node (vert_cent) {\hspace{-13pt}$\phantom{-}$};
  \draw
        (5:1cm) arc (5:175:1cm) ;
\draw (-1,0) circle [radius=0.1];
\draw (1,0) circle [radius=0.1];
\filldraw (135:1) circle [radius=0.1];
\filldraw (0.5,0) circle [radius=0.1];
\draw [bend left=40] (0,0) to (0,1);
\draw [bend right=40] (0,0) to (0,1);
\draw (-0.9,0) -- (0.9,0);
\draw (0.9,0) -- (0.1,0);
}\right)_{IR}\left\{\tikz[scale=0.6,baseline=(vert_cent.base)]{
  \node (vert_cent) {\hspace{-13pt}$\phantom{-}$};
  \draw (-0.1,0)--(0.1,0)
        (0.7,0) ++(0:0.6cm) arc (0:360:0.6cm and 0.4cm)
        (1.3,0)--(1.5,0);
}+\Delta\left(\tikz[scale=0.6,baseline=(vert_cent.base)]{
  \node (vert_cent) {\hspace{-13pt}$\phantom{-}$};
  \filldraw (1.3,0) circle [radius=0.1]; \filldraw (0.1,0) circle [radius=0.1];\draw
        (0.7,0) ++(0:0.6cm) arc (0:360:0.6cm and 0.4cm);
}\right)\right\}
\Bigr].
\end{align}
Using Eq.~\eqref{irdiffs} we find
\begin{subequations}
\begin{align}
I^{5,73}_1=&\,L(1,1)^2L(1+\tfrac12\epsilon,2)L(2+\epsilon,\tfrac12\epsilon)-\tfrac{2}{\epsilon}L(1,1)L(1,2)L(2+\tfrac12\epsilon,\tfrac12\epsilon)\nn
&+\Delta_{4,24}\nn
=&\,-\tfrac{8}{3}\tfrac{1}{\epsilon^3}+\tfrac43\tfrac{1}{\epsilon^2}+\ldots,\\
I^{5,73}_2=&\,L(1,1)^2L(2+\tfrac12\epsilon,\tfrac12\epsilon)-\tfrac{2}{\epsilon}L(1,1)L(2,\tfrac12\epsilon)+\Delta_{3,6}=\tfrac{4}{\epsilon^2}+\tfrac{2}{\epsilon}+\ldots,\\
I^{5,73}_3=&\,-\tfrac{2}{\epsilon}L(1,1)+\left(\tfrac{2}{\epsilon}\right)^2=-\tfrac{2}{\epsilon}I_1^{2,2},\\
I^{5,73}_4=&\,L(1,1)-\tfrac{2}{\epsilon}=I_1^{2,2},
\end{align}
\end{subequations}
so that using Eq.~\eqref{irdiffs}, we find
\be
\delta_{5,73}=\Kcal\left(2\delta^{IR}_1I^{5,73}_1+(\ncal_1^{\prime2}-\ncal_1^2)I^{5,73}_2+\delta^{IR}_{3,3}I^{5,73}_3+\delta^{IR}_{4,4}I^{5,73}_4\right)=0.
\label{del573}
\ee

Diagram 93 is similar, and we have
\begin{align}
\Delta_{5,93}=&\,-\Kcal\Rbar^*\left(\tikz[scale=0.6,baseline=(vert_cent.base),scale=1]{
	%\draw[help lines] (-3,-3) grid (3,3);
	\node (vert_cent) {\hspace{-13pt}$\phantom{-}$};
	\draw (0,0) circle[radius=1cm];
\draw [bend left=40] (-1,0) to (1,0);
\draw [bend left=40] (0,0) to (0,-1);
\draw [bend right=40] (0,0) to (0,-1);
\draw (-1.25,0) -- (1.25,0);
\filldraw (-45:1) circle [radius=0.1];
\filldraw (0.5,0) circle [radius=0.1];
}\right)
=-\Kcal\Bigl[\ldots +2\left(\tikz[scale=0.6,baseline=(vert_cent.base)]{
  \node (vert_cent) {\hspace{-13pt}$\phantom{-}$};
  \draw (0,-.4)--(0,.4);
        \draw (0,.5) circle [radius=0.1];
\draw (0,-.5) circle [radius=0.1];
\filldraw (0,0) circle [radius=0.1];
}\,\right)_{IR}
\Bigl\{\tikz[scale=0.6,baseline=(vert_cent.base),scale=1]{
	%\draw[help lines] (-3,-3) grid (3,3);
	\node (vert_cent){\hspace{-13pt}$\phantom{-}$};
	\draw (0,0) circle [radius=1cm];
\draw [bend left=40] (-1,0) to (1,0);
\draw [bend left=40] (0,0) to (0,-1);
\draw [bend right=40] (0,0) to (0,-1);
\draw (-1.25,0) --(-1,0) (0,0) -- (1.25,0);
\filldraw (-135:1) circle [radius=0.1];
}
+\Delta\left(\tikz[scale=0.6,baseline=(vert_cent.base)]{
  \node (vert_cent) {\hspace{-13pt}$\phantom{-}$};
  \filldraw (1.3,0) circle [radius=0.1]; \filldraw (0.1,0) circle [radius=0.1];\draw
        (0.7,0) ++(0:0.6cm) arc (0:360:0.6cm and 0.4cm);
}\right)\tikz[scale=0.6,baseline=(vert_cent.base),scale=1]{
	%\draw[help lines] (-3,-3) grid (3,3);
	\node (vert_cent) {\hspace{-13pt}$\phantom{-}$};
	\draw (0,0) circle[radius=1cm];
\draw [bend right=40] (1,0) to (0,-1);
\draw [bend left=40] (-1,0) to (1,0);
\draw (-1.25,0) --(-1,0) (1,0) -- (1.25,0);
\filldraw (-135:1) circle [radius=0.1];
}\nn
&+\Delta\left(\tikz[scale=0.6,baseline=(vert_cent.base),scale=1]{
	%\draw[help lines] (-3,-3) grid (3,3);
	\node (vert_cent) {\hspace{-13pt}$\phantom{-}$};
	\draw (0,0) circle[radius=1cm];
           \draw [bend left=40] (1,0) to (0,1);
\draw (-1.25,0) -- (-1,0);
\draw (1.25,0) -- (1,0);
}\right)\tikz[scale=0.6,baseline=(vert_cent.base),scale=1]{
	%\draw[help lines] (-3,-3) grid (3,3);
	\node (vert_cent) {\hspace{-13pt}$\phantom{-}$};
	\draw (0,0) circle[radius=1cm];
\draw (-1.25,0) -- (1.25,0);
\filldraw (0,-1) circle [radius=0.1];
}+\Delta\left(\tikz[scale=0.6,baseline=(vert_cent.base),scale=1]{
	%\draw[help lines] (-3,-3) grid (3,3);
	\node (vert_cent){\hspace{-13pt}$\phantom{-}$};
	\draw (0,0) circle [radius=1cm];
\draw [bend left=40] (-1,0) to (1,0);
\draw [bend left=40] (0,0) to (0,-1);
\draw [bend right=40] (0,0) to (0,-1);
\draw (0,0) -- (1,0);
\filldraw (-135:1) circle [radius=0.1];
}\right)\Bigl\}\nn
&+\left(\tikz[scale=0.6,baseline=(vert_cent.base)]{
  \node (vert_cent) {\hspace{-13pt}$\phantom{-}$};
  \draw (0,-.4)--(0,.4);
        \draw (0,.5) circle [radius=0.1];
\draw (0,-.5) circle [radius=0.1];
\filldraw (0,0) circle [radius=0.1];
}\,\right)_{IR}\left(\tikz[scale=0.6,baseline=(vert_cent.base)]{
  \node (vert_cent) {\hspace{-13pt}$\phantom{-}$};
  \draw (0,-.4)--(0,.4);
        \draw (0,.5) circle [radius=0.1];
\draw (0,-.5) circle [radius=0.1];
\filldraw (0,0) circle [radius=0.1];
}\,\right)_{IR}\Bigl\{\Delta\left(\tikz[scale=0.6,baseline=(vert_cent.base),scale=1]{
	%\draw[help lines] (-3,-3) grid (3,3);
	\node (vert_cent) {\hspace{-13pt}$\phantom{-}$};
	\draw (0,0) circle[radius=1cm];
           \draw [bend left=40] (1,0) to (0,1);
\draw (-1.25,0) -- (-1,0);
\draw (1.25,0) -- (1,0);
}\right)\tikz[scale=0.6,baseline=(vert_cent.base)]{
  \node (vert_cent) {\hspace{-13pt}$\phantom{-}$};
  \draw (-0.1,0)--(0.1,0)
        (0.7,0) ++(0:0.6cm) arc (0:360:0.6cm and 0.4cm)
        (1.3,0)--(1.5,0);
}+\Delta\left(\tikz[scale=0.6,baseline=(vert_cent.base),scale=1]{
	%\draw[help lines] (-3,-3) grid (3,3);
	\node (vert_cent) {\hspace{-13pt}$\phantom{-}$};
	\draw (0,0) circle[radius=1cm];
           \draw [bend left=40] (1,0) to (0,1);
\draw (-1.25,0) -- (-1,0);
\draw (1.25,0) -- (1,0);
}\right)\Delta\left(\tikz[scale=0.6,baseline=(vert_cent.base)]{
  \node (vert_cent) {\hspace{-13pt}$\phantom{-}$};
  \filldraw (1.3,0) circle [radius=0.1]; \filldraw (0.1,0) circle [radius=0.1];\draw
        (0.7,0) ++(0:0.6cm) arc (0:360:0.6cm and 0.4cm);
}\right)\Bigr\}\nn
&+\left(\tikz[scale=0.6,baseline=(vert_cent.base),scale=1]{
\filldraw  (0.5,0.5) circle [radius=0.1];
\filldraw  (0.5,-0.5) circle [radius=0.1];
\draw (-1,0) circle [radius=0.1];
\draw [bend right=40] (0,0) to (-0.5,0.5);
\draw [bend left=40] (-1,0.07) to (-0.5,0.5);
\draw [bend left=40] (0,0) to (-0.5,-0.5);
\draw [bend right=40] (-1,-0.07) to (-0.5,-0.5);
\draw (1,0) circle [radius=0.1];
\draw [bend left=40] (0,0) to (0.5,0.5);
\draw [bend right=40] (1,0.07) to (0.5,0.5);
\draw [bend right=40] (0,0) to (0.5,-0.5);
\draw [bend left=40] (1,-0.07) to (0.5,-0.5);
}\right)_{IR}\Bigl\{\Delta\left(\tikz[scale=0.6,baseline=(vert_cent.base)]{
  \node (vert_cent) {\hspace{-13pt}$\phantom{-}$};
  \filldraw (1.3,0) circle [radius=0.1]; \filldraw (0.1,0) circle [radius=0.1];\draw
        (0.7,0) ++(0:0.6cm) arc (0:360:0.6cm and 0.4cm);
}\right)\tikz[scale=0.6,baseline=(vert_cent.base)]{
  \node (vert_cent) {\hspace{-13pt}$\phantom{-}$};
  \draw (-0.1,0)--(0.1,0)
        (0.7,0) ++(0:0.6cm) arc (0:360:0.6cm and 0.4cm)
        (1.3,0)--(1.5,0);
}+\Delta\left(\tikz[scale=0.6,baseline=(vert_cent.base)]{
  \node (vert_cent) {\hspace{-13pt}$\phantom{-}$};
  \filldraw (1.3,0) circle [radius=0.1]; \filldraw (0.1,0) circle [radius=0.1];\draw
        (0.7,0) ++(0:0.6cm) arc (0:360:0.6cm and 0.4cm);
}\right)\Delta\left(\tikz[scale=0.6,baseline=(vert_cent.base)]{
  \node (vert_cent) {\hspace{-13pt}$\phantom{-}$};
  \filldraw (1.3,0) circle [radius=0.1]; \filldraw (0.1,0) circle [radius=0.1];\draw
        (0.7,0) ++(0:0.6cm) arc (0:360:0.6cm and 0.4cm);
}\right)\Bigl\}\nn
&+\left(\tikz[scale=0.6,baseline=(vert_cent.base)]{
  \node (vert_cent) {\hspace{-13pt}$\phantom{-}$};
  \draw
        (5:1cm) arc (5:175:1cm) ;
\draw (-1,0) circle [radius=0.1];
\draw (1,0) circle [radius=0.1];
\filldraw (45:1) circle [radius=0.1];
\filldraw (0.5,0) circle [radius=0.1];
\draw [bend left=40] (0,0) to (0,1);
\draw [bend right=40] (0,0) to (0,1);
\draw (-0.9,0) -- (0.9,0);
\draw (0.9,0) -- (0.1,0);
}\right)_{IR}\left\{\tikz[scale=0.6,baseline=(vert_cent.base)]{
  \node (vert_cent) {\hspace{-13pt}$\phantom{-}$};
  \draw (-0.1,0)--(0.1,0)
        (0.7,0) ++(0:0.6cm) arc (0:360:0.6cm and 0.4cm)
        (1.3,0)--(1.5,0);
}+\Delta\left(\tikz[scale=0.6,baseline=(vert_cent.base)]{
  \node (vert_cent) {\hspace{-13pt}$\phantom{-}$};
  \filldraw (1.3,0) circle [radius=0.1]; \filldraw (0.1,0) circle [radius=0.1];\draw
        (0.7,0) ++(0:0.6cm) arc (0:360:0.6cm and 0.4cm);
}\right)\right\}\Bigr].
\end{align}
We have 
\begin{subequations}
\begin{align}
I^{5,93}_1=&\,L(1,1)^2L(1,1+\tfrac12\epsilon)L(2+\epsilon,\tfrac12\epsilon)-\tfrac{2}{\epsilon}L(1,1)^2L(2+\tfrac12\epsilon,\tfrac12\epsilon)\nn
&+\left(\tfrac{2}{\epsilon^2}-\tfrac{1}{\epsilon}\right)L(1,1)L(2,\tfrac12\epsilon)+\Delta_{4,23}=I^{5,73}_1+O\left(\tfrac{1}{\epsilon}\right)\nn
=&-\tfrac{8}{3}\tfrac{1}{\epsilon^3}+\tfrac43\tfrac{1}{\epsilon^2}+\ldots,\\
I^{5,93}_2=&\,\left(\tfrac{2}{\epsilon^2}-\tfrac{1}{\epsilon}\right)L(1,1)-\left(\tfrac{2}{\epsilon^2}-\tfrac{1}{\epsilon}\right)\tfrac{2}{\epsilon}=I^{5,73}_2+O(1)=\tfrac{4}{\epsilon^2}+\tfrac{2}{\epsilon}+\ldots,\\
I^{5,93}_3=&\,I^{5,73}_3,\\
I^{5,93}_4=&\,I^{5,73}_4.
\end{align}
\end{subequations}
\be
\delta_{5,93}=\Kcal\left(2\delta^{IR}_1I^{5,93}_1+(\ncal_1^{\prime2}-\ncal_1^2)I^{5,93}_2+\delta^{IR}_{3,3}I^{5,93}_3+\delta^{IR}_{4,4}I^{5,93}_4\right)=0.
\label{del593}
\ee

To summarise, we have shown in this Appendix that our AIR counterterms give the same UV divergences as the standard versions for every IR-divergent diagram (with one or more double propagators) considered in Ref.~\cite{klein}. We shall complete the picture by considering the small number of IR divergent diagrams in Ref.~\cite{klein} with no double propagator, in Appendix C.

\section{Special cases}
So far we have focussed on IR-divergent structures with at least one double propagator, for which we have a well-defined alternative proposal for the IR counterterms. In this section we turn to the case of IR counterterms for IR divergent structures with no double propagators; and this will lead us to an alternative approach to the more familiar structures with one or more double propagators. At the five-loop level we are concerned with, the IR divergent structures with no double propagators are simply $\Ical_{3,4}$, $\Ical_{4,5}$ and $\Ical_{4,6}$. As we have emphasised, we have obtained our proposal for the AIR counterterms by a process of informed guesswork. Nevertheless our success in so far reproducing known results gives us some confidence that we are correct. However, in the current case we have far less scope for such comparisons. All the same, a sensible approach would seem to be to assume a similar ansatz; namely to break one of the propagators to obtain a severed graph $G^{\rm{sev}}$ as in Section 3, and then to define the IR counterterm as in Eq.~\eqref{Xdef}. An $(L-1)$-loop $G^{\rm{sev}}$ is now $O(1/(p^2)^{1+(L-1)/2\epsilon})$ and combines with the external propagator to give a $O(1/(p^2)^{2+(L-1)/2\epsilon})$ momentum dependence which is responsible for generating $C_L$ in the same way as before. In the case of $\Ical_{3,4}$, all propagators in the CVG are equivalent and we obtain $L(1,1,1,1,1)C_3$ as shown in Table~\ref{IRres4}\footnote{In the diagrams depicted in Table~\ref{IRres4} for $G^{\rm{sev}}$ corresponding to $\Ical_{3,4}$, $\Ical_{4,5}$ and $\Ical_{4,6}$, the external momentum is fed in along the single external propagator and emerges at the point indicated by a black square.}. As we see in Eq.~\eqref{irdiffs:f}, this agrees up to $O(1)$ with the standard IR counterterm. At the next order we seem to have a dilemma, since clearly the propagators in the CVG corresponding to $\Ical_{4,5}$ are not all equivalent. However, whichever propagator is selected to sever, the result for the IR counterterm is $L(1,1)L(1+a_1,\ldots, 1+a_5)C_3$ where either all, or all but one of $a_1\ldots a_5$ is zero. It was shown by Kazakov\cite{kaz1} that 
\be
L(1+a_1,\ldots, 1+a_5)=\tfrac{1}{1-\tfrac12\epsilon}[6\zeta_3+\tfrac94\zeta_4\epsilon+\ldots],
\ee
independent of $a_1\ldots a_5$, which is sufficient to guarantee that the result for the IR counterterm is independent of the choice of severed propagator up to $O(1)$ terms. Of course we have used this result to obtain our alternative expressions for $\ncal'_{3,4}$ and $\ncal'_{4,5}$ which we then compared with Eq.~\eqref{IRold} to obtain $\delta^{IR}_{3,4}$ and $\delta^{IR}_{4,5}$ in Eqs.~\eqref{irdiffs:f} and \eqref{irdiffs:k}.

Up to five loops, the only diagram involving the IR divergences $\Ical_{3,4}$ and $\Ical_{4,5}$ is the five-loop Diagram 34 in Ref.~\cite{klein}. The result for the UV counterterm is 
\begin{align}
\Delta_{5,34}=&\,-\Kcal\Rbar^*\left(
\tikz[scale=0.6,baseline=(vert_cent.base)]{
  \node (vert_cent) {\hspace{-13pt}$\phantom{-}$};
  \draw (0,0) circle [radius=1] ;
\draw [bend right=40] (-1,0) to (1,0);
\draw (0,0.5)--(45:1);
\draw (0,0.5)--(135:1);
\draw (0,0) -- (0,0.5);
\draw (-1,0) -- (1,0);
}\right)\nn
=&\,-\Kcal\Bigl[\ldots \left(\tikz[scale=0.6,baseline=(vert_cent.base)]{
  \node (vert_cent) {\hspace{-13pt}$\phantom{-}$};
  \draw
        (5:1cm) arc (5:175:1cm) ;
\draw (0,0.5)--(45:1);
\draw (0,0.5)--(135:1);
\draw (0,0.1) -- (0,0.5);
\draw (-1,0) circle [radius=0.1];
\draw (1,0) circle [radius=0.1];
\draw (0,0) circle [radius=0.1];
}\right)_{IR}\left\{\tikz[scale=0.6,baseline=(vert_cent.base),scale=1]{
	%\draw[help lines] (-3,-3) grid (3,3);
	\node (vert_cent) {\hspace{-13pt}$\phantom{-}$};
	\draw (0,0) circle[radius=1cm];
\draw (-1.25,0) -- (1.25,0);
\filldraw (0,-1) circle [radius=0.1];
}+\Delta\left(\tikz[scale=0.6,baseline=(vert_cent.base),scale=1]{
	%\draw[help lines] (-3,-3) grid (3,3);
	\node (vert_cent) {\hspace{-13pt}$\phantom{-}$};
	\draw (0,0) circle[radius=1cm];
\draw (-1,0) -- (1,0);
\filldraw (0,-1) circle [radius=0.1];
}\right)\right\}\nn
&+\left(\tikz[scale=0.6,baseline=(vert_cent.base)]{
  \node (vert_cent) {\hspace{-13pt}$\phantom{-}$};
  \draw
        (5:1cm) arc (5:175:1cm) ;
\draw (0,0.5)--(45:1);
\draw (0,0.5)--(135:1);
\draw (0,0) -- (0,0.5);
\draw (-1,0) circle [radius=0.1];
\draw (1,0) circle [radius=0.1];
\draw (-0.9,0) -- (0.9,0);
}\right)_{IR}\left\{\tikz[scale=0.6,baseline=(vert_cent.base)]{
  \node (vert_cent) {\hspace{-13pt}$\phantom{-}$};
  \draw (-0.1,0)--(0.1,0)
        (0.7,0) ++(0:0.6cm) arc (0:360:0.6cm and 0.4cm)
        (1.3,0)--(1.5,0);
}+\Delta\left(\tikz[scale=0.6,baseline=(vert_cent.base)]{
  \node (vert_cent) {\hspace{-13pt}$\phantom{-}$};
  \filldraw (1.3,0) circle [radius=0.1]; \filldraw (0.1,0) circle [radius=0.1];\draw
        (0.7,0) ++(0:0.6cm) arc (0:360:0.6cm and 0.4cm);
}\right)\right\}
\Bigr],
\end{align}
where, as in Section B, we focus on the contribution from the IR counterterms (n.b. there is a minor misprint in the corresponding equation in Ref.~\cite{klein}) 

We have (using the same notation as in Section B)
\be
I^{5,34}_3=I^{5,53}_3,\quad I^{5,34}_4=I^{5,53}_4,
\ee
and so $I^{5,34}_3$, $I^{5,34}_4$ may be read off from Eq.~\eqref{i553}, and then using also Eqs.~\eqref{irdiffs:f}, \eqref{irdiffs:k} we find that the expression for the difference between using the standard and the alternative counterterms is given by
\be
\delta_{5,34}=\delta^{IR}_{3,4}I^{5,34}_3+\delta^{IR}_{4,5}I^{5,34}_4=O(1).
\ee
Hence the use of the AIR counterterms once again has no effect on the UV counterterm. If we follow the procedure of Section 5, we find that this behaviour is guaranteed by the identity
\be
\delta^{IR}_{3,4}+\Delta_{1,1}\delta^{IR}_{4,5}=O(1)
\ee
which is easily verified.

The remaining IR counterterm with no double propagator relevant up to five loops is $\Ical_{4,6}$, which only appears in the UV counterterm computation for one diagram, namely the five-loop Diagram 25. The UV counterterm for this diagram is given by
\be
\Delta_{5,25}=-\Kcal\Rbar^*\left(\tikz[scale=0.6,baseline=(vert_cent.base)]{
  \node (vert_cent) {\hspace{-13pt}$\phantom{-}$};
\draw (-0.5,0.5) -- (0.5,0.5) (-0.5,0.5) -- (-0.5,-0.5)  (-0.5,-0.5) -- (0.5,-0.5) (0.5,-0.5) -- (0.5,0.5)
(-1,0) -- (-0.5,0.5) (-1,0) -- (-0.5,-0.5) (1,0) -- (0.5,0.5) (1,0) -- (0.5,-0.5);
\draw (0:1cm) arc (0:360:1cm) ;
}\right)=-\Kcal\left[\ldots+\left(\tikz[scale=0.6,baseline=(vert_cent.base)]{
  \node (vert_cent) {\hspace{-13pt}$\phantom{-}$};
\draw (-1,0) circle [radius=0.1];
\draw (1,0) circle [radius=0.1];
\draw (-0.5,0.5) -- (0.5,0.5) (-0.5,0.5) -- (-0.5,-0.5)  (-0.5,-0.5) -- (0.5,-0.5) (0.5,-0.5) -- (0.5,0.5)
(-0.92,0.08) -- (-0.5,0.5) (-0.92,-0.08) -- (-0.5,-0.5) (0.92,0.08) -- (0.5,0.5) (0.92,-0.08) -- (0.5,-0.5);
}\right)_{IR}\left\{\tikz[scale=0.6,baseline=(vert_cent.base)]{
  \node (vert_cent) {\hspace{-13pt}$\phantom{-}$};
  \draw (-0.1,0)--(0.1,0)
        (0.7,0) ++(0:0.6cm) arc (0:360:0.6cm and 0.4cm)
        (1.3,0)--(1.5,0);
}+\Delta\left(\tikz[scale=0.6,baseline=(vert_cent.base)]{
  \node (vert_cent) {\hspace{-13pt}$\phantom{-}$};
  \filldraw (1.3,0) circle [radius=0.1]; \filldraw (0.1,0) circle [radius=0.1];\draw
        (0.7,0) ++(0:0.6cm) arc (0:360:0.6cm and 0.4cm);
}\right)\right\}\right],
\ee
once again picking out the IR counterterm contribution. Clearly $I^{5,25}_4=I^{5,53}_4=O(1)$, and we see immediately from Eq.~\eqref{irdiffs:l} that
\be
\delta_{5,25}=\delta^{IR}_{4,6}I^{5,25}_4=O(1),
\ee
so that the use of the AIR counterterms yet again has no effect on the UV counterterm.

However, we now find ourselves facing a potential dilemma. We have presented evidence that we can derive viable AIR counterterms (i.e. which produce the correct UV counterterms) in cases where the IR-divergent structures have no double propagators by severing an ordinary $1/p^2$ propagator in the CVG and applying the procedure of Section 3. But the question which then arises is whether we can also obtain viable AIR counterterms by severing single propagators even in the usual case where there are also double propagators.  A moment's reflection shows that this cannot be the case without some qualification, since if one considers the CVG for $\Ical_2$ in Table~\ref{IR12} and severs a single propagator rather than the double one, one obtains $L(1,2)$ rather than $L(1,1)$; and these have simple poles of opposite signs. One salient difference between this situation and the earlier examples where severing the single propagator gave the correct result, is that in the current case, the severed diagram has an IR subdivergence. One is therefore motivated to subtract the IR subdivergences.  If we do this, we find that severing one of the single propagators in the CVG for $\Ical_2$ gives a result which agrees with $\ncal'_2$ up to $O(\epsilon)$. Severing one of the single propagators in the CVG for $\Ical_{3,1}$ gives a result which agrees with $\ncal'_{3,1}$ up to $O(1)$, and similarly for $\Ical_{3,2}$. However, even this is not necessarily sufficient, since the IR counterterms potentially multiply divergent quantities and therefore even finite differences in the IR counterterms might have an impact on the final results for the UV counterterms. We then recall that in Section 5 it appeared that the equivalence of our AIR counterterms to the standard ones is guaranteed by a set of simple conditions, which essentially correspond to requiring that at each order the AIR counterterms are given by a similar equation to the standard ones up to finite terms. It seems natural to expect that this requirement will also apply to the IR countertems obtained by severing single propagators and will guarantee that these too are equivalent to the standard ones.  The $\delta^{IR}$ in equations such as Eq.~\eqref{delrel1} will now become the differences between the standard IR counterterms and the new ones obtained by severing a single propagator. Our next task is therefore to check that the conditions Eqs.~\eqref{delrel1a}, \eqref{delrel1}, \eqref{delrel}, \eqref{delrel2}, \eqref{delrel3} are satisfied with these new $\delta^{IR}$.

We present the alternative results for the IR counterterms in the form of a Table where we show the CVG, the severed diagram and the corresponding counterterm. As we have commented already, we see here that we have added IR counterterm contributions to the results for the severed diagrams. The new IR counterterms are denoted by $\Jcal_2$ etc. Useful results for $L(a_1,a_2,a_3,a_4,a_5)$ are presented in Eq.~\eqref{L4res}. In the current situation there are potentially several distinct ways of severing a single propagator in a given CVG and these are labelled $\alpha$, $\beta$, etc. The differences between each new possible expression for the IR counterterm and the standard one are denoted by $\dtil_2$ etc. Those of interest for our current purposes are listed in Eq.~\eqref{listdtil}.
\begin{table}
\begin{tabular}{|c|c|c|c|}
\hline
Label&CVG&$G^{\rm{sev}}$&IR counterterm\nn
\hline
$\Jcal_2$&\tikz[scale=0.6,baseline=(vert_cent.base),scale=1]{
	%\draw[help lines] (-3,-3) grid (3,3);
	\node (vert_cent) {\hspace{-13pt}$\phantom{-}$};
\node at (0,-1.1) () {};
	\draw (0,0) circle[radius=1cm];
\draw (0,-1) -- (0,1);
\filldraw (0,0) circle [radius=0.1];
}&\tikz[scale=0.6,baseline=(vert_cent.base),scale=1]{
\node at (0,1.1) () {};
\draw (180:1cm) arc (180:0:1cm) ;
\filldraw (90:1) circle [radius=0.1];
\draw (-1.25,0) -- (1,0);
\vsq at (1,0) {};
}&$L(1,2)C_2+\Jcal_1C_1$\nn
\hline
$\Jcal_{3,1}$&
\tikz[scale=0.6,baseline=(vert_cent.base),scale=1]{
	%\draw[help lines] (-3,-3) grid (3,3);
	\node (vert_cent) {\hspace{-13pt}$\phantom{-}$};
\node at (0,-1.1) () {};
	\draw (0,0) circle[radius=1cm];
	\draw [bend right=40] (-1,0) to (0,1);
           \draw [bend left=40] (1,0) to (0,1);
\filldraw (0,-1) circle [radius=0.1];
}&\tikz[scale=0.6,baseline=(vert_cent.base),scale=1]{
\node at (0,1.1) () {};
\draw (180:1cm) arc (180:0:1cm) ;
\draw [bend left=40] (0,1) to (-1,0);
\filldraw (45:1) circle [radius=0.1];
\draw (-1.25,0) -- (1,0);
\vsq at (1,0) {};
}&$L(1,1)L(1,2+\tfrac12\epsilon)C_3+\Jcal_2C_1$\nn 
\hline
$\Jcal_{3,2\alpha}$&
\tikz[scale=0.6,baseline=(vert_cent.base),scale=1]{
	%\draw[help lines] (-3,-3) grid (3,3);
	\node (vert_cent) {\hspace{-13pt}$\phantom{-}$};
	\draw (0,0) circle[radius=1cm];
	\draw [bend right=40] (-1,0) to (0,1);
           \draw [bend left=40] (1,0) to (0,1);
\filldraw (135:1) circle [radius=0.1];
}&\tikz[scale=0.6,baseline=(vert_cent.base),scale=1]{
\node at (0,1.1) () {};
\draw (180:1cm) arc (180:0:1cm) ;
\draw [bend left=40] (0,1) to (-1,0);
\filldraw (135:1) circle [radius=0.1];
\draw (-1.25,0) -- (1,0);
\vsq at (1,0) {};
}&$L(1,2)L(1,2+\tfrac12\epsilon)C_3+\Jcal_1L(1,2)C_2+\Jcal_2C_1$\nn 
$\Jcal_{3,2\beta}$&
&\tikz[scale=0.6,baseline=(vert_cent.base),scale=1]{
\draw (180:1cm) arc (180:0:1cm) ;
\draw [bend left=40] (0,1) to (-1,0);
\filldraw (0,0) circle [radius=0.1];
\draw (-1.25,0) -- (1,0);
\vsq at (1,0) {};
}&$L(1,1)L(2,1+\tfrac12\epsilon)C_3+\Jcal_1L(1,1)C_2$\nn 
$\Jcal_{3,2\gamma}$&
&\tikz[scale=0.6,baseline=(vert_cent.base),scale=1]{
\node at (0,-1.1) () {};
\draw (-0.5,0) circle [radius=0.5];
\draw (0.5,0) circle [radius=0.5];
\draw (-1.25,0) -- (-1,0);
\vsq at (1,0) {};
\filldraw (-0.5,0.5) circle [radius=0.1];
}&$L(1,1)L(1,2)C_3+\Jcal_1L(1,1)C_2$\nn
\hline
$\Jcal_{3,3}$&\tikz[scale=0.6,baseline=(vert_cent.base),scale=1]{
	%\draw[help lines] (-3,-3) grid (3,3);
	\node (vert_cent) {\hspace{-13pt}$\phantom{-}$};
	\draw (0,0) circle[radius=1cm];
         \draw [bend right=40] (0,1) to (0,-1);
           \draw [bend left=40] (0,1) to (0,-1);
\filldraw (180:1) circle [radius=0.1];
\filldraw (0:1) circle [radius=0.1];
}&\tikz[scale=0.6,baseline=(vert_cent.base),scale=1]{
	%\draw[help lines] (-3,-3) grid (3,3);
	\node (vert_cent) {\hspace{-13pt}$\phantom{-}$};
\node at (0,1.1) () {};
\node at (0,-1.1) () {};
	\draw (0,0) circle[radius=1cm];
\draw (-1.2,0) -- (1,0);
\filldraw (0,0) circle [radius=0.1];
\filldraw (0,1) circle [radius=0.1];
\vsq at (1,0) {};
}&$L(1,2)L(2,1+\tfrac12\epsilon)C_3+2\Jcal_1L(1,2)C_2+(\Jcal_1)^2C_1$\nn
\hline
$\Jcal_{4,1\alpha}$&\tikz[scale=0.6,baseline=(vert_cent.base),scale=1]{
	%\draw[help lines] (-3,-3) grid (3,3);
	\node (vert_cent) {\hspace{-13pt}$\phantom{-}$};
	\draw (0,0) circle[radius=1cm];
           \draw (0,0) -- (45:1);
           \draw (0,0) -- (135:1);
	\draw [bend right=40] (0,0) to (0,-1);
           \draw [bend left=40] (0,0) to (0,-1);
\filldraw (0,1) circle [radius=0.1];
}&\tikz[scale=0.6,baseline=(vert_cent.base),scale=1]{
	%\draw[help lines] (-3,-3) grid (3,3);
	\node (vert_cent) {\hspace{-13pt}$\phantom{-}$};
\node at (0,1.1) () {};
	\draw (0,0) circle[radius=1cm];
           \draw [bend left=40] (1,0) to (0,1);
\draw (0,-1) -- (0,1);
\draw (-1.25,0) -- (-1,0);
\vsq at (1,0) {};
\filldraw (225:1) circle [radius=0.1];
}&$L(1,1)L(1,2,\tfrac12\epsilon,1,1)C_4+\Jcal_1L(1,1)L(2,\tfrac12\epsilon)C_3+\Jcal_2L(1,1)C_2$\nn
$\Jcal_{4,1\beta}$&&\tikz[scale=0.6,baseline=(vert_cent.base),scale=1]{
	%\draw[help lines] (-3,-3) grid (3,3);
	\node (vert_cent) {\hspace{-13pt}$\phantom{-}$};
\node at (0,1.1) () {};
	\draw (0,0) circle[radius=1cm];
\draw (-1.25,0) -- (1,0);
\vsq at (1,0) {};
\draw (0,0) -- (0,-1);
\filldraw (0,-0.5) circle [radius=0.1];
}&$L(1,2+\epsilon)L(1,1,1,1,2)C_4+\Jcal_1L(1,2)L(2,1+\tfrac12\epsilon)C_3+2\Jcal_2L(1,2)C_2$\nn
&&&$+\Jcal_{3,1}C_1$\\
\hline
$\Jcal_{4,2\alpha}$&\tikz[scale=0.6,baseline=(vert_cent.base),scale=1]{
	%\draw[help lines] (-3,-3) grid (3,3);
	\node (vert_cent) {\hspace{-13pt}$\phantom{-}$};
	\draw (0,0) circle[radius=1cm];
           \draw (0,0) -- (45:1);
           \draw (0,0) -- (135:1);
	\draw [bend right=40] (0,0) to (0,-1);
           \draw [bend left=40] (0,0) to (0,-1);
\filldraw (-0.5,0.5) circle [radius=0.1];
}&\tikz[scale=0.6,baseline=(vert_cent.base),scale=1]{
	%\draw[help lines] (-3,-3) grid (3,3);
	\node (vert_cent) {\hspace{-13pt}$\phantom{-}$};
\node at (0,1.1) () {};
	\draw (0,0) circle[radius=1cm];
           \draw [bend right=40] (-1,0) to (0,1);
\draw (0,-1) -- (0,1);
\draw (-1.25,0) -- (-1,0);
\vsq at (1,0) {};
\filldraw (225:1) circle [radius=0.1];
}&$L(1,1)L(2,\tfrac12\epsilon,1,1,1)C_4+\Jcal_1L(1,1)L(1,2)C_3+\Jcal_2L(1,1)C_2$\nn
$\Jcal_{4,2\beta}$&&\tikz[scale=0.6,baseline=(vert_cent.base),scale=1]{
	%\draw[help lines] (-3,-3) grid (3,3);
	\node (vert_cent) {\hspace{-13pt}$\phantom{-}$};
\node at (0,1.1) () {};
	\draw (0,0) circle[radius=1cm];
           \draw [bend right=40] (-1,0) to (0,1);
\draw (0,-1) -- (0,1);
\draw (-1.25,0) -- (-1,0);
\vsq at (1,0) {};
\filldraw (0,0) circle [radius=0.1];
}&$L(1,1)L(\tfrac12\epsilon,1,1,1,2)C_4+\Jcal_1L(1,1)L(2,1+\tfrac12\epsilon)C_3+\Jcal_2L(1,1)C_2$\\
$\Jcal_{4,2\gamma_{\alpha}}$&&\tikz[scale=0.6,baseline=(vert_cent.base),scale=1]{
	%\draw[help lines] (-3,-3) grid (3,3);
	\node (vert_cent) {\hspace{-13pt}$\phantom{-}$};
\node at (0,1.1) () {};
	\draw (0,0) circle[radius=1cm];
\draw (-1.25,0) -- (1,0);
\vsq at (1,0) {};
\draw (0,0) -- (0,-1);
\filldraw (-135:1) circle [radius=0.1];
}&$L(1,2+\epsilon)L(1,2,1,1,1)C_4+\Jcal_1L(1,2)L(1,2+\tfrac12\epsilon)C_3+\Jcal_2L(1,2)C_2$\\
&&&$+\Jcal_{3,2\alpha}C_1$\\
$\Jcal_{4,2\delta}$&&\tikz[scale=0.6,baseline=(vert_cent.base),scale=1]{
	%\draw[help lines] (-3,-3) grid (3,3);
	\node (vert_cent) {\hspace{-13pt}$\phantom{-}$};
\node at (0,-1.1) () {};
	\draw (0,0) circle[radius=1cm];
           \draw [bend right=40] (1,0) to (0,-1);
\draw (0,-1) -- (0,1);
\draw (-1.25,0) -- (-1,0);
\vsq at (1,0) {};
\filldraw (225:1) circle [radius=0.1];
}&$L(1,1)L(1,2,1,\tfrac12\epsilon,1)C_4+\Jcal_1L(1,1)L(1,1+\tfrac12\epsilon)C_3$\\
\hline

\end{tabular}
\caption{\label{IRnew}Alternative results for IR counterterms (severing single propagators)}
\end{table}
\begin{table}
\begin{tabular}{|c|c|c|c|}
\hline
Label&CVG&$G^{\rm{sev}}$&IR counterterm\nn
\hline
$\Jcal_{4,3\alpha}$&\tikz[scale=0.6,baseline=(vert_cent.base),scale=1]{
	%\draw[help lines] (-3,-3) grid (3,3);
	\node (vert_cent) {\hspace{-13pt}$\phantom{-}$};
	\draw (0,0) circle[radius=1cm];
           \draw (0,0) -- (45:1);
           \draw (0,0) -- (135:1);
	\draw [bend right=40] (0,0) to (0,-1);
           \draw [bend left=40] (0,0) to (0,-1);
\filldraw (-0.2,-0.5) circle [radius=0.1];
}&\tikz[scale=0.6,baseline=(vert_cent.base),scale=1]{
	%\draw[help lines] (-3,-3) grid (3,3);
	\node (vert_cent) {\hspace{-13pt}$\phantom{-}$};
\node at (0,1.1) () {};
	\draw (0,0) circle[radius=1cm];
\draw (-1.25,0) -- (1,0);
\vsq at (1,0) {};
\draw (0,0) -- (0,-1);
\filldraw (0,1) circle [radius=0.1];
}&$L(2,1+\epsilon)L(1,1,1,1,1)C_4+\Jcal_1L(1,1,1,1,1)C_3$\nn
$\Jcal_{4,3\beta}$&&\tikz[scale=0.6,baseline=(vert_cent.base),scale=1]{
	%\draw[help lines] (-3,-3) grid (3,3);
	\node (vert_cent) {\hspace{-13pt}$\phantom{-}$};
\node at (0,1.1) () {};
	\draw (0,0) circle[radius=1cm];
           \draw [bend left=40] (1,0) to (0,1);
\draw (0,-1) -- (0,1);
\draw (-1.25,0) -- (-1,0);
\vsq at (1,0) {};
\filldraw (45:1) circle [radius=0.1];
}&$L(1,2)L(1+\tfrac12\epsilon,1,1,1,1)C_4+\Jcal_1L(1,1,1,1,1)C_3$\\
$\Jcal_{4,3\gamma}$&&\tikz[scale=0.6,baseline=(vert_cent.base),scale=1]{
	%\draw[help lines] (-3,-3) grid (3,3);
	\node (vert_cent) {\hspace{-13pt}$\phantom{-}$};
\node at (0,1.1) () {};
\node at (0,-1.1) () {};
	\draw (0,0) circle[radius=1cm];
         \draw [bend right=40] (0,1) to (0,-1);
           \draw [bend left=40] (0,1) to (0,-1);
\draw (-1.25,0) -- (-1,0);
\vsq at (1,0) {};
\filldraw (-0.4,0) circle [radius=0.1];
}&$L(1,2)L(1,1,1,1,1+\tfrac12\epsilon)C_4+\Jcal_1L(1,1,1,1,1)C_3$\\
\hline
$\Jcal_{4,4\alpha}$&\tikz[scale=0.6,baseline=(vert_cent.base),scale=1]{
	%\draw[help lines] (-3,-3) grid (3,3);
	\node (vert_cent) {\hspace{-13pt}$\phantom{-}$};
	\draw (0,0) circle[radius=1cm];
           \draw [bend left=40] (1,0) to (0,1);
\draw [bend right=40] (-1,0) to (0,1);
\draw (-1.25,0) -- (1,0);
\vsq at (1,0) {};
\filldraw (135:1) circle [radius=0.1];
\filldraw (45:1) circle [radius=0.1];
}&
\tikz[scale=0.6,baseline=(vert_cent.base),scale=1]{
\node at (0,1.1) () {};
\draw (180:1cm) arc (180:0:1cm) ;
\draw [bend right=40] (0,1) to (1,0);
\draw [bend left=40] (0,1) to (-1,0);
\filldraw (135:1) circle [radius=0.1];
\filldraw (45:1) circle [radius=0.1];
\draw (-1.25,0) -- (1,0);
\vsq at (1,0) {};
}&$L(1,2^2)L(1,2+\epsilon)C_4+2\Jcal_1L(1,2)L(1,2+\tfrac12\epsilon)C_3+(\Jcal_1)^2L(1,2)C_2$\\
&&&$+\Jcal_{3,3}C_1$\\
$\Jcal_{4,4\beta}$&&
\tikz[scale=0.6,baseline=(vert_cent.base),scale=1]{
\draw (180:1cm) arc (180:0:1cm) ;
\draw [bend right=40] (0,1) to (1,0);
\draw [bend left=40] (0,1) to (-1,0);
\filldraw (135:1) circle [radius=0.1];
\filldraw (0,0) circle [radius=0.1];
\draw (-1.25,0) -- (1,0);
\vsq at (1,0) {};
}&$L(1,1)L(1,2)L(2,1+\epsilon)C_4+\Jcal_1L(1,1)L(2,1+\tfrac12\epsilon)C_3$\\
&&&$+\Jcal_1L(1,1)L(1,2)C_3+(\Jcal_1)^2L(1,1)C_2$\\
\hline$\Jcal_{4,23\alpha}$&\tikz[scale=0.6,baseline=(vert_cent.base),scale=1]{
	%\draw[help lines] (-3,-3) grid (3,3);
	\node (vert_cent){\hspace{-13pt}$\phantom{-}$};
	\draw (0,0) circle [radius=1cm];
\draw [bend left=40] (-1,0) to (1,0);
\draw [bend left=40] (0,0) to (0,-1);
\draw [bend right=40] (0,0) to (0,-1);
\draw (-1.25,0) --(-1,0) (0,0) -- (1,0);
\vsq at (1,0) {};
\filldraw (-135:1) circle [radius=0.1];
}&\tikz[scale=0.6,baseline=(vert_cent.base),scale=1]{
\node at (0,1.1) () {};
\draw (180:1cm) arc (180:0:1cm) ;
\draw [bend left=40] (0,1) to (-1,0);
\draw [bend left=40] (0,1) to (135:1);
\filldraw (157:1) circle [radius=0.1];
\draw (-1.25,0) -- (1,0);
\vsq at (1,0) {};
}&$L(1,1)L(1,2+\tfrac12\epsilon)L(1,2+\epsilon)C_4+\Jcal_2L(1,2)C_2+\Jcal_{3,1}C_1$\\
$\Jcal_{4,23\beta}$&
&\tikz[scale=0.6,baseline=(vert_cent.base),scale=1]{
\draw (180:1cm) arc (180:0:1cm) ;
\draw [bend left=40] (0,1) to (-1,0);
\draw [bend right=40] (1,0) to (0,0);
\filldraw (45:1) circle [radius=0.1];
\draw (-1.25,0) -- (1,0);
\vsq at (1,0) {};
}&$L(1,1)^2L(1+\tfrac12\epsilon,2+\tfrac12\epsilon)C_4+\Jcal_2C_2$\\
$\Jcal_{4,23\gamma}$&
&\tikz[scale=0.6,baseline=(vert_cent.base),scale=1]{
\draw (-0.5,0) circle [radius=0.5];
\draw (0.5,0) circle [radius=0.5];
\draw (-1.25,0) -- (-1,0);
\vsq at (1,0) {};
\draw [bend left=40] (1,0) to (0.5,0.5);
\filldraw (0.15,0.35) circle [radius=0.1];
}&$L(1,1)^2L(1,2+\tfrac12\epsilon)C_4+\Jcal_2C_2$\\
$\Jcal_{4,23\delta_{\alpha}}$&
&\tikz[scale=0.6,baseline=(vert_cent.base),scale=1]{
\node at (0,1.1) () {};
\draw (180:1cm) arc (180:0:1cm) ;
\draw [bend left=40] (0,1) to (-1,0);
\draw [bend left=40] (0,1) to (135:1);
\filldraw (45:1) circle [radius=0.1];
\draw (-1.25,0) -- (1,0);
\vsq at (1,0) {};
}&$L(1,1)L(1,1+\tfrac12\epsilon)L(1,2+\epsilon)C_4+\Jcal_{3,2\alpha}C_1$\\
\hline
\end{tabular}
\caption{\label{IRnewa}Alternative results for IR counterterms (severing single propagators) continued}
\end{table}

\begin{subequations}\label{listdtil}
\begin{align}
\dtil_2=&\,\tfrac12+\left(\tfrac14+\tfrac12\zeta_3\right)\epsilon+\ldots,\\
\dtil_{3,1}=&\,\tfrac{2}{\epsilon}+\tfrac43+\tfrac83\zeta_3+\ldots,\\
\dtil_{3,2\alpha}=&\,\tfrac{1}{\epsilon}+\tfrac12+\tfrac13\zeta_3+\ldots,\\
\dtil_{3,2\beta}=&\,\tfrac{1}{\epsilon}+\tfrac52+\tfrac73\zeta_3+\ldots,\\
\dtil_{3,2\gamma}=&\,\tfrac{1}{\epsilon}+\tfrac52+\tfrac13\zeta_3+\ldots,\\
\dtil_{3,3}=&\,\tfrac13-\tfrac43\zeta_3+\ldots,\\
\dtil_{4,1\alpha}=\dtil_{4,1\beta}+O(1)=&\,\tfrac{2}{\epsilon^2}+\left(\tfrac83+\tfrac{10}{3}\zeta_3\right)\tfrac{1}{\epsilon}+\ldots,\\
\dtil_{4,2\alpha}=\dtil_{4,2\beta}+O(1)=\dtil_{4,2\gamma_1}+O(1)=&\,\tfrac{1}{\epsilon^2}+\left(1-\tfrac{1}{3}\zeta_3\right)\tfrac{1}{\epsilon}+\ldots,\label{listdtil:h}\\
\dtil_{4,2\gamma_2}=\dtil_{4,2\delta}+O(1)=&\,\tfrac{1}{\epsilon^2}+\left(5+\tfrac{11}{3}\zeta_3\right)\tfrac{1}{\epsilon}+\ldots,\\
\dtil_{4,2\gamma_3}=&\,\tfrac{1}{\epsilon^2}+\left(5-\tfrac{1}{3}\zeta_3\right)\tfrac{1}{\epsilon}+\ldots,\label{listdtil:j}\\
\dtil_{4,3\alpha}=\dtil_{4,3\beta}+O(1)=&\,O(1),\\
\dtil_{4,4\alpha}=\dtil_{4,4\beta}+O(1)=&\,\left(\tfrac23-\tfrac{8}{3}\zeta_3\right)\tfrac{1}{\epsilon}+\ldots,\label{listdtil:l}\\
\dtil_{4,23\alpha}=\dtil_{4,23\delta_1}+O(1)=&\,\tfrac{3}{\epsilon^2}+\left(\tfrac83+3\zeta_3\right)\tfrac{1}{\epsilon}+\ldots,\\
\dtil_{4,23\beta}=\dtil_{4,23\delta_2}+O(1)=&\,\tfrac{3}{\epsilon^2}+\left(\tfrac{10}{3}+7\zeta_3\right)\tfrac{1}{\epsilon}+\ldots,\\
\dtil_{4,23\gamma}=\dtil_{4,23\delta_3}+O(1)=&\,\tfrac{3}{\epsilon^2}+\left(\tfrac{20}{3}+3\zeta_3\right)\tfrac{1}{\epsilon}+\ldots,\\
\end{align}
\end{subequations}
We have already implied that Eq.~\eqref{del2a} generalises to the current situation, and this is manifest in Eq.~\eqref{listdtil}. We also see that $\dtil_{4,3\alpha}=\dtil_{4,3\beta}=O(1)$ which immediately verifies the analogue of Eq.~\eqref{delrel1a}. We now turn to Eq.~\eqref{delrel1}. This is a simple case where every single propagator is equivalent.
We find that $\dtil_2$ and $\dtil_{3,1}$ are the same at leading order as the corresponding $\delta^{IR}_2$ and $\delta_{3,1}^{IR}$ given in Eq.~\eqref{irdiffs}, and this is sufficient to guarantee
\be
\dtil_{3,1}+2\Delta_{1,1}\dtil_{2}=O(1),
\label{del31a}
\ee
which is the counterpart of Eq.~\eqref{delrel1}. The case of $\dtil_{3,2}$, and therefore the analogue of Eq.~\eqref{delrel:a}, is slightly more complicated. Here there are three inequivalent ways of severing a single propagator. However, the corresponding $\dtil_{3,2\alpha}$, $\dtil_{3,2\beta}$ and $\dtil_{3,2\gamma}$ can be seen to be the same at leading order as $\delta_{3,2}$ and hence we have
\be
\dtil_{3,2\alpha}+\Delta_{1,1}\dtil_{2}=O(1),
\ee
and similarly for $\dtil_{3,2\beta}$ and $\dtil_{3,2\gamma}$, which is the counterpart of Eq.~\eqref{delrel:a}.

Now we turn to the analogue of Eq.~\eqref{delrel2}. Here there is only one inequivalent choice of single propagator to sever in $G_{3,1}$, but two in $G_{4,1}$. Ar first sight this seems likely to lead to a problem, but then we see that $\dtil_{4,1\alpha}=\dtil_{4,1\beta}$ at leading order (LO) and next-to-leading order (NLO) and in fact we find that  
\be
\dtil_{4,1\phi}+\Delta_{1,1}\dtil_{3,1}+2\Delta_{2,2}\dtil_{2}=O(1),
\label{delrel2a}
\ee
for $\phi\in\{\alpha,\beta\}$, which is the counterpart of Eq.~\eqref{delrel2}.
Next we examine the analogue of Eq.~\eqref{delrel:b}. Here there are yet more subtleties. Now there are four inequivalent ways of severing a propagator in $G_{4,2}$ but only three inequivalent ways in $G_{3,2}$. Again this seems likely to lead to a contradiction, but in fact we see in Eq.~\eqref{listdtil:h} - \eqref{listdtil:j} that (at LO and NLO) there are only three different inequivalent results $\dtil_{4,2\alpha}$ etc, which potentially can match the three inequivalent results for $\dtil_{3,2\alpha}$. A further subtlety within this is that the result for $\Jcal_{4,2\gamma}$, and therefore for $\dtil_{4,2\gamma}$, depends on what we choose for the three-loop IR counterterm corresponding to $G_{3,2}$ - as we see in Table~\ref{IRnew} where we have chosen $\Jcal_{3,2\alpha}$. If we make this choice of $\Jcal_{3,2\alpha}$ we get the same result at LO and NLO for $\Jcal_{4,2}$ as for $\Jcal_{4,2\alpha}$ and $\Jcal_{4,2\beta}$; we have denoted this by $\Jcal_{4,2\gamma_{\alpha}}$ . If we choose $\Jcal_{3,2\beta}$ for the three-loop IR counterterm in $\Jcal_{4,2\gamma}$, we get the same result at LO and NLO for $\Jcal_{4,2}$ as for $\Jcal_{4,2\delta}$ (we denote this version of $\Jcal_{4,2\gamma}$ by $\Jcal_{4,2\gamma_{\beta}})$. Finally, if we choose $\Jcal_{3,2\gamma}$ we get a different result from either of these, even at NLO; we call this $\Jcal_{4,2\gamma_{\gamma}}$. Then it seems natural when evaluating the analogue of Eq.~\eqref{delrel:b} to combine the results for $\dtil_{4,2}$ from the 1st group of choices $\{\alpha,\beta,\gamma_{\alpha}\}$, with $\dtil_{3,2\alpha}$; the results for the 2nd group $\{\gamma_{\beta},\delta\}$ with $\dtil_{3,2\beta}$; and the result for $\gamma_{\gamma}$ with $\dtil_{3,2\gamma}$. When we do this, we find that the  analogue of Eq.~\eqref{delrel:b} is valid in every case. In summary, we have
\be
\dtil_{4,2\phi}+\Delta_{1,1}\dtil_{3,2\psi}+\Delta_{2,2}\dtil_{2}=O(1),
\label{delrelaa}
\ee
when either $\phi\in\{\alpha,\beta,\gamma_{\alpha}\}$ and $\psi=\alpha$; or $\phi\in\{\gamma_{\beta},\delta\}$ and $\psi=\beta$; or $\phi=\gamma_{\gamma}$ and $\psi=\gamma$.

Now looking at Eq.~\eqref{delrel3}, we see that there are two inequivalent ways to sever a single propagator in $G_{4,4}$ but only one way for $G_{3,3}$. However, as we see in Eqs.~\eqref{listdtil:l}, $\Jcal_{4,4\alpha}$ and $\Jcal_{4,4\beta}$ agree at LO and NLO, and we have 
\be
\dtil_{4,4\phi}+\Delta_{1,1}\dtil_{3,3}=O(1),
\label{delrel3a}
\ee
for $\phi\in\{\alpha,\beta\}$. This completes our check that Eqs.~\eqref{delrel1a}, \eqref{delrel1}, \eqref{delrel}, \eqref{delrel2}, \eqref{delrel3} are still satisfied in this more general setting where we can sever any propagator in the CVG.

An interesting question is whether in equations like Eq.~\eqref{delrelaa}, there is a simple way of determining in advance which values of $\psi$ correspond to which values of $\phi$. In fact, if one considers the diagrams for $\Jcal_{4,2\alpha}$ and $\Jcal_{4,2\beta}$ and shrinks the one-loop bubble to a point, one obtains the diagram for $\Jcal_{3,2\alpha}$. If one considers the diagram for $\Jcal_{4,2\delta}$  and shrinks the one-loop bubble to a point, one obtains the diagram for $\Jcal_{3,2\beta}$. In each case this association of a $\Jcal_{4,2\phi}$ and a $\Jcal_{3,2\psi}$ leads to the correct pair of $\dtil_{4,2\phi}$ and $\dtil_{3,2\psi}$ to ensure validity of Eq.~\eqref{delrelaa}. On the other hand, in the case of $\Jcal_{4,2\gamma}$, there is no one-loop bubble available for shrinking; but there is a choice of $\Jcal_{3,2\psi}$ for the 3-loop IR counterterm, and that choice of $\psi$ is then the  correct choice for $\dtil_{3,2\psi}$ to satisfy Eq.~\eqref{delrelaa}. The obvious question is whether there always some such rule to determine the appropriate single propagators which should be severed to satisfy relations like Eq.~\eqref{delrelaa}. We shall present one more example which shows a similar rule in operation. The CVG is shown in Table~\ref{IRnewa}. We refer to it as $G_{4,23}$ in accord with the notation of Ref.~\cite{klein}. The relation to be satisfied by the $\delta_{4,23}^{IR}$ (i.e. using the AIR counterterms obtained by severing the double propagator) may be obtained in the usual manner and is
\be
\delta^{IR}_{4,23}+\Delta_{1,1}\delta^{IR}_{3,1}+\Delta_{1,1}\delta^{IR}_{3,2}+\Delta_{1,1}^2\delta^{IR}_2=O(1).
\label{delrel23a}
\ee
However, there are four inequivalent ways of severing a single propagator in $G_{4,23}$, labelled by $\alpha$-$\delta$ and depicted in Table~\ref{IRnewa}; and as we know, three inequivalent ways of severing a single propagator in $G_{3,2}$. The relation to be satisfied by the $\dtil_{4,23\alpha}$ etc may be deduced as in the case of Eq.~\eqref{delrelaa} and is 
\be
\dtil_{4,23\phi}+\Delta_{1,1}\dtil_{3,1}+\Delta_{1,1}\dtil_{3,2\psi}+\Delta_{1,1}^2\dtil_2=O(1).
\label{delrel23}
\ee
This is valid when either $\phi\in\{\alpha,\delta_{\alpha}\}$ and $\psi=\alpha$; or $\phi\in\{\beta,\delta_{\beta}\}$ and $\psi=\beta$; or $\phi\in\{\gamma,\delta_{\gamma}$ and $\psi=\gamma$. Here again the subscript on $\delta$ reflects the possible choices of three-loop IR counterterm corresponding to $G_{3,2}$ - as we see in Table~\ref{IRnewa} where we have chosen $\Jcal_{3,2\alpha}$.
We have of course chosen our notation for the greek subscripts in $\delta_{4,23\phi}$ {\it post hoc} after an explicit computation, in order to match those in the corresponding $\delta_{3,2\psi}$; but again we ask if this matching may be determined in advance.
In fact, we see that if we shrink the one-loop bubble in $G_{4,23\alpha}$ we obtain $G_{3,2\alpha}$; if we shrink one of the one-loop bubbles in $G_{4,23\beta}$ we obtain $G_{3,2\beta}$, while if we shrink the other we get $G_{3,1}$; and if we shrink one of the one-loop bubbles in $G_{4,23\gamma}$ we obtain $G_{3,2\gamma}$, while if we shrink the other we get $G_{3,1}$. In each of these cases there is a clear indication that in Eq.~\eqref{delrel23} we should choose $\phi=\psi$ for $\phi\in\{\alpha,\beta,\gamma\}$. However, if we shrink the one-loop bubble in $G_{4,23\delta}$ we only obtain $G_{3,1}$, rather than one of the $G_{3,2\phi}$. But $\Jcal_{4,23\delta}$ is the only one of the four choices of $\alpha-\delta$ where there is a choice of $\Jcal_{3,2\psi}$ for the 3-loop IR counterterm, and that choice of $\psi$ is then once again, as in the case of $\dtil_{4,2}$, the  correct choice for $\dtil_{3,2\psi}$ to satisfy Eq.~\eqref{delrel23}. Therefore in every case there is a natural way to predict the matching of $\phi$ with $\psi$ which we noted in Eq.~\eqref{delrel23}. It is not yet clear whether there will always be a similar way to determine which combinations of choices of severed propagators are compatible in relations such as Eq.~\eqref{delrelaa}, but there appear to be grounds for optimism.

We might mention here that so far in cases of two double propagators such as $\Ical_{3,3a}$ and $\Ical_{4,4a}$ etc, the two double propagators appeared symmetrically in the CVG (as may be seen in Tables~\ref{IRres3} and \ref{IRres4}), so that the same result was obtained whichever was severed. However, in general the situation might be similar to the case of several single propagators; individual results for $\dtil$ quantities might depend on which of the double propagators is severed, but there might be a natural way of choosing the severed propagators at each loop order which ensures that relations like Eq.~\eqref{delrelaa} are satisfied. Once again, this needs further investigation.

\section{UV Counterterm results}
In this Appendix we summarise known results for one, two, three and four loop counterterms in scalar $\phi^4$ theory which have been used in Appendix B. Again for ease of comparison we adopt the labelling of Ref.~\cite{klein}, but the four-loop results were first derived in Ref.~\cite{kaz}.
\begin{subequations}\label{divs}
\begin{align}
\Delta_{1,1}=&\,-\tfrac{2}{\epsilon},\label{divs:a}\\
\Delta_{2,2}=&\,\tfrac{1}{\epsilon^2}(2-\epsilon),\label{divs:b}\\
\Delta_{3,5}=\Delta_{3,6}=&\,\tfrac{1}{\epsilon^3}\left(-\tfrac83+\tfrac43\epsilon+\tfrac23\epsilon^2\right),\label{divs:c}\\
\Delta_{3,7}=&\,\tfrac{1}{\epsilon^3}\left(-\tfrac43+2\epsilon-\tfrac43\epsilon^2\right),\\
\Delta_{4,8}=&\,\tfrac{1}{\epsilon^4}\left(\tfrac43-\tfrac{10}{3}\epsilon+\tfrac{13}{3}\epsilon^2-\left(\tfrac{11}{6}-\zeta_3\right)\epsilon^3\right),\\
\Delta_{4,9}=&\,\tfrac{1}{\epsilon^4}\left(\tfrac{10}{3}-\tfrac{10}{3}\epsilon-\tfrac16\epsilon^2+\tfrac12\epsilon^3\right),\\
\Delta_{4,14}=&\,\tfrac{1}{\epsilon^4}\left(\tfrac83-\tfrac83\epsilon-\tfrac43\epsilon^2-(2\zeta_3-2)\epsilon^3\right),\\
\Delta_{4,18}=&\,\tfrac{1}{\epsilon^4}\left(4-2\epsilon-\epsilon^2
-\left[\tfrac12-\zeta_3\right]\epsilon^3\right),\label{divs:h}\\
\Delta_{4,22}=&\,\tfrac{1}{\epsilon^4}\left(\tfrac43-2\epsilon+\tfrac13\epsilon^2+\tfrac56\epsilon^3\right),\\
\Delta_{4,23}=&\,\tfrac{1}{\epsilon^4}\left(2-\tfrac83\epsilon+\tfrac56\epsilon^2+\tfrac23\epsilon^3\right),\\
\Delta_{4,24}=&\,\tfrac{1}{\epsilon^4}\left(\tfrac43-2\epsilon+\tfrac13\epsilon^2-\left[\zeta_3-\tfrac56\right]\epsilon^3\right),\\
\Delta_{4,25}=&\,\tfrac{1}{\epsilon^4}\left(\tfrac23-2\epsilon+\tfrac{19}{6}\epsilon^2-\left[\tfrac52-2\zeta_3\right]\epsilon^3\right),\label{divs:l}\\
\Delta_{4,26}=&\,\tfrac{1}{\epsilon^4}\left(\tfrac23-2\epsilon+\tfrac{19}{6}\epsilon^2-\tfrac52\epsilon^3\right),
\end{align}
\end{subequations}
In the following equation we have collected useful results for various instances of $L(a_1,a_2,a_3,a_4,a_5)$ (defined in Eq.~\eqref{L4def}), derived using the Integration by Parts technique.
\begin{subequations}
\begin{align}
L(a_1,a_2,1,1,1)=&\,\tfrac{L(1,1)}{d-2-a_1-a_2}\Bigl\{a_1[L(a_1+1,a_2)-L(a_1+1,a_2+\tfrac12\epsilon)]\nn
&+a_2[L(a_1,a_2+1)-L(a_1+\tfrac12\epsilon,a_2+1)]\Bigr\},\\
L(1,2,1,a,1)=&\,\tfrac{1}{d-3-a}\Bigl\{L(1,2)[L(2,a)-L(2,a+1+\tfrac12\epsilon)]\nn
&+a[L(1,2)L(1,1+a)-L(a+1,1,1,1,1)]\Bigr\},\\
L(1,2,a,1,1)=&\,\tfrac{1}{d-3-a}\Bigl\{aL(1,2)[L(1,1+a)-L(2+\tfrac12\epsilon,1+a)]\nn
&+L(1,2)L(2,a)-L(a,2,1,1,1)\Bigr\},\\
L(a,1,1,1,2)=&\,\tfrac{1}{d-5-a}\Bigl\{a[L(a+1,1,1,1,1)-L(1,2)L(2+\tfrac12\epsilon,a+1)]\nn
&+L(a,2,1,1,1)-L(1,2)L(2,a+1+\tfrac12\epsilon)\Bigr\},\\
L(1,1,1,1,2)=&\,\tfrac{2}{d-6}[L(2,1,1,1,1)-L(1,2)L(2,2+\tfrac12\epsilon)].
\label{L4res}
\end{align}
\end{subequations}

\end{document}